\documentclass{jfm}

\usepackage{amsmath,amssymb}
\usepackage{graphicx}
\usepackage{natbib}
\usepackage{epstopdf,textcomp} 
\usepackage{color} 

\ifCUPmtlplainloaded \else
  \checkfont{eurm10}
  \iffontfound
    \IfFileExists{upmath.sty}
      {\typeout{^^JFound AMS Euler Roman fonts on the system,
                   using the 'upmath' package.^^J}%
       \usepackage{upmath}}
      {\typeout{^^JFound AMS Euler Roman fonts on the system, but you
                   dont seem to have the}%
       \typeout{'upmath' package installed. JFM.cls can take advantage
                 of these fonts,^^Jif you use 'upmath' package.^^J}%
      }
  \else
  \fi
\fi

\ifCUPmtlplainloaded \else
  \checkfont{msam10}
  \iffontfound
    \IfFileExists{amssymb.sty}
      {\typeout{^^JFound AMS Symbol fonts on the system, using the
                'amssymb' package.^^J}%
       \usepackage{amssymb}%
         \let\leq=\leqslant
         
      }{}
  \fi
\fi

\ifCUPmtlplainloaded \else
  \IfFileExists{amsbsy.sty}
    {\typeout{^^JFound the 'amsbsy' package on the system, using it.^^J}%
     \usepackage{amsbsy}}
    {\providecommand\boldsymbol[1]{\mbox{\boldmath $##1$}}}
\fi

\providecommand\bnabla{\boldsymbol{\nabla}}
\providecommand\bcdot{\boldsymbol{\cdot}}

\newcommand\Rey{\mbox{\textit{Re}}}  
\newcommand\Pen{\mbox{\textit{Pe}}}  

\newsavebox{\astrutbox}
\sbox{\astrutbox}{\rule[-5pt]{0pt}{20pt}}

\title[Flow and streaming potential, non-isothermal]{Flow and streaming potential of an electrolyte in a channel with an axial temperature gradient}

\author[M. Dietzel and S. Hardt]%
{M\ls A\ls T\ls H\ls I\ls A\ls S\ns D\ls I\ls E\ls T\ls Z\ls E\ls L%
  \thanks{Email address for correspondence: mdietzel.edu@e.mail.de}\ns
\and S\ls T\ls E\ls F\ls F\ls E\ls N\ns H\ls A\ls R\ls D\ls T}

\affiliation{Center of Smart Interfaces, TU Darmstadt, Alarich-Weiss-Strasse 10, 64287 Darmstadt, Germany\\[\affilskip]}

\pubyear{2010}
\volume{650}
\pagerange{119--126}
\date{?; revised ?; accepted ?. - To be entered by editorial office}
\begin{document}

\maketitle

\begin{abstract}
The effect of an axial temperature gradient on the flow profile and the induced streaming potential of a pressure-driven 
symmetric electrolyte in a slit channel is investigated. Based on the non-isothermal Nernst-Planck equations as well 
as the Poisson equation in the lubrication approximation, expressions for the ion distribution in the electric double 
layer (EDL) are derived. It is found that thermophoretic ion motion and a temperature-dependent electrophoretic 
ion mobility increase the local EDL thickness with temperature, whereas a temperature-dependent permittivity 
shrinks the EDL. Within the Debye-H\"uckel approximation, the Navier-Stokes equation with the corresponding 
electric body force terms is solved. Analytical expressions for the flow profile and the induced (streaming) field 
under non-isothermal conditions are derived. It is shown that for such a situation the induced electric field is the 
linear superposition of at least seven individual contributions. For very wide channels, only the thermoelectric field 
typically present in bulk electrolytes when subjected to a temperature gradient (Soret equilibrium) as well as the 
conventional pressure-induced streaming field are of importance. Counterintuitively, for the latter, while still being 
affected by the temperature dependence of the dielectric permittivity and local salt concentration, the 
temperature dependencies of the viscosity, Fickian diffusion coefficients and ion electro-mobilities exactly cancel each other. 
For narrow channels, five additional contributions become relevant, which -similar to the Soret voltage- do not vanish in 
the case that the externally applied pressure gradient is removed. The first is caused by selective 
thermo-electro-migration driven by the interplay between the temperature-dependent electrophoretic ion 
mobility and the interaction of the ions with the surface wall charge. This non-advective effect is at its maximum under 
extreme confinement. For channels whose widths are of the same order as the EDL thickness, four thermoosmotic effects 
become significant. Besides the well-known thermoosmosis due to the temperature dependence of the dielectric 
permittivity in the (extended) Korteweg-Helmholtz force, it is demonstrated that -by contrast to isothermal 
conditions- a thermal gradient renders the ion cloud in the EDL out of mechanical equilibrium. In this context 
it is shown that a thermophoretic ion motion (i.e. the intrinsic Soret effect of the ions) and a 
temperature-dependent ion electro-mobility as well as a temperature-dependent permittivity not only cause an axial gradient 
of the EDL potential, but simultaneously lead to a pressure of thermal origin, which sets the fluid into an 
advective motion. Corresponding phenomena were not previously discussed in the literature and may be 
interpreted as an apparent, thermally induced slip velocity within the EDL. Subsequently, the ion advection affiliated 
with such thermoosmotic flow may induce a thermoelectric field of similar order of magnitude as the one caused by more 
conventional thermal effects. \\ \\
This is the post-print authors' version of the manuscript, which was published in the
Journal of Fluid Mechanics. doi:10.1017/jfm.2016.844  \copyright \ Cambridge University Press 2017
\end{abstract}

\begin{keywords}
Authors should not enter keywords on the manuscript, as these must be chosen by the author during the online submission process and will then be added during the typesetting process (see http://journals.cambridge.org/data/\linebreak[3]relatedlink/jfm-\linebreak[3]keywords.pdf for the full list)
\end{keywords}

\section{Introduction}\label{sec:intro}

Over the last couple of decades, electrokinetic flow phenomena have received significant attention by the scientific community. 
Within the general framework of electrohydrodynamics \citep{Castellanos:Springer1998}, the motion of fluids carrying dissolved 
electric charges in an electric field as well as the transport of these charges relative to the carrier fluid is considered. 
Electrokinetics is of crucial importance in the stabilization and motion of particles in colloidal suspensions 
\citep{Russel:CambrUnivPress1989} and is relevant in electrospray-based fabrication methods \citep{Salata:CurrNanoscie2005} 
or DNA-manipulation/separation techniques \citep{Viovy:RevModPhys2000}, to name a few. Ion transport in dilute electrolytes is commonly 
captured by the Nernst-Planck equation. At moderate to high ion concentrations though, the finite size and the discrete nature 
of the ions have to be taken into account \citep{Nadler:JPhysCondMatter2004}. To capture the momentum transfer between ions 
and the solvent, the conventional stress tensor in the Navier-Stokes equation is supplemented by the Maxwell stresses. Being 
a manifestation of the Onsager reciprocal principle 
\citep{Onsager:PhysRev1931}, the interaction between the dissolved ions and the liquid carrier goes along with two distinct 
types of electrokinetic coupling with single-phase fluids: one where an electric field drives a fluid motion such as in 
electroosmotic flow (EOF), and another where ions advected along with the fluid generate an electric field. While EOF and 
induced-charge EOF is well suited to propel fluid \citep{Stone:AnnRevFluidMech2004,Squires:JFM2004,Kim:ExpFluids2002,Yossifon:POF2006} 
or to enhance mixing in microchannels \citep{Wang:BioSenseElectr2006,Barz:JFM2011}, ion advection is described by the so-called 
streaming potential (SP) \citep{Dukhin:AdvCollIntScie1993}. It is relevant in a number of physical phenomena related to the 
advection of charged interfaces such as in the electro-viscous drag enhancement observed in particle suspensions 
\citep{Sherwood:JFM1980}. In addition, it can be used to convert mechanical (and as will be shown also 
thermal) energy into electric energy \citep{Yang:JMicroMechEng2003}. This is, among others, in the focus of the current study.

Electrokinetic phenomena are commonly associated with the excess of one ion species in the vicinity of an interfacial charge of 
opposite polarity carried by submerged solid bodies or walls. The ions form a diffusion-dominated electric double layer (EDL) 
which screens the surface charge. Unlike the ions in the Stern layer, the ions in the EDL remain mobile and, as in electrokinetic 
streaming applications, can be advected with the flow. Depending on the bulk ion concentration, the EDL is typically only a few 
to a couple of hundred nm thick so that many studies of electrokinetic phenomena do not resolve the EDL, but assume an effective 
slip velocity (Smoluchowski limit). The liquid outside the EDL is irrelevant for the momentum source term driving the 
flow. By contrast, it contributes to the usually undesired ion flux by means of electro-migration (i.e. the bulk conduction current) 
caused by the applied potential difference. Therefore, to minimize the detrimental influence of the bulk fluid, many studies 
on electrokinetics focus on system dimensions of the same order as the EDL 
thickness (the Debye parameter is of order unity) \citep{vdHeyden:PRL2005,Daiguji:ChemSocRev2010,Xie:LabChip2011}. 
Most studies of the electrokinetic streaming potential are concerned with pressure-driven flow, while -in comparison- few investigations 
were performed on shear-driven flow \citep{Soong:JCollIntSci2004} or other sources of fluid propulsion. Owing to 
the superposition principle in Stokes flow, electrokinetic and pressure-driven flow fields can be linearly superimposed. 
As a result, the streaming potential becomes a linear function of the driving pressure difference.

To date, studies on thermal effects in electrokinetic flow are comparably scarce. Nevertheless, within the general scope of recent energy 
sustainability efforts, it is of interest to investigate thermally driven electrokinetic charge separation 
\citep{Grosu:SurfEngApplElectrochem2010} based on using waste heat (generated, for example, by the central processing unit (CPU) of a computer). 
Most of the thermally induced fluid propulsion (by buoyancy, thermocapillarity or evaporation) can be formulated 
-at least in the Stokes limit- as an effective pressure difference or shear force, which can be subsequently combined with the conventional electrokinetic 
theory to estimate the streaming potential generated by a thermally propelled liquid. In this case, a combined study of thermal, 
fluid mechanical and electrokinetic effects does not appear to be necessary. This holds as long as other effects induced by a variation of 
temperature are negligible. Roughly, four different non-isothermal contributions may enter the problem formulation: firstly, most of 
the bulk properties such as viscosity, diffusivities, electric conductivity and permittivity are temperature-dependent. Under the 
application of direct (DC) \citep{Wong:PoF1969} or alternating current (AC) \citep{Gonzalez:JFM2006} voltages this may lead, 
for instance, to electrothermal convection. Secondly, dissipative effects occurring in the bulk, such as viscous dissipation and Joule heating 
\citep{Zhao:JMicroMechEng2002,Maynes:IJHMT2004,Sadeghi:IJHMT2010}, should be included in the energy equation as well. Thirdly, the formation of a wall 
($\zeta$-) potential is strongly dependent on the dissociation processes of surface groups and ion absorption at the wall 
\citep{Revil:JGeoPhysRes1999}. More specifically, the wall potential is determined, at least under quasi-equilibrated conditions, by a 
temperature-dependent equilibrium constant, leading in turn to a temperature-dependent $\zeta$ potential. This becomes particularly 
important for flow through porous media at elevated temperatures, as, for instance, treated in geophysical research studies 
\citep{Ishido:Tectonophys1983,Reppert:JGeoPhysRes2003}. Lastly, similar to the thermal diffusion of 
colloidal particles in a non-isothermal liquid \citep{Piazza:JPhysCM2004,Wuerger:RepProgPhys2010}, also the charge carriers in an 
electrolyte are set into a thermally induced diffusive motion. Under the condition of local charge neutrality, steady-state 
and the absence of any external pressure gradient, this leads to a well-known thermoelectric potential in bulk electrolytes 
\citep{Guthrie:JChemPhys1949}. As briefly summarized in appendix \ref{sec:app_thermodiffpot}, this Seebeck-type of thermoelectric 
potential vanishes if the thermal mobilities of the ion species do not differ from each other. The effect is typically quantified 
in terms of a bulk Soret coefficient, $\sigma_T$, and can be enhanced, for instance, by the utilization of more exotic electrolytes 
\citep{Bonetti:JChemPhys2011}. In the context of such thermodiffusive processes, there has been a long standing interest in ion-selective 
membranes exposed to a temperature gradient \citep{Hill:Nature1957,Gaeta:JPhysChem1992}, which are also relevant for the sensation 
of heat felt by humans \citep{Tyrrell:Nature1954}. This interest was recently renewed in the realm of unconventional thermoelectric 
energy conversion \citep{Sandbakk:JMembrScie2013}. The classical description of membrane potentials is based on 
nonequilibrium thermodynamics and averaged transport numbers \citep{Tasaka:BiophyChem1978,Tasaka:PureApplChem1986}, 
without making specific reference to the ion distribution inside the pore. In fact, virtually all of these studies involving the 
Soret effect assume local charge neutrality. While this is valid in the bulk, the fluid within the EDL is not electroneutral. 
This is particularly relevant for electrokinetic flows through non-isothermal nanochannels, e.g. employed as 
electrochemical thermal energy harvester \citep{Kang:AdvFunctMat2012}. Numerical work on pressure-driven 
electrokinetic flow in a slit microchannel, which was ten times wider than the EDL and exhibited a $\zeta$ potential of 
approximately $50 \cdot 10^{-3} \textrm{V}$, while the wall temperature increased along its length by approximately $17\:\%$, 
indicates that in comparison to isothermal conditions thermodiffusive and electrothermal effects may reduce or enhance the overall 
volumetric flow through the channel \citep{Ghonge:PhysRevE2013}. Recent numerical simulations of a nanochannel of width twice 
the EDL thickness and with a $\zeta$ potential of $50$-$100 \cdot 10^{-3} \textrm{V}$ suggests that viscous dissipation and Joule 
heating play a role only at higher salt concentrations for which -however- no EDL overlap and thus no ion selectivity of the 
nanochannel is present. It was shown that applying a temperature difference between both channel ends modifies the ion selectivity 
of the channel \citep{Wood:JPhysCondMatt2016}.
\\
\indent The present work focuses on the implications and significance of ion motion induced by a gradient in temperature 
in symmetric electrolytes in confined geometry. Of special interest is the comparison between semi-analytical and full numerical models 
to identify the dominant non-isothermal effects within the different regimes of the characteristic parameters, 
especially of the EDL thickness scaled to the slit height (Debye parameter) and the wall $\zeta$ potential. 
As a model system, a pressure-driven slit channel flow of a fully dissociated binary electrolyte, subject to 
a temperature gradient along the channel center plane, is chosen. Unlike the classical treatments of thermoelectricity in bulk 
electrolytes, the condition of local charge neutrality is not enforced. In the course of the derivation, thermophoretic 
ion motion as well as the temperature dependencies of the electrophoretic ion mobility (equally referred to as ion electromobility) and the 
dielectric permittivity of the solvent are shown to be the dominant non-isothermal effects. A modified Boltzmann 
distribution is derived from the scaled, non-isothermal Nernst-Planck equation, taking advantage of the disparate ratio 
between channel height and channel length. This expression for the ion distribution leads to a correlation between 
the local EDL thickness and the local temperature. While a uniform growth of the EDL generally increases the streaming potential, 
the thermal gradient along the channel center plane gives rise not only to a corresponding gradient in the EDL potential but 
-together with the temperature dependence of the ion electromobility and the space charge of the EDL- to a non-advective ion 
transport by means of selective (i.e. polarity-depending) migration. In a recent work focusing on non-advective effects 
\citep{Dietzel:PRL2016}, the latter was shown to be dominant under extreme confinement, inducing for a vanishing Debye 
parameter a thermoelectric field of order $\zeta/T$, with $T$ being the absolute temperature. The current work explicitly 
includes advective effects and the gradient of the EDL potential along the channel (developing due to the combination 
of the relatively weak temperature-induced modification of the EDL thickness and the strong electric field 
within the EDL) leads to a gradient in electrohydrostatic (or, equivalently, in the electroosmotic) pressure 
and an additional axial Maxwell stress, whose superposition triggers an advective ion transport. While thermally 
induced osmotic pressure gradients are known to be responsible for the thermoosmotic propulsion of colloidal particles in 
a thermal gradient \citep{Wuerger:RepProgPhys2010} as well as for the thermoosmotic transport across semi-permeable (but not 
explicitly charged) membranes \citep{Dariel:JPhysChem1975}, thermoosmotically induced electric fields have been attributed 
to the additional term arising in the Korteweg-Helmholtz force when the dielectric permittivity is not constant 
\citep{Derjaguin:NewYork1987}. The accurate description of the temperature-dependent EDL potential and its effect not 
only on the electroosmotic pressure but also on the Maxwell stress has not been achieved in any of these works. Considering 
that at isothermal conditions the electroosmotic ion pressure exactly cancels the Maxwell stress exerted by the EDL potential 
(so that the ion cloud of the EDL is in mechanical equilibrium), at non-uniform temperature the simultaneous consideration 
of the effect of temperature variations on the EDL potential, on the electroosmotic pressure and on the Maxwell stress 
is deemed imperative. \citet{Sasidhar:JCollIntScie1982} investigated a related scenario in the osmotic transport of an electrolyte 
through charged cylindrical pores, which are kept at uniform temperature but are subjected to an externally imposed axial 
concentration gradient. Nevertheless, so far and to the best of our knowledge, the thermoosmotically driven transport and 
the implications of such a flux on the electrokinetic streaming have never been explicitly and systematically addressed. 
This is the focus of the present study. 
\\
\indent In \S \ref{sec:model_eqn}, an analytical model based on the Debye-H\"uckel (DH) approximation at low $\zeta$ potential 
and a full numerical model of the local, non-isothermal streaming potential (in form of a scaled induced electric field) are 
developed. The latter is a function of the pressure and temperature difference, nominal Debye parameter, $\zeta$ potential, 
Soret coefficient as well as of the temperature dependencies of dielectric permittivity and ion mobility. Counterintuitively, 
to leading order in the temperature difference, the temperature dependencies of viscosity, heat conductivity, heat capacity and 
Fickian diffusion coefficients have no effect. The thermally induced modifications of the EDL potential are discussed. In \S \ref{sec:results}, 
the induced streaming potential is analyzed for specific limiting cases. Subsequently, in the limit of a vanishing external pressure 
difference, the thermoosmotically induced electric field is obtained by numerical integration along the channel center 
plane for the case that both thermophoretic ion mobilities are the same and discussed for a realistic range of parameters. 
The full numerical model is compared to semi-analytical expressions valid within the DH approximation. 

\section{Model equations and perturbative solutions}\label{sec:model_eqn}

\begin{figure}
	\centerline{\includegraphics[width=13cm]{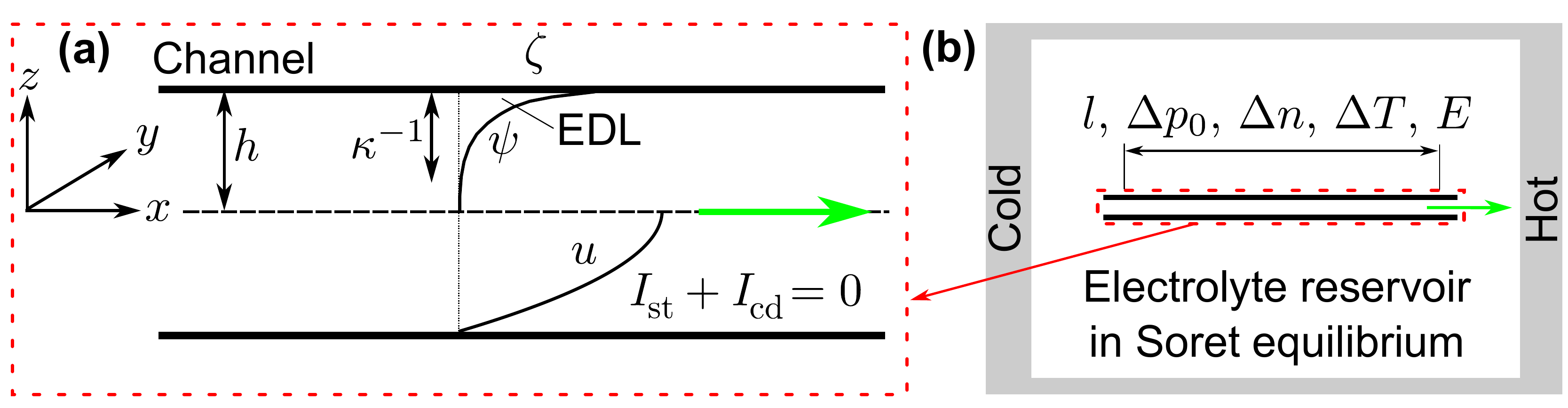}}
	\caption{(a) Schematic drawing of a slit channel of length $l$ and half gap width $h$. The problem is assumed 
	to be symmetric about the $x,y$-plane and translationally invariant in $y$-direction. The flow profile $u$, mainly 
	driven by the axial pressure difference $\Delta p_0$, is approximately parabolic with zero velocity at the shear plane, 
	where the electric potential is equal to the $\zeta$ potential. The latter is screened by ions in the electric double 
	layer (EDL), leading to an internal potential distribution $\psi$. No net charge is transported in the system, so 
	that advective ion streaming generates a streaming current $I_\textrm{st}$, which is exactly compensated by a conduction current 
	$I_\textrm{cd}$ due to an induced (streaming) electric field $E$. Apart from the pressure gradient, uniform axial 
	gradients in temperature, $\Delta T/l$, and in salt concentration, $\Delta n/l$ are imposed. It is assumed that the salt 
	concentrations at both ends of the channel, $\Delta n$, are determined by thermodiffusion equivalent to the Soret 
	equilibrium in bulk electrolytes. As illustrated in (b), this can be considered as if the channel were submerged 
	in a large reservoir filled with the electrolyte at rest and subjected to the same temperature difference $\Delta T$.}
	\label{Fig:problem_schematic}
\end{figure}

In the following, the governing equations are summarized and simplified to obtain analytical solutions for the flow field, 
along with its thermoosmotically driven contributions, and the streaming potential inside a parallel-plate slit channel 
exposed to the combined action of an externally applied pressure ($\Delta p_0$) and temperature difference ($\Delta T$) between 
both ends of the channel, with the channel length denoted by $l$ and one half of its gap width denoted by $h$, see figure 
\ref{Fig:problem_schematic} (a). In such channels, the order-of-magnitude of the pressure-driven (isothermal) flow can be 
approximated by $u_0=\Delta p_0 A h /(3\eta_0)$ (Poiseuille), where $A=h/l$ and $A^2 \ll 1$. The dynamic viscosity of the electrolyte 
$\eta_0$ is determined at the ambient reference temperature $T_0$, where $\Delta T/T_0 < 1$ is assumed. The electrokinetic 
response of the channel is commonly a strong function of the non-dimensional Debye parameter $\overline{\kappa} = \kappa h$, 
where $\kappa^{-1} = \sqrt{\epsilon k_\textrm{B} T/(2e^2 \nu^2 n)}$ is the Debye length of the electric double layer 
(EDL) near charged interfaces. The Boltzmann constant is denoted by $k_\textrm{B}$, $e$ is the elementary charge, $\nu$ is the valence 
of the symmetric $\nu:\nu$ electrolyte, and $n$ is the number concentration of the dissolved salt. Within the electroneutral regime, 
i.e. sufficiently far away from any charged wall, this equals the local ion number concentrations. With $\varepsilon_0$ being 
the vacuum dielectric permittivity and $\varepsilon_r$ being the relative permittivity of the liquid, its permittivity is 
given by $\epsilon = \varepsilon_0 \varepsilon_r$. Furthermore, the bulk electric conductivity of liquid electrolytes with 
dissolved salt ions as charge carriers of ionic mobility $\omega \approx D/(k_\textrm{B} T)$ can be approximated by 
$\sigma^{(\infty)} = 2 e^2 \nu^2 n \omega$, where $D$ is a suitably averaged reference diffusion coefficient.

Herein, unlike conventional considerations of electrokinetic streaming in microchannels, the temperature is not constant and its 
distribution is governed by the energy equation. As also demonstrated in the Supplemental Material of \citep{Dietzel:PRL2016}, viscous 
dissipation and Joule heating are negligibly small \citep{Wood:JPhysCondMatt2016}. Neglecting also the kinetic energy of the flow, 
the energy equation is given by $\rho c_p d_t T = \bnabla \cdot \left(k \bnabla T\right)$, with $T$ being the absolute 
temperature. The substantial derivative $d_t \equiv d/dt$ is denoted by $d_t (.)\equiv\partial_t (.)+ \boldsymbol{v} \bcdot \bnabla(.)$ 
with the velocity vector $\boldsymbol{v}=(u,w)$. The slit geometry implies a vanishingly small velocity component in $y$-direction, 
so that this is readily omitted from the description. The fluid density, thermal conductivity and heat capacity at constant pressure 
are labeled by $\rho$, $k$ and $c_p$, respectively. All thermophysical parameters of the solvent 
($\eta$, $\rho$, $c_p$, $k$, $\epsilon$) as well as of the dissolved ions ($D$, $\omega$) are, in general, 
temperature-dependent. Table \ref{Tbl:thermoprops_relvaria} summarizes typical relative variations for aqueous solutions. To 
calculate the relative variations of $D$ and $\omega$ with temperature, the Stokes-Einstein relation 
$D \approx k_\textrm{B} T /(6 \pi \eta R_\textrm{h})$ was used, where $R_\textrm{h}$ is the hydrodynamic radius of the ions. 
At least for $\textrm{Na}^+$, $\textrm{K}^+$ and $\textrm{Cl}^-$, $R_\textrm{h}$ is practically unaffected by temperature (the change 
is between $O(10^{-4})$ to low $O(10^{-3})\:\textrm{K}^{-1}$ \citep{Oelkers:JSolChem1989}). With this one finds 
$\partial_T D/D = 1/T - \partial_T \eta/\eta$ and $\partial_T \sigma^{(\infty)}/\sigma^{(\infty)} = \partial_T \omega/\omega = -\partial_T \eta/\eta$. 
Given the very small concentration of dissociated water molecules in comparison to the salt ions, the electric conductivity of pure water, 
$\sigma^{(\textrm{DI})}$, (stemming from the dissociated $\textrm{H}^+$- and $\textrm{OH}^-$-ions), is ignored. Therefore, we neglect the 
change of the electric conductivity of pure water with temperature. Hence, for the present purpose, the variation of the diffusivity 
and viscosity with temperature has the largest effect, while the change of the heat capacity and of the liquid density with 
temperature can be readily neglected. Furthermore, since liquids are virtually incompressible not only $c_p \equiv c_{p,0}$ 
is a constant but also $\rho \equiv \rho_0$.
\begin{table}
	\begin{center}			
		\begin{tabular}{c|c|c|c|c|c|c|c}
			$\partial_T \eta/\eta$ & $\partial_T \rho/\rho$ & $\partial_T c_p/c_p$ & $\partial_T k/k$ & $\partial_T \epsilon/\epsilon$ 
			& $\partial_T \sigma^{(\textrm{DI})}/\sigma^{(\textrm{DI})}$ & $\partial_T D/D$ & $\partial_T \omega/\omega$\\
			\hline
      $-15.7$ & $-0.357$ & $4.45\cdot 10^{-3}$ & $2.41$ & $-4.35$ & 83.6 & $19.1$ & $15.7$ \\
		\end{tabular}
	\end{center}
\caption{Relative variation of thermophysical properties with temperature in units of $10^{-3}\:\textrm{K}^{-1}$, determined at 
$25\:^{o}\textrm{C}$ ($T_0 = 298\:\textrm{K}$). The derivatives with respect to temperature of $\eta$, $\rho$, $c_p$, $k$ 
and $\epsilon$ were calculated by evaluating the respective parameter at $T_0 + \Delta T$ minus the value at $T_0$ 
divided by $\Delta T = 25\:\textrm{K}$ (forward scheme). Tabulated data of \citet{Lide:CRCPubComp2009} was used, except for 
$\sigma^{(\textrm{DI})}$, which was taken from \citet{Light:AnalChem1987}. In addition, 
$\partial_T D/D \approx 1/T - \partial_T \eta/\eta$ and $\partial_T \omega/\omega \approx -\partial_T \eta/\eta$.}
\label{Tbl:thermoprops_relvaria}
\end{table}

The energy equation can be brought in a dimensionless form by scaling the axial $x$-direction by $l$ while the lateral 
$z$-coordinate is scaled by $h$, i.e. $\boldsymbol{X} = (X,Z) = (x/l,z/h)$. With $\alpha_0 = k_0/(c_{p,0} \rho_0)$ 
as the thermal (reference) diffusivity and $\Pen_T = h u_0/\alpha_0$ as the thermal P\'eclet number, one finds
\begin{equation}
\label{Eq:Energy} A \Pen_T \left(d_\tau \Theta\right) - A^2 \left[\partial_X \left(\frac{k}{k_0} \partial_X \Theta\right)\right] = 
\partial_Z \left(\frac{k}{k_0} \partial_Z \Theta\right),
\end{equation}
where $\Theta = (T-T_0)/\Delta T$ is the non-dimensional temperature and 
$\partial_\varphi \equiv \partial/\partial \varphi$ ($\varphi = \tau,X,Z$). In non-dimensional notation, the substantial 
derivative is denoted by $d_{\tau} (.) = \partial_{\tau}(.) + \boldsymbol{V} \bcdot \overline{\bnabla}(.) = (l/u_0)d_t(.)$, 
with $\tau=t u_0/l$, $\overline{\bnabla} = (\partial_X,\partial_Z)$ and $\boldsymbol{V}=(U,W)=(u/u_0,w/w_0)$. As implied by the non-dimensional 
continuity equation, $\overline{\bnabla} \cdot \boldsymbol{V}$, the vertical scaling velocity is $w_0 = A u_0$. It is assumed that $h$ 
is much smaller than the axial extent $l$ and thus $A^2 \ll 1$. As will be shown later using table \ref{Tbl:thermoprops_ratios} 
in \S \ref{subsec:estimates}, for simple electrolytes in typical applications of pressure-driven streaming in microchannels 
$\Pen_T \lesssim O(A)$. Hence, the left-hand side (LHS) of (\ref{Eq:Energy}) can be neglected up to first order in $A$. Using 
the symmetry condition along the center plane at $Z=0$, it follows that the temperature is identical to the local wall 
temperature. It is assumed here that a linear temperature profile along $X$ is imposed on the channel wall, so that also the 
liquid temperature varies according to $\partial_X \Theta = \textrm{constant}$, while 
$\partial_Z \Theta = 0$. Hence, in the present limit, the temperature dependence of the thermal conductivity of the liquid 
is of no importance.

\subsection{General axial velocity profile}\label{sec:axial_flow}

The motion of an incompressible electrolyte of density $\rho \equiv \rho_0$ and mass averaged velocity vector $\boldsymbol{v}$ is 
described by the Navier-Stokes equation $\rho d_t \boldsymbol{v} = \bnabla \bcdot \mathsfbi{\sigma}^V + \bnabla \bcdot \mathsfbi{\sigma}^M$ 
with Maxwell stresses added as a source term. The mechanical stress tensor is denoted by 
$\mathsfbi{\sigma}^V = -p \mathsfbi{I} + \eta [\bnabla \boldsymbol{v}+(\bnabla \boldsymbol{v})^T]$ 
with $\mathsfbi{I}$ being the unit tensor, $p$ the liquid pressure, and where the dynamic viscosity $\eta$ does not necessarily 
need to be constant but might vary with temperature. The assumed small channel size allows to omit hydrostatic contributions, and the channel is aligned 
orthogonal to the direction of the gravitational acceleration $\boldsymbol{g}$ to remove buoyancy effects ($\bnabla T \cdot \boldsymbol{g} = 0$). 
For a homogeneous, incompressible fluid, the Maxwell stress tensor reads 
$\mathsfbi{\sigma}^M = \epsilon(\bnabla \phi \bnabla \phi - \bnabla \phi \bcdot \bnabla \phi \mathsfbi{I}/2)$, where $\phi$ is the electric 
potential. The force contribution due to the Maxwell stress tensor can be expressed in terms of the Korteweg-Helmholtz electric force 
per volume given by $\bnabla \bcdot \mathsfbi{\sigma}^M = -\rho_f \bnabla \phi -(\bnabla \phi)^2 \epsilon_T \bnabla T/2$ 
\citep{Russel:CambrUnivPress1989}, where the temperature dependence of the dielectric permittivity was incorporated by 
$\bnabla \epsilon = \epsilon_T \bnabla T$, with $\epsilon_T \equiv d\epsilon/dT$. The charge density $\rho_f$ is related to $\phi$ 
according to the Poisson equation $\bnabla \bcdot (\epsilon \bnabla \phi) = -\rho_f$. The common assumption of electrokinetics is made that 
$\phi = \psi + \phi_0$ is a linear superposition of the electric potential due to the ion double layer at the interfaces, $\psi(x,z)$, 
measuring the departure from electro-neutrality \citep{Fair:JChemPhys1971}, while $\phi_0$ is an induced electric potential with vanishing 
affiliated charge density (source free). Hence, the Laplace operator acting on $\phi_0$ vanishes. Along with the symmetry condition at the 
channel center plane, this implies that $\phi_0 = \phi_0(x)$ and $d_x \phi_0 \equiv -E$ is a constant. As will be further discussed in \S \ref{subsec:EDL}, 
$E$ is calculated by integrating the axial velocity profile over the channel cross section. In non-dimensional form, the Poisson equation reads
\begin{equation}
\label{Eq:Poisson} A^2 \left(\partial^2_X \Phi + M \Delta T \partial_X \Phi \partial_X \Theta \right) 
+ \partial^2_Z \Psi + M \Delta T \partial_Z \Psi \partial_Z \Theta 
= -\frac{1}{2} \overline{\kappa}^2_M \frac{\rho_f}{e \nu n_0},
\end{equation}
where $M = \epsilon_T/\epsilon$, while $(\Phi,\Psi) = (\phi,\psi)e\nu/(k_\textrm{B} T_0)$ and $\overline{\kappa}_M$ are 
the dimensionless potentials and the Debye parameter, respectively. In (\ref{Eq:Poisson}), it was already taken into account that $E$ 
is a constant so that $\partial_Z \Phi \equiv \partial_Z \Psi$ (and thus, $\partial^2_Z \Phi \equiv \partial^2_Z \Psi$ as well). 
Expression (\ref{Eq:Poisson}) implies that, to leading order in $A$, the charge density only affects the electric field components 
in $Z$-direction. Since the electric potentials are non-dimensionalized with $e\nu/(k_\textrm{B} T_0)$, the reference temperature 
in $\overline{\kappa}_M$ equals $T_0$. Nevertheless, $\overline{\kappa}_M$ is not a constant but depends on the local value of 
$\epsilon$ and can be expressed by
\begin{align}
\label{Eq:local_Debye_thickness} \overline{\kappa}_M &= \overline{\kappa}_0 \sqrt{\frac{\epsilon_0}{\epsilon}} 
= \overline{\kappa}_0 (1+M_0 \Delta T \Theta)^{-1/2} \nonumber \\
&\approx \overline{\kappa}_0 (1-\frac{1}{2} M_0 \Delta T \Theta).
\end{align}
where $\overline{\kappa}_0 = \kappa_0 h$, $M_0 = \epsilon_T/\epsilon_0$, and 
$\kappa^{-1}_0 = \sqrt{\epsilon_0 k_\textrm{B} T_0/(2e^2 \nu^2 n_0)}$ denotes the nominal EDL thickness 
for which all parameters are evaluated at the reference temperature $T_0$ so that $\epsilon=\epsilon_0$ and $n=n_0$. 
Furthermore, the velocity field is obtained from the Navier-Stokes equation. The non-dimensional axial velocity 
component fulfills
\begin{eqnarray}
\label{Eq:momentum_x} A \Rey \left(d_{\tau} U\right) - 
A^2\left(\overline{\eta}\partial^2_X U + 2\partial_X U\partial_X \overline{\eta} + 
\partial_X W \partial_Z \overline{\eta}\right) \nonumber \\
- A^2\frac{Ha}{\overline{\kappa}^2_M}\left\{\partial^2_X \Phi \partial_X \Phi + 
M \Delta T\left[\partial_X \Phi \partial_X \Phi - 
\frac{1}{2}\left(\partial_X \Phi\right)^2\right] \partial_X \Theta\right\} = -\partial_X P \nonumber \\
+ \partial_Z\left(\overline{\eta} \partial_Z U\right) + 
\frac{Ha}{\overline{\kappa}^2_M} \left\{\partial^2_Z \Psi \partial_X \Phi + 
M \Delta T\left[\partial_Z \Psi \partial_X \Phi \partial_Z \Theta - 
\frac{1}{2}\left(\partial_Z \Phi\right)^2\partial_X \Theta\right]\right\},
\end{eqnarray}
while the lateral component is obtained from
\begin{eqnarray}
\label{Eq:momentum_z} A^3 \Rey \left(d_{\tau} W\right)
-A^2\left(A^2 \overline{\eta} \partial^2_X W 
+ \overline{\eta} \partial^2_Z W + A^2 \partial_X W \partial_X \overline{\eta} + 
2 \partial_Z W \partial_Z \overline{\eta} + 
\partial_Z U \partial_X \overline{\eta}\right) \nonumber \\
- A^2 \frac{Ha}{\overline{\kappa}^2_M}\left\{\partial^2_X \Phi \partial_Z \Phi + 
M \Delta T\left[\partial_X \Phi \partial_Z \Phi \partial_X \Theta - 
\frac{1}{2}\left(\partial_X \Phi\right)^2 \partial_Z \Theta\right]\right\} = -\partial_Z P \nonumber \\
+ \frac{Ha}{\overline{\kappa}^2_M} \left\{\partial^2_Z \Psi \partial_Z \Phi + 
M \Delta T\left[\partial_Z \Psi \partial_Z \Phi - 
\frac{1}{2}\left(\partial_Z \Phi\right)^2\right] \partial_Z \Theta\right\}. \: \: \:
\end{eqnarray}

In this formulation, the charge density $\rho_f$ was expressed by (\ref{Eq:Poisson}), $\Rey = \rho u_0 h/\eta_0$ 
is the Reynolds number and $\overline{\eta} = \eta/\eta_0$ is the non-dimensional local viscosity. 
Further, $Ha = 2 A h n_0 k_\textrm{B} T_0/(u_0 \eta_0)$ is the (scaled) Hartmann number, which compares the velocity 
induced by the osmotic reference pressure, $n_0 k_\textrm{B} T_0$, to the characteristic velocity scale. The fluid pressure is 
non-dimensionalized according to $P = A h p/(u_0 \eta_0)$. In Poiseuille-type flow with constant $\Delta p$, the 
characteristic velocity $u_0$ is proportional to $A$ (see expression at the beginning of \S 2). For sufficiently viscous 
fluids and sufficiently small values of $\Delta p$ and $h$, one obtains $u_0 \leq {\cal O}(A)$. Hence, with $A<1$, $u_0$ 
can be expected to be a sufficiently small quantity, suggesting that the $\Rey$-number is small at least to order $A$ as well. In addition 
it follows that $Ha$ and $P$ do not depend on $A$. For low-$\Rey$-number flows in small-scale geometries, diffusion 
processes typically dominate over advective transport so that the ionic P\'eclet number $\Pen_k = l u_0/D_k$ of ion species 
$k=(+,-)$ can be assumed to be not larger than ${\cal O}(1)$. As discussed by \citep{Yariv:JFM2011}, $Ha$ cannot be of the same order 
as $\Pen_k$. In fact, also with the present scaling one has $Ha = (\varsigma/\Pen_k) \overline{\kappa}^2_M$, where 
$\varsigma = \epsilon/(\eta_0 D)[k_\textrm{B} T_0/(\nu e)]^2$ is the intrinsic P\'eclet number \citep{Saville:AnnRevFluidMech1977}. 
For typical aqueous solutions $\varsigma \approx 0.5$. Consequently, 
$Ha/\overline{\kappa}^2_M \approx Ha/\overline{\kappa}^2_0 \leq {\cal O}(1)$ is a consistent, 
and for the present purpose sufficient scaling. As will be detailed in \S \ref{sec:results}, table \ref{Tbl:thermoprops_ratios} 
summarizes the validity of the assumptions made. While $\partial_Z \Phi \equiv \partial_Z \Psi$, it was shown earlier that 
herein $\partial_Z \Theta = 0$. Neglecting terms of order $A^2$ and higher orders, one can deduce from (\ref{Eq:momentum_z}) 
that $P \approx Ha/(2\overline{\kappa}^2_M) (\partial_Z \Psi)^2 + c_1(X)$ where $c_1(X)$ is an integration constant, which 
simply equals the externally applied pressure $P_0(X)$. The first term proportional to the square of the lateral electric 
field is the electrostatic pressure contribution, representing the electroosmotic pressure of the ion cloud, 
$p_\textrm{osm} = n k_\textrm{B} T$. In common studies of electrokinetic streaming in long (micro-) channels, this term 
is not a function of the axial coordinate $X$. By contrast, as will be shown later, the thickness of the EDL and the EDL potential are 
herein a function of temperature. Therefore, the electrostatic pressure varies in axial direction, and the overall axial pressure 
gradient reads $\partial_X P = Ha/(2\overline{\kappa}^2_M) [\partial_X(\partial_Z \Psi)^2 
- 2(\partial_Z \Psi)^2 \partial_X \overline{\kappa}_M /\overline{\kappa}_M] + \partial_X P_0$. With $\Rey = {\cal O}(A)$ 
and $Ha/\overline{\kappa}^2_M \leq {\cal O}(1)$, the LHS of (\ref{Eq:momentum_x}) is small to order $A^2$ and one can write
\begin{equation}
\label{Eq:momentum_x2} \partial_Z \left(\overline{\eta} \partial_Z U\right) \approx 
\frac{Ha}{\overline{\kappa}^2_M}\left[\partial_X \left(\partial_Z \Psi\right)^2 - 
\partial_Z\left(\partial_Z \Psi \partial_X \Phi\right) + 
\left(\partial_Z \Psi \right)^2 M \Delta T \partial_X \Theta \right] + \partial_X P_0.
\end{equation}

In general, the viscosity is a function of the shear rate, concentration of dissolved species as well as of temperature. 
Shear rates are assumed to be sufficiently small so that shear thinning or thickening behavior is of no importance, 
i.e. a Newtonian fluid behavior is assumed. Furthermore, significant relative concentration changes of the dissolved ions are 
only present in the EDL. Absolute values of ion concentration are proportional to $n_0$ which is typically very small in dilute electrolytes. 
Therefore, even within the EDL, the dependence of the viscosity on the local ion concentration is expected to be negligibly small. Finally, the temperature varies only 
in axial direction and (\ref{Eq:momentum_x2}) can be integrated twice in $Z$ even without explicit knowledge of the viscosity-temperature relationship. 
Symmetry is assumed at $Z=0$ (subscript $\textrm{c}$), while the no-slip condition has to be fulfilled at the wall ($Z=1$, subscript $\textrm{s}$). With 
$\partial_X \Phi = \partial_X \Psi - \overline{\textrm{E}}$, this leads to the expression for the axial velocity
\begin{equation}
\label{Eq:vel_ax1} U = -\frac{\partial_X P_0}{2\overline{\eta}}(1-Z^2) 
+ \frac{Ha}{\overline{\eta} \: \overline{\kappa}^2_M} \overline{\textrm{E}}(\Psi-\overline{\zeta})  
+ \frac{Ha}{\overline{\eta} \: \overline{\kappa}^2_M}\left[\Omega-\Omega_\textrm{s} + \partial_Z \Omega_\textrm{c}(1-Z)\right],
\end{equation}
where $\overline{\textrm{E}} = e \nu l E/(k_\textrm{B} T_0)$. Note that the local dimensionless viscosity still depends on 
the axial coordinate, i.e. $\overline{\eta} = f(X)$. In expression (\ref{Eq:vel_ax1}), the $\zeta$ potential has been made dimensionless by 
$\overline{\zeta} = \zeta e \nu /(k_\textrm{B} T_0)$. Subsequently, at the present level of approximation, 
the $\zeta$ potential will be assumed to be unaffected by temperature. This important issue and the suitability of that assumption will be 
discussed in more detail in the last paragraph of section \ref{subsubsec:poisson}. Furthermore, one has
\begin{equation}
\label{Eq:Ips_integral} \Omega = \int\! \int \partial_X \left(\partial_Z \Psi \right)^2 d^2Z - 
\int \partial_Z \Psi \partial_X \Psi dZ + \partial_X \Theta M \Delta T\int\! \int\left(\partial_Z \Psi\right)^2d^2Z,
\end{equation}
where the integration symbols denote primitives of the corresponding functions and $d^2 Z$ denotes double integration of the integrand 
with respect to $Z$. In (\ref{Eq:Ips_integral}), the first two integrals are the sum of the electrohydrostatic (EHS) contribution and 
the electro-migration force (EMF), while the EHS alone is just one half of the first integral. For reasons of symmetry, one has 
$\partial_Z \Psi_\textrm{c} = 0$ and thus also $\partial_Z \Omega_\textrm{c} = 0$. The expression for $\Omega$ vanishes if the EDL potential 
$\Psi$ does not depend on the axial coordinate and the temperature dependence of the permittivity is neglected. In this case, equation 
(\ref{Eq:vel_ax1}) resembles the well-known isothermal result. For further evaluation, a closing relation between the axial velocity 
$U$, the electric potential $\Psi$ and the induced electric field $\overline{\textrm{E}}$ is required. 

\subsection{Double-layer potential}\label{subsec:EDL}

\subsubsection{Ion distribution}\label{subsubsec:ion_distribution}

Despite known mathematical inconsistencies \citep{Dreyer:PCCP2013}, for ion concentrations distinctively 
below $1\:\textrm{M}$ ($\textrm{M} \equiv \textrm{mol}\:\textrm{dm}^{-3}$) \citep{Levine:JChemSoc1975} and 
sufficiently small temperature gradients and electric fields, the transport of $k=1,..,K$ ion species in 
liquids is commonly described by the Nernst-Planck equations reading
\begin{equation} 
\label{Eq:NPE_dim} d_t n_k = \bnabla \bcdot (D_k \bnabla n_k + n_k D_{T,k} \bnabla{T} + e\nu_k n_k \omega_k \bnabla \phi).
\end{equation}
The number concentration of positive or negative ions labeled with $k=(+,-)$ is referred to by $n_k$. The 
diffusion coefficients affiliated with concentration gradients are $D_k$, while 
$D_{T,k}$ are the thermophoretic mobility (i.e. thermodiffusion) coefficients of the ions when subjected to a temperature 
difference. The ionic mobilities under the action of a gradient in the electric potential, $\bnabla \phi$, are given 
by $\omega_k\approx D_k/(k_\textrm{B} T)$.

Thermal diffusion in multi-component fluids is frequently described in terms of so-called heats of transport, $Q_k$ 
\citep{Helfand:JChemPhys1960}, or, equivalently, entropies of transport $s_k=Q_k/T$ \citep{Tasaka:PureApplChem1986}. As summarized 
in appendix \ref{sec:app_thermodiffpot}, these quantities are accurately defined within the phenomenological theory of non-equilibrium 
thermodynamics \citep{deGrootMazur:Dover1984,Fitts:McGrawHill1962} and emerge from a cross correlation between heat transport due to 
material fluxes on the one hand and matter transport due to thermal gradients on the other. Related phenomena are commonly termed heat-matter 
cross effects. Equivalently, a concentration gradient of one species may lead, apart from the regular Fickian diffusion 
of that species, to diffusion of another, which is commonly termed cross-diffusion. The problem under study comprises a tertiary mixture of 
electrically neutral solvent with two ion species. In such a setting, cross diffusion can be omitted if electroneutrality is assumed 
throughout the fluid domain, simplifying the problem to that of an effective binary mixture involving only a single linearly independent 
concentration field and a single non-advective ion flux \citep{Haase:Dover1969}. By contrast, as particularly relevant for nanochannel flow, 
electrolytes close to walls carrying a surface charge are not electrically neutral. The problem remains tertiary with linearly independent 
species concentrations, so that, according to the theoretical framework referred to, in principle cross-diffusional effects need to be considered explicitly. 
This is routinely neglected in most studies of 
electrokinetic streaming involving wall effects, in which ion transport is described by the (isothermal) Nernst-Planck equation. 
Adding thermophoretic ion diffusion to this equation in terms of $Q_k$ might give the misleading impression that the resulting equation 
is fully consistent with non-equilibrium thermodynamics (needed to define the $Q_k$), although this would only be the case if cross-diffusional 
fluxes between different ion species were included [see for instance (\ref{Eq:mass_flux5}) of appendix \ref{sec:app_thermodiffpot}]. 
To avoid this source of confusion and to emphasize the (still) limited validity of (\ref{Eq:NPE_dim}), instead of $Q_k$, effective thermal 
diffusion coefficients $D_{T,k}$ are used herein. Nevertheless, within the present approximation, the distinction between 
$D_{T,k}$ and $Q_k$ has solely a cosmetic character. In fact, in appendix \ref{sec:app_thermodiffpot} it is shown that (herein) 
$S_k \equiv Q_k/(k_\textrm{B} T^2)$ \citep{Wuerger:RepProgPhys2010}, where 
$S_k = D_{T,k}/D_k$ are the intrinsic Soret coefficients (other authors would call it the thermal diffusion ratio) of the ions in 
units of $\textrm{K}^{-1}$ \citep{Vigolo:Langmuir2010}. In this context, it is important to point out that all of these parameters 
equivalently quantifying the thermomobility of individual ion species ($Q_k$, $D_{T,k}$ or $S_k$) can be determined experimentally
only relative to each other \citep{Hill:Nature1957} and not on an absolute scale. Typically, as a point of reference, the thermophoretic 
mobility of the $\textrm{Cl}^{-}$-ion is arbitrarily set to zero. Table \ref{Tbl:Soret_coeff} summarizes values of intrinsic 
Soret coefficients of common monovalent ions as derived from data found in the literature. Experimental data are usually reported in terms 
of ionic heats of transport $Q_{m,0,k}$, in units of $\textrm{J}(\textrm{mol})^{-1}$ and for $T=T_0 = 298\:\textrm{K}$. To 
arrive at $S_{0,k}$ in units of $\textrm{K}^{-1}$, the following conversion was used
\begin{equation} 
\label{Eq:Soret_conversion} S_{0,k} = \frac{Q_{m,0,k}}{N_A k_\textrm{B} T^2_0},
\end{equation}
where $N_A$ is the Avogadro constant. The experiments to determine $Q_{m,0,k}$ were conducted at a salt concentration of 
$n_0 = 0.01\:\textrm{M}$. Given that a full theoretical treatment is achievable only in the limit of infinite dilution, 
in some cases the experimental values were extrapolated towards $n_0 \rightarrow 0$, either using graphical 
extrapolation \citep{Snowdon:TransFaradaySoc1960II} or a reduction rule \citep{Takeyama:JSolChem1988}. As apparent from 
table \ref{Tbl:Soret_coeff}, $S_{0,k}$ are generally ${\cal O}(10^{-3})$ - ${\cal O}(10^{-2})\:\textrm{K}^{-1}$. Values taken from 
\citet{Takeyama:JPhysSocJap1983} and listed in the first row are an order of magnitude larger than corresponding values reported 
by other authors. Considering entries listed in rows $2$-$5$ only, the $S_{0,k}$ of one ion species in cation/anion-combinations 
of $\textrm{Na}^+$ or $\textrm{K}^+$ with $\textrm{F}^-$ as well as $\textrm{Li}^+$ with $\textrm{Cl}^-$ or $\textrm{Br}^-$ is 
similar to that of the respective counter ion. By contrast, the corresponding values of each ion species in combinations of $\textrm{Na}^+$ or 
$\textrm{K}^+$ with $\textrm{Cl}^-$ or $\textrm{Br}^-$ as well as $\textrm{Li}^+$ with $\textrm{F}^-$ differ substantially.
\begin{table}
	\begin{center}			
		\begin{tabular}{c|c||c|c|c||c|c|c}
			Source & $n_0$ & $\textrm{Li}^{+}$ & $\textrm{Na}^{+}$ & $\textrm{K}^{+}$ & $\textrm{F}^{-}$ & $\textrm{Cl}^{-}$ & $\textrm{Br}^{-}$ \\
			\hline
			$[1]^{\textrm{rd}}$ & $\rightarrow 0$ & $19.5$ & $19.1$ & $13.1$ & $18.3$ & $9.90$ & $9.76$ \\
			$[2]^{\textrm{rd}}$ & $\rightarrow 0$ & $0.718$ & $4.69$ & $3.51$ & $5.33$ & $0.718$ & $0.813$ \\
			$[3]^{\textrm{ex}}$ & $\rightarrow 0$ & $-$ & $4.82$ & $3.91$ & $-$ & $0$ & $-$ \\
			$[4] \: \: \: \:$ & $0.01 \textrm{M}$ & $-0.142$ & $4.37$ & $2.86$ & $4.89$ & $0$ & $0.227$ \\
			$[5] \: \: \: \:$ & $0.01 \textrm{M}$ & $0.0397$ & $4.00$ & $2.77$ & $4.64$ & $0$ & $0.119$ \\
		\end{tabular}
	\end{center}
\caption{Literature values for the intrinsic Soret coefficients $S_{0,k}$ in units of $10^{-3}\:\textrm{K}^{-1}$. All values 
were determined at $T_0 = 298\:\textrm{K}$. Experiments were conducted at a concentration of $n_0 = 0.01\:\textrm{M}$. 
Values reported at infinite dilution ($n_0 \rightarrow 0$) are determined from experimental values obtained at finite concentration either 
by graphical extrapolation ('ex') or using a reduction rule ('rd'). $[1]$ - \citet{Takeyama:JPhysSocJap1983}, $[2]$ - \citet{Takeyama:JSolChem1988}, 
$[3]$ - \citet{Snowdon:TransFaradaySoc1960II}, $[4]$ - \citet{Agar:ProcRsSocLondA1960} $[5]$ - \citet{Snowdon:TransFaradaySoc1960}}
\label{Tbl:Soret_coeff}
\end{table}

With $N_k = n_k/n_0$ and $\overline{\nu}_k = \nu_k/\nu$, the dimensionless form of (\ref{Eq:NPE_dim}) reads
\begin{eqnarray} 
\label{Eq:NPE} A^2\left\{\frac{u_0 l}{D_k}U \partial_X N_k 
- \frac{D_0}{D_k}\partial_X \left[\frac{D_k}{D_0}\left(\partial_X N_k 
+ N_k S_k \Delta T \partial_X \Theta + \frac{\overline{\nu}_k N_k}{1+\Theta \Delta T/T_0} \partial_X \Phi \right) \right] \right\} \nonumber \\
= \frac{D_0}{D_k}\partial_Z \left[\frac{D_k}{D_0}\left(\partial_Z N_k + N_k S_k \Delta T \partial_Z \Theta 
+ \frac{\overline{\nu}_k N_k}{1+\Theta \Delta T/T_0} \partial_Z \Phi \right) \right], \: \: \: \: \: \: \: \: \: \: \: \: \:
\end{eqnarray}
where $D_0$ denotes a reference diffusion coefficient determined at $T=T_0$, and a stationary-state situation was assumed. 
For diffusion-dominated problems as the present one, the ionic P\'eclet numbers $\Pen_k = u_0 l/D_k$ are not larger than ${\cal O}(1)$, 
so that within the lubrication approximation (i.e. to first order in $A$) the LHS of (\ref{Eq:NPE}) can be neglected. With 
$\partial_Z \Phi \equiv \partial_Z \Psi$ and $\partial_Z \Theta = 0$, one finds
\begin{equation} 
\label{Eq:NPE_2} D_k \left[\partial_Z N_k + \overline{\nu}_k N_k \partial_Z \left(\frac{\Psi}{1+\Theta \Delta T/T_0}\right)\right] = c_2,
\end{equation}
where $c_2$ is an integration constant. Given the symmetry boundary condition at the channel center plane, the latter is zero. Division 
by $D_k$ and again integrating in lateral direction leads to an ion distribution resembling the Boltzmann distribution 
\begin{equation} 
\label{Eq:Boltzmann} N_k = N^{(\infty)}_k \textrm{exp} \left(-\frac{\overline{\nu}_k \Psi}{1 + \widehat{\Theta}}\right),
\end{equation}
where for shorter notation $\widehat{\Theta} = \Theta \Delta T/T_0$ was used. The reference ion concentrations at $\Psi=0$ 
(i.e. typically found far away from charged walls) are denoted by $N^{(\infty)}_k$. In dimensional form expression (\ref{Eq:Boltzmann}) reads
\begin{equation} 
\label{Eq:Boltzmann_dim} n_k = n^{(\infty)}_k \textrm{exp} \left(-\frac{e \nu_k \psi}{k_\textrm{B} T}\right).
\end{equation}
Hence, under the assumptions made, the ion distribution has the same structure as in the isothermal case. By inserting (\ref{Eq:Boltzmann}) 
the derivatives with respect to $Z$ (right hand side) are removed from (\ref{Eq:NPE}), while the axial derivative of $N^{(\infty)}_k$ as well 
as the change of the ion mobility along the channel enter the equation. Furthermore, the flow is assumed to be fully developed, 
so that $U \partial_X N_k = \partial_X (U N_k)$. With this, integration of (\ref{Eq:NPE}) in X leads to
\begin{equation} 
\label{Eq:NPE_ax} \frac{u_0 l}{D_0} U N_k - \frac{D_k}{D_0} N_k \left(\partial_X \textrm{ln}(N^{(\infty)}_k) + S_k T_0 \partial_X \widehat{\Theta} 
+ \frac{\overline{\nu}_k \Psi}{(1+\widehat{\Theta})^2} \partial_X \widehat{\Theta}
- \frac{\overline{\nu}_k}{1+\widehat{\Theta}} \overline{\textrm{E}} \right) = c_k(Z),
\end{equation}
where $c_k(Z)$ are integration constants identical to $j_{x,k} l/(n_0 D_0)$, i.e. proportional to the ion flux densities 
transported along the channel. The total electric field $\partial_X \Phi = \partial_X \Psi - \overline{\textrm{E}}$ was expressed as 
the sum of the EDL field and an induced electric field $\overline{\textrm{E}}$, which is uniform across the channel. At isothermal conditions, 
$\widehat{\Theta} = 0$ and $\partial_X \textrm{ln}(N^{(\infty)}_k) = 0$ so that $\overline{\textrm{E}} \equiv \overline{\textrm{E}}_\textrm{st}$, 
where $\overline{\textrm{E}}_\textrm{st}$ is the conventional (induced) streaming field caused by ion advection. The latter is determined 
by multiplying (\ref{Eq:NPE_ax}) with $e \nu_k$, summing over all ion species, integrating over the channel cross section and setting this net 
current to zero. At non-isothermal conditions, where the axial gradients of the ion concentrations are not necessarily vanishing, an equivalent 
procedure can be used once the full description of the non-isothermal ion distribution has been found.

With the given set of equations, assumptions and boundary conditions, one has $2K+4$ equations but $3K+4$ unknowns, with $K$ being 
the number of ion species. Thus, the $K$ values of $n^{(\infty)}_k$ are independent parameters which need to be given as additional 
constraints. At isothermal conditions, it is typically postulated that these concentrations are constants and not affected by e.g the advective 
ion motion. Under non-isothermal conditions, constant $n^{(\infty)}_k$ imply that, in comparison to an isothermal liquid, the ion 
distribution is modified only by the temperature-dependent electric mobility of the ions whereas the distribution remains independent of the 
intrinsic Soret coefficients. This has the unphysical effect that for a local $T$ arbitrarily exceeding $T_0$, $n_k$ would never get smaller 
than a reference concentration $n^{(\infty)}_k > 0$, no matter whether the ions are thermophobic ($S_k > 0$) or thermophilic ($S_k < 0$). In 
the related problem of electrolyte osmosis in a channel with an axial concentration gradient, the approach taken by \citet{Sasidhar:JCollIntScie1982} 
(equation (3) in that paper) suggests that the $n^{(\infty)}_k$ should not be constants but vary with the axial coordinate, i.e. 
$n^{(\infty)}_k = f^2_{S,k}(x) n_0$, where $f_{S,k}(x)$ are functions to be determined. An identical approach was taken by \citet{Fair:JChemPhys1971}.

By definition, $\psi$ (respectively $\Psi$) measures the deviation from electroneutrality \citep{Fair:JChemPhys1971} and the $n^{(\infty)}_k$ 
must therefore be the corresponding ion concentrations under the condition that the charge density vanishes. For a symmetric electrolyte this 
implies that these ion concentrations are identical for each ion species, i.e. $n^{(\infty)}_k \equiv n$, with $n$ being the overall salt 
concentration. The latter remains a free parameter, and -similar to the externally applied pressure or the induced potential- its values at 
the channel entrance and exit determine the driving force of osmotic transport in the slit channel \citep{Fair:JChemPhys1971}. In the following, 
all governing equations will be developed in terms of the unknown value of $n$ (respectively $N = n/n_0$ in dimensionless form). Subsequently, 
the special case is considered where the channel is assumed to be submerged in a large reservoir filled with electrolyte and subjected to a 
temperature gradient aligned with the channel center plane, see figure \ref{Fig:problem_schematic} (b). Under such conditions, the salt concentration 
$n$ (respectively $N$) in the (electroneutral) bulk of the electrolyte can be approximated by the salt redistribution characterizing the 
conventional Soret equilibrium (obtained under the assumption of electroneutrality). Based on (\ref{Eq:Soret-equilibrium}) derived in appendix 
\ref{sec:app_thermodiffpot} one has
\begin{equation} 
\label{Eq:SalinitySoret} N^{(\infty)} = N = \textrm{exp}\left(-\overline{S} T_0 \widehat{\Theta}\right),
\end{equation}
with $\overline{S} = \sum^2_{k=1}{n^{(\infty)}_k S_k}/(2 n) = \sum^2_{k=1}{S_k}/2$ being the average (intrinsic) Soret coefficient 
($\overline{S}$ is identical to $\alpha$ in (\ref{Eq:Soret-equilibrium})) and $n \equiv n_0$ at $\widehat{\Theta} = 0$. 
For a symmetric electrolyte under electroneutral conditions, one has $n^{(\infty)}_k = n$ for both ion species. 
In the following, temperature-independent intrinsic Soret coefficients are assumed so that $\overline{S} \approx \overline{S}_0$. 

Equation (\ref{Eq:SalinitySoret}), explicitly valid for an extended (electroneutral) bulk electrolyte subjected to a temperature gradient, 
needs to be recovered in a confined system for the special case where $\overline{\zeta} \rightarrow 0$ so that also $\Psi \rightarrow 0$. 
The most straightforward way to see that is to express $N^{(\infty)}_k$ in (\ref{Eq:Boltzmann}) by (\ref{Eq:SalinitySoret}). Under 
this assumption, the redistribution of ions in the vicinity of charged walls due to thermodiffusion as well as a temperature-dependent 
ion mobility can be described by
\begin{equation} 
\label{Eq:ion_distribution} N_k = \textrm{exp}\left(-\overline{S}_0 T_0 \widehat{\Theta}\right)
\textrm{exp} \left(-\frac{\overline{\nu}_k \Psi}{1+\widehat{\Theta}} \right).
\end{equation}
Hence the sought after function $f_{S}$ equals $\textrm{exp}(-\overline{S}_0 T_0 \widehat{\Theta}/2)$.

\subsubsection{Poisson equation}\label{subsubsec:poisson}

In conventional treatments of thermoelectricity in bulk electrolytes, the ion number distribution (\ref{Eq:Soret-equilibrium}) is derived 
under the condition of charge neutrality, i.e. the charge density vanishes, $\rho_f \equiv 0$. Here, in general, the latter does 
not vanish but is -in general- governed by the Nernst-Planck equations. Within the present approximation and with (\ref{Eq:Boltzmann}), 
$N^{(\infty)}_k \equiv N(X)$ as well as for a symmetric $\nu:\nu$-electrolyte $\rho_f$ reads 
\begin{equation} 
\label{Eq:charge_dens} \frac{\rho_f}{e \nu n_0} = \sum^{2}_{k=1} \overline{\nu}_k N_k = 
-2 N \textrm{sinh} (\widetilde{\Psi}),
\end{equation}
where $\widehat{\Psi} = \Psi/(1+\widehat{\Theta})$ ($\psi$ is scaled to the constant temperature $T_0$ when employing $\Psi$, 
whereas $\psi$ is scaled to the local temperature $T$ when employing $\widehat{\Psi}$). Inserting (\ref{Eq:charge_dens}) into the 
dimensionless Poisson equation (\ref{Eq:Poisson}) leads to
\begin{equation}
\label{Eq:Poisson_1D} \partial^2_Z \widehat{\Psi} = \frac{\overline{\kappa}^2_M N}{1+\widehat{\Theta}} \textrm{sinh} (\widehat{\Psi}),
\end{equation}
where terms of ${\cal O} (A^2)$ were neglected and $\partial_Z \Theta \equiv 0$ was used. Furthermore, 
\begin{equation}
\label{Eq:EDLparam_local} \overline{\kappa}_M \sqrt{\frac{N}{1+\widehat{\Theta}}} 
= h \sqrt{\frac{2 e^2 \nu^2 n}{\epsilon k_\textrm{B} T}} \equiv \overline{\kappa}
\end{equation}
denotes the local dimensionless Debye parameter evaluated at the local temperature, i.e. besides the dielectric permittivity 
the salt concentration is evaluated at the local temperature as well. 

Hence, for $\partial_X \Theta \neq 0$ the equation to determine the EDL potential distribution is 
qualitatively similar to the one found under isothermal conditions (denoted by $\Psi^{(\textrm{e})}$), which is 
given by
\begin{equation}
\label{Eq:Poisson_1D_isoth} \partial^2_Z \Psi^{(\textrm{e})} =\overline{\kappa}^2_0 \textrm{sinh}(\Psi^{(\textrm{e})}).
\end{equation}
Given the similarities, (\ref{Eq:Poisson_1D}) can be solved with the same well-known methodologies available to solve 
for $\Psi^{(\textrm{e})}$. By contrast, for the case of a slit channel for which the temperature gradient is applied vertically 
to the channel center plane (i.e. $\partial_Z \Theta = \textrm{constant}$ while $\partial_X \Theta = 0$), the conditional equation 
to determine the EDL potential differs qualitatively from the isothermal case: instead of being symmetric the EDL potential is 
found to be asymmetric with respect to the channel center plane \citep{Dietzel:muFlu2012,Zhou:JHT2015}.

As a physical interpretation of (\ref{Eq:Poisson_1D}), the temperature dependencies of permittivity and 
electrophoretic mobility as well as the intrinsic Soret effect alter the local EDL thickness according to
\begin{equation}
\label{Eq:EDLparam_local_Soret} \overline{\kappa} = 
h\: \textrm{exp}\left(-\frac{1}{2}\overline{S}_0 T_0 \widehat{\Theta}\right) \sqrt{\frac{2 e^2 \nu^2 n_0}{\epsilon k_\textrm{B} T}}
\end{equation}
instead of its isothermal value $\kappa^{-1}_0$. For small $\widehat{\Theta}$, linearization leads to
\begin{equation}
\label{Eq:EDL_param_local_approx} \overline{\kappa} \approx \overline{\kappa}_0 
\left\{ 1 - \frac{1}{2}\left[1 + \left(\overline{S}_0 + M_0\right)T_0 \right] \widehat{\Theta} \right\}.
\end{equation}

To further understand the effects of a temperature-dependent electrophoretic mobility, thermophoretic ion motion 
and a temperature-dependent permittivity, equation (\ref{Eq:Poisson_1D}) has to be 
solved, using the symmetry condition $\partial_Z \Psi_\textrm{c} = 0$ at $Z=0$ (channel center plane) as well as the fixed potential 
at $Z=1$ (channel wall) given by $\Psi_\textrm{s} \cong \overline{\zeta} = e \nu \zeta/(k_\textrm{B} T_0)$, where $\zeta$ is the 
original zeta potential. Apart from $\overline{\zeta}$, the solution is a function of $\overline{\kappa}_0$, $N$ 
(respectively $\overline{S}_0$), $\widehat{\Theta}$ and $M_0$.
 
At low $\zeta$ potentials the Debye-H\"uckel approximation (DH) can be applied, and the analytical solution of 
(\ref{Eq:Poisson_1D_isoth}) reads
\begin{equation}
\label{Eq:EDLpot_isoth_DH} \Psi^{(\textrm{e,DH})} = \overline{\zeta} \frac{\textrm{cosh}(\overline{\kappa}_0 Z)}
{\textrm{cosh}(\overline{\kappa}_0)},
\end{equation}
while the solution of (\ref{Eq:Poisson_1D}) is given by
\begin{equation}
\label{Eq:EDLpot_DH} \Psi^{(\textrm{DH})} = \overline{\zeta} \frac{\textrm{cosh}(\overline{\kappa} Z)}
{\textrm{cosh}(\overline{\kappa})}.
\end{equation}
Note that the factor $1/(1+\widehat{\Theta})$, used in the definition of $\widehat{\Psi}$, drops out since, 
within the DH limit, the Poisson equation is linear in $\widehat{\Psi}$. Thus, the isothermal and 
non-isothermal EDL potentials have qualitatively the same form. The difference is that $\Psi^{(e)}$ is identical 
in each cross section of the channel, whereas $\Psi$ varies in axial direction due to a variation of the local 
EDL thickness with temperature.

\begin{figure}
  \centerline{\includegraphics[width=13cm]{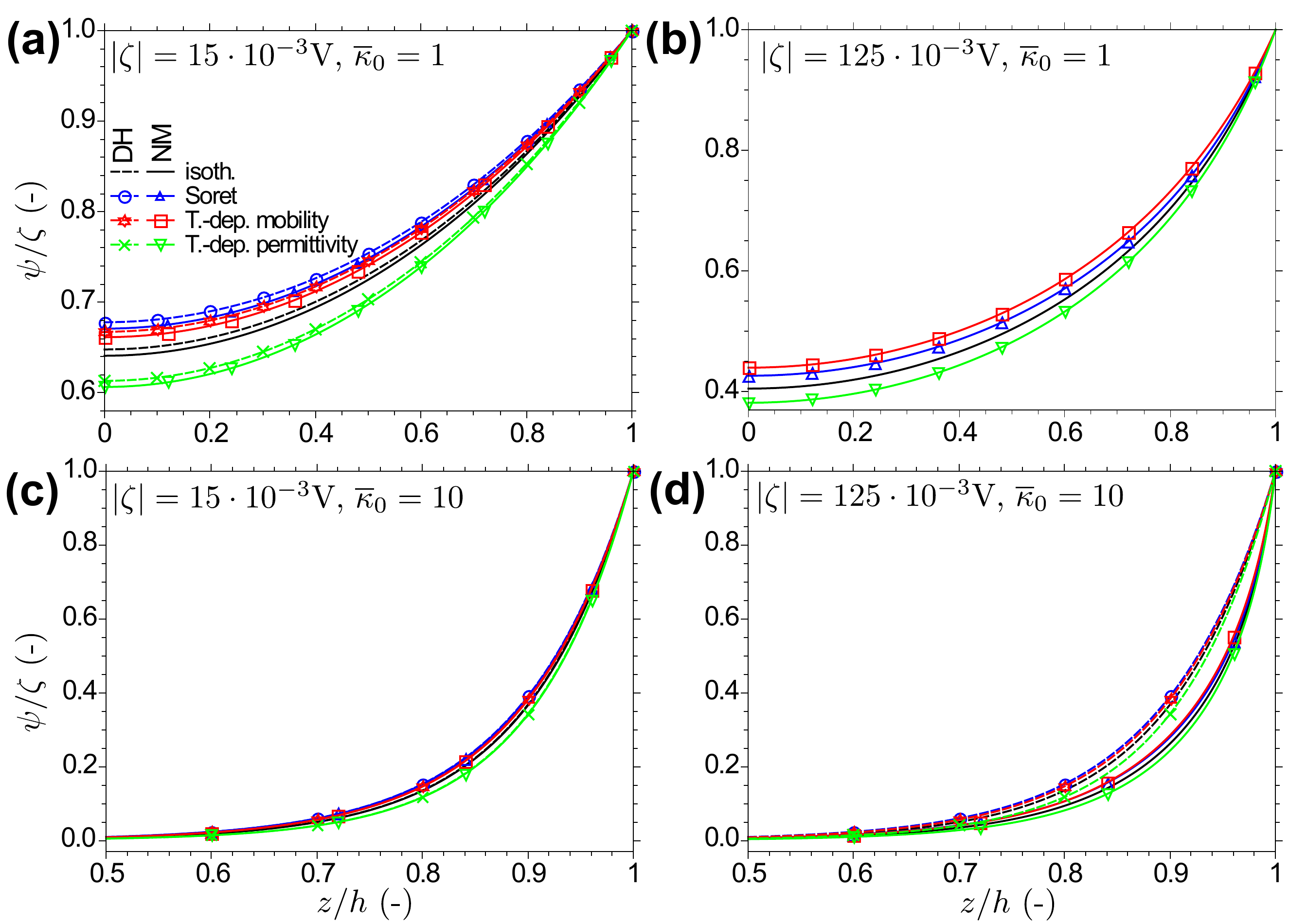}}
  \caption{Comparison of the relative EDL potential $\Psi(Z)/\overline{\zeta} = \psi(Z)/\zeta$ for the isothermal case ('isoth') 
	with the potentials obtained for cases where a temperature dependence is introduced by means of either the Soret effect ('Soret', 
	$\overline{S}_0 = 5 \cdot 10^{-3}\:\textrm{K}^{-1}$),	a temperature-dependent electrophoretic mobility ('T.-dep. mobility') or 
	a temperature-dependent dielectric permittivity ('T.-dep. permittivity', $M_0 = -5.1 \cdot 10^{-3}\:\textrm{K}^{-1}$). Each 
	non-isothermal effect was analyzed isolated from the other ones for $\Delta T = 25\:\textrm{K}$. 
	Within the Debye-H\"uckel ('DH') approximation, the results 
	were computed from the analytical expressions (\ref{Eq:EDLpot_isoth_DH}) and (\ref{Eq:EDLpot_DH}), respectively, while the 
	numerical results were obtained by solving (\ref{Eq:Poisson_1D_isoth}) or (\ref{Eq:Poisson_1D}) ('NM'). The nominal Debye 
	parameter is either $\overline{\kappa}_0 = 1$ [(a), (b)] or $\overline{\kappa}_0 = 10$ [(c), (d)]; the $\zeta$ potential 
	is set either to $|\zeta| = 15 \cdot 10^{-3}\:\textrm{V}$ [(a), (c)] or to $|\zeta| = 125 \cdot 10^{-3}\:\textrm{V}$ [(b), (d)]. 
	The legend depicted in (a) is valid for (b)-(d) as well. In (b), the solution obtained from the DH approximation is not 
	shown as it deviates significantly from the (more accurate) numerical solution.}
\label{Fig:PsifZ}
\end{figure}

In figure \ref{Fig:PsifZ} the relative EDL potential $\Psi/\overline{\zeta} = \psi/\zeta$, either affected by the 
intrinsic Soret effect ('Soret'), the temperature-dependent 
electrophoretic mobility ('T.-dep. mobility') or the temperature-dependent dielectric permittivity ('T.-dep. permittivity'), 
is shown. In the cases capturing the Soret effect, $N$ as given by (\ref{Eq:SalinitySoret}) was used; otherwise $N = 1$. 
Solutions obtained with the DH approximation (given by (\ref{Eq:EDLpot_DH})) are compared with the corresponding numerical 
solutions ('NM', determined by solving (\ref{Eq:Poisson_1D})) as well as with the isothermal result. The nominal 
Debye parameter was either $\overline{\kappa}_0 = 1$ [(a), (b)] or $\overline{\kappa}_0 = 10$ 
[(c), (d)], while the $\zeta$ potential was either $|\zeta| = 15 \cdot 10^{-3}\:\textrm{V}$ ($|\overline{\zeta}| \approx 0.6$, 
(a) and (c)) or $|\zeta| = 125 \cdot 10^{-3}\:\textrm{V}$ ($|\overline{\zeta}| \approx 5$, 
(b) and (d)). For all plots, $\overline{S}_0 = 5 \cdot 10^{-3}\:\textrm{K}^{-1}$ ($\overline{S}_0 T_0 = 1.49$), 
$M_0 = -5.1 \cdot 10^{-3}\:\textrm{K}^{-1}$ ($M_0 T_0 = -1.52$) and $\Delta T = 25\:\textrm{K}$ 
($\widehat{\Theta} = 8.39 \cdot 10^{-2}$) were used. Hence the plots compare the EDL potential at the channel 
exit ($x=l$, local temperature equals $T=T_0 + \Delta T$) with the one at the channel entrance ($x=0$, $T=T_0$). 
The boundary value problem described by (\ref{Eq:Poisson_1D}), 
respectively by (\ref{Eq:Poisson_1D_isoth}), along with $\Psi(\pm 1) = \pm \overline{\zeta}$ at the walls, was solved 
by collocation with the BVP4C-function implemented in Matlab (Version 8.0.0.783, R2012b) on a Dell Precision 
T7500 workstation operated with Ubuntu 12.04 LTS. As verified in a grid independence study, the solutions shown are 
practically indistinguishable from those obtained when the default mesh density of approximately $500$ grid points 
in $Z$-direction was reduced more than ten-fold and the default relative tolerance of $0.1\%$ was increased more 
than hundred-fold, respectively.

For $\overline{\kappa}_0 = 10$ and $\zeta = 15 \cdot 10^{-3}\:\textrm{V}$, 
the DH approximations and numerical solutions agree very well. By contrast, for 
$\zeta = 125 \cdot 10^{-3}\:\textrm{V}$ the (expected) difference can be as large as $40\%$. For $\overline{\kappa}_0 = 1$ 
the well-known tendency of the DH approximation to overpredict the EDL overlap is already visible at low $\zeta$ potentials. 
At $\zeta = 125 \cdot 10^{-3}\:\textrm{V}$, shown in figure \ref{Fig:PsifZ} (b), the mismatch between the analytical and the 
(more accurate) numerical solution is too severe, so that only the numerical results are depicted. For positive values of $\Delta T$, a 
temperature-dependent electrophoretic ion mobility always expands the EDL, i.e. the relative electric potential is always larger than the 
corresponding isothermal value at the same location. Since $\overline{S}_0$ is commonly positive, the intrinsic 
Soret effect typically enhances this behavior. On the other hand, since the dielectric permittivity typically decreases with 
temperature, the corresponding effect reduces the relative EDL potential, i.e. it appears to 
shrink the EDL. In fact, if
\begin{equation}
\label{Eq:EDLeffect_cancel} \overline{S}_0 \approx \overline{S}_{0,\textrm{eq}} = -\left(\frac{1}{T_0}+M_0\right),
\end{equation}
the characteristics of the EDL are unaffected by the presence of a small thermal gradient. For an aqueous electrolyte at 
$T_0 = 298\:\textrm{K}$ one has $\overline{S}_{0,\textrm{eq}} \approx 1.7 \cdot 10^{-3}\:\textrm{K}^{-1}$. According to table 
\ref{Tbl:Soret_coeff}, this is within the possible range of Soret coefficients. Thus, if not completely compensated, the 
effects of an intrinsic thermophoretic ion motion and a temperature-dependent electrophoretic mobility on the EDL 
potential are at least weakened to a comparable extent by a temperature-dependent dielectric permittivity.

As apparent from figure \ref{Fig:PsifZ}, the effect of a temperature gradient on the (relative) EDL potential 
$\Psi/\overline{\zeta}$ is relatively weak for any values of $\overline{\zeta}$ and $\overline{\kappa}_0$. At 
$\overline{\kappa}_0 = 1$ and $\zeta = 15 \cdot 10^{-3}\:\textrm{V}$ (shown in figure \ref{Fig:PsifZ} (a)) the largest relative change 
of the EDL potential in comparison to isothermal conditions is approximately $4.6\:\%$. At 
$\zeta = 125 \cdot 10^{-3}\:\textrm{V}$, (shown in (b)) this difference is approximately $5.5\:\%$. At 
$\zeta = 15 \cdot 10^{-3}\:\textrm{V}$ the increase of the relative EDL potential (compared its isothermal value) due to the 
intrinsic Soret effect is slightly larger than that caused by a temperature-dependent electrophoretic mobility. At 
$\zeta = 125 \cdot 10^{-3}\:\textrm{V}$ the corresponding behavior is reversed. The same applies at $\overline{\kappa}_0 = 10$. 
For that case, the maximal change of the relative EDL potential in comparison with isothermal conditions is a little 
less than $10\:\%$ for $\zeta = 15 \cdot 10^{-3}\:\textrm{V}$ and approximately $11\:\%$ at $\zeta = 125 \cdot 10^{-3}\:\textrm{V}$.

In the previous section it was argued that for $\Delta T \neq 0$ the velocity distribution (\ref{Eq:vel_ax1}) differs from the isothermal 
result because there may exist an axial gradient in $\Psi$. Under the condition that the $\zeta$ potential along the channel walls 
is constant, an expression for $\Psi$ was derived above, indicating that it is a function of the local value of the 
EDL thickness (described by (\ref{Eq:EDL_param_local_approx})), which changes with $\widehat{\Theta}(X)$. Thus, indeed $\Psi = f(X)$. It 
can be shown that this is also the case if - instead of a constant $\zeta$ - a constant surface charge density is assumed. Consequently, 
the discussed non-isothermal effects will affect the axial velocity distribution no matter whether a constant $\zeta$ potential or 
a constant surface charge density along the channel wall is present.

In any case, the assumption that either the $\zeta$ potential or the surface charge density as well as the charges in the Stern layer remain 
unaffected by the temperature gradient is to some extent questionable. However, without compromising the rigor of the present derivation, 
a complete treatment would also involve the temperature-dependent dissociation process of surface groups. In the supplemental material 
of \citet{Dietzel:PRL2016} such a non-isothermal charge regulation model was developed for an aqueous electrolyte in a silicate channel, 
which indicated that the surface charge density appears to be only weakly affected by a non-uniform temperature. Nevertheless, 
employing other wall materials might give different results, about which, to the best of our knowledge, only little work is available 
in the literature. This is further complicated by the circumstance that many of the problem parameters are not accessible experimentally, 
or only in a time- and space-averaged fashion. Hence, experimentally distinguishing between these various effects is a formidable challenge. 
In the light of the focus of this work, these questions must be left for future investigations. 

\subsection{Axial velocity profile for specific $\Psi$}\label{subsec:axial_flow_SoretEDL}

With the knowledge of $\Psi$ and $\Omega$ in (\ref{Eq:Ips_integral}), the axial velocity (\ref{Eq:vel_ax1}) can be further 
worked out. In $X$-direction, $\Psi$ and its derivatives with respect to $Z$ vary only with $\widehat{\Theta}$, so that 
$\partial_X (.) = \partial_{\widehat{\Theta}} (.) \partial_X \widehat{\Theta}$. Integrating (\ref{Eq:Poisson_1D}) and using the 
definition of $\widehat{\Psi}$ provides
\begin{equation}
\label{Eq:ddZPsi2} (\partial_Z \Psi)^2 = 2 (1+\widehat{\Theta})^2 \overline{\kappa}^2 
\left[\textrm{cosh}(\widehat{\Psi}) - \textrm{cosh}(\widehat{\Psi}_\textrm{c})\right],
\end{equation}
where, according to (\ref{Eq:EDLparam_local}), $(1+\widehat{\Theta}) \overline{\kappa}^2$ can be replaced by 
$\overline{\kappa}^2_M N$. The axial derivative of this equation reads
\begin{eqnarray}
\label{Eq:ddX_ddZPsi2} \partial_X (\partial_Z \Psi)^2 = (\partial_Z \Psi)^2 
\left(\frac{1}{1+\widehat{\Theta}} + 2\frac{\partial_{\widehat{\Theta}} \overline{\kappa}_M}{\overline{\kappa}_M} 
+ \frac{\partial_{\widehat{\Theta}} N}{N}\right)\partial_X \widehat{\Theta} \nonumber \\
+ 2(1+\widehat{\Theta}) \overline{\kappa}^2_M N \left[\textrm{sinh}(\widehat{\Psi}) \partial_{\widehat{\Theta}} \widehat{\Psi} 
- \textrm{sinh}(\widehat{\Psi}_\textrm{c})\partial_{\widehat{\Theta}} \widehat{\Psi}_\textrm{c}\right] \partial_X \widehat{\Theta}.
\end{eqnarray}
With this, $\Omega$ can be written as
\begin{eqnarray}
\label{Eq:Ips_integral_2} \Omega = 2 (1+\widehat{\Theta}) \overline{\kappa}^2_M N \partial_X \widehat{\Theta} 
\int\! \int \left[\textrm{sinh}(\widehat{\Psi}) \partial_{\widehat{\Theta}} \widehat{\Psi} 
- \textrm{sinh}(\widehat{\Psi}_\textrm{c})\partial_{\widehat{\Theta}} \widehat{\Psi}_\textrm{c}\right] d^2Z 
- \partial_X \widehat{\Theta} \int \partial_Z \Psi \partial_{\widehat{\Theta}} \Psi dZ  \nonumber \\
+ \left(\frac{1}{1+\widehat{\Theta}} + \frac{\partial_{\widehat{\Theta}} N}{N}
+ 2\frac{\partial_{\widehat{\Theta}} \overline{\kappa}_M}{\overline{\kappa}_M} 
+ M T_0\right) \partial_X \widehat{\Theta} \int\! \int\left(\partial_Z \Psi \right)^2d^2Z.\:\:\:\:\:\:\:\:\:\:\:\:\:\:\:\:\:\:\:\:\:\:
\end{eqnarray}

In the following, the derivatives with respect to $\widehat{\Theta}$ need be expressed. Considering the definition of 
$\overline{\kappa}_M$ one has
\begin{equation}
\label{Eq:ddTheta_KP_permit} \frac{\partial_{\widehat{\Theta}} \overline{\kappa}_M}{\overline{\kappa}_M} = -\frac{1}{2} M T_0,
\end{equation}
i.e. the last two terms in the round bracket of (\ref{Eq:Ips_integral_2}) (related to the temperature dependence of 
the dielectric permittivity) exactly cancel each other. Furthermore, given the definition of $\widehat{\Psi}$, one has
\begin{equation}
\label{Eq:ddThetaPsi} \partial_{\widehat{\Theta}} \Psi = (1+\widehat{\Theta}) \partial_{\widehat{\Theta}} \widehat{\Psi} + \widehat{\Psi},
\end{equation}
with $\partial_{\widehat{\Theta}} \widehat{\Psi} = 
\partial_{\overline{\kappa}} \widehat{\Psi} \partial_{\widehat{\Theta}} \overline{\kappa} 
+ \partial_{\widehat{\overline{\zeta}}} \widehat{\Psi} \partial_{\widehat{\Theta}} \widehat{\overline{\zeta}}$ and 
$\partial_{\widehat{\Theta}} \widehat{\overline{\zeta}} = -\widehat{\overline{\zeta}}/(1+\widehat{\Theta})$. Thus
\begin{equation}
\label{Eq:ddThetaTildePsi} \partial_{\widehat{\Theta}} \widehat{\Psi} = \partial_{\overline{\kappa}} \widehat{\Psi} \partial_{\widehat{\Theta}} \overline{\kappa} 
- \frac{\widehat{\overline{\zeta}}}{1+\widehat{\Theta}} \partial_{\widehat{\overline{\zeta}}} \widehat{\Psi}
\end{equation}
and
\begin{equation}
\label{Eq:ddThetaPsi_2} \partial_{\widehat{\Theta}} \Psi = (1+\widehat{\Theta}) \partial_{\overline{\kappa}} \widehat{\Psi} \partial_{\widehat{\Theta}} \overline{\kappa} 
- \widehat{\overline{\zeta}} \partial_{\widehat{\overline{\zeta}}} \widehat{\Psi} + \widehat{\Psi}.
\end{equation}
Moreover, with the definition of $\overline{\kappa}$ one has
\begin{equation}
\label{Eq:ddTheta_KP_Soret} \frac{\partial_{\widehat{\Theta}} \overline{\kappa}}{\overline{\kappa}} = 
-\frac{1}{2}\left(\frac{1}{1+\widehat{\Theta}} + M T_0 - \frac{\partial_{\widehat{\Theta}} N}{N}\right).
\end{equation}
In the case that the redistribution of ions in the electroneutral area is approximated by the Soret equilibrium, expression (\ref{Eq:SalinitySoret}) 
provides
\begin{equation} 
\label{Eq:SalinityGrad} \frac{\partial_{\widehat{\Theta}} N}{N} = -\overline{S}_0 T_0.
\end{equation}

An analytical solution exists for equations of the form (\ref{Eq:Poisson_1D}), which is well-described in the literature 
\citep{Burgreen:JPhysChem1964,Levine:JChemSoc1975,Keh:JCollIntScie2001}. From this, it is 
in principal possible to find analytical expression for $\partial_{\overline{\kappa}} \widehat{\Psi}$ and 
$\partial_{\widehat{\overline{\zeta}}} \widehat{\Psi}$  as well. However, in the analytical solution 
$\widehat{\Psi}$ is given implicitly, involving the incomplete elliptic integral of the first kind 
\citep{Langmuir:JChemPhys1938}. In preliminary tests utilizing standard Matlab routines it was seen that this implicit 
character impedes obtaining $\widehat{\Psi}$ in a reliable fashion, especially at larger values of $\overline{\kappa}_0$. 
Therefore, in practice, it is preferable to evaluate these gradients numerically. To this end, (\ref{Eq:Poisson_1D}) is 
solved numerically for several values $\overline{\kappa} = \overline{\kappa}_w$ in the vicinity of a mean value 
$\overline{\kappa}_*$, where the difference $\overline{\kappa}_w - \overline{\kappa}_* = \pm w \: \Delta \overline{\kappa}$ 
is a multiple $w$ of a small increment $\Delta \overline{\kappa} \ll 1$. For instance (and for simplicity), the derivative 
$\partial_{\overline{\kappa}} \widehat{\Psi}$ can be approximated by a finite-difference (FD) scheme, using the 
discrete values of $\widehat{\Psi}(Z)_{|\overline{\kappa}_w}$ obtained in the previous step. An equivalent strategy 
can be followed to obtain a numerical approximation of $\partial_{\widehat{\overline{\zeta}}} \widehat{\Psi}$.

In the case that the EDL potential is so small that the DH approximation holds, $\Psi^{(\textrm{DH})}$ is 
given by (\ref{Eq:EDLpot_DH}). In this case, (\ref{Eq:ddThetaTildePsi}) and (\ref{Eq:ddThetaPsi_2}), 
respectively, can be simplified to read
\begin{equation}
\label{Eq:ddThetaTildePsi_2} \partial_{\widehat{\Theta}} \widehat{\Psi}^{(\textrm{DH})} 
= \partial_{\overline{\kappa}} \widehat{\Psi}^{(\textrm{DH})} \partial_{\widehat{\Theta}} \overline{\kappa} 
- \frac{\widehat{\Psi}^{(\textrm{DH})}}{1+\widehat{\Theta}}
\end{equation}
and
\begin{equation}
\label{Eq:ddThetaPsi_3} \partial_{\widehat{\Theta}} \Psi^{(\textrm{DH})} 
= (1+\widehat{\Theta}) \partial_{\overline{\kappa}} \widehat{\Psi}^{(\textrm{DH})} \partial_{\widehat{\Theta}} \overline{\kappa} 
= \partial_{\overline{\kappa}} \Psi^{(\textrm{DH})} \partial_{\widehat{\Theta}} \overline{\kappa},
\end{equation}
where
\begin{equation}
\label{Eq:ddkappa_Psi_DH_Soret} \frac{\partial_{\overline{\kappa}} \widehat{\Psi}^{(\textrm{DH})}}{\widetilde{\overline{\zeta}}} = 
\frac{\partial_{\overline{\kappa}} \Psi^{(\textrm{DH})}}{\overline{\zeta}} =
\frac{\textrm{cosh}(\overline{\kappa} Z)}{\textrm{cosh}(\overline{\kappa})} 
\left[\textrm{tanh}(\overline{\kappa} Z)Z-\textrm{tanh}(\overline{\kappa})\right].
\end{equation}
In appendix \ref{sec:app_variaEDLtemp}, the numerical evaluation of the partial derivatives 
$\partial_{\overline{\kappa}} \Psi(Z)$ and $\partial_{\widehat{\overline{\zeta}}} \widehat{\Psi}(Z)$ 
is further discussed. Results are presented for several values of the nominal Debye parameter and compared with 
solutions obtained from the DH approximation. It is shown that $\partial_{\widehat{\Theta}} \Psi$ takes significant 
values only inside the EDL, i.e. in the non-electroneutral portion of the channel. 

Within the DH approximation, after some algebra, one finds for the $\Omega$-integral expressed by 
(\ref{Eq:Ips_integral}) 
\begin{eqnarray}
\label{Eq:Ips_integral_DH} \Omega^{(\textrm{DH})} = -\frac{\overline{\zeta}^2 \partial_X \widehat{\Theta}}{8 \textrm{cosh}^2(\overline{\kappa})}
\left\{\left[\frac{1}{2}\textrm{cosh}(2\overline{\kappa} Z)-\overline{\kappa}^2 Z^2 
+ 2\overline{\kappa}^3 Z^2 \textrm{tanh}(\overline{\kappa})\right]\left(\frac{1}{1+\widehat{\Theta}}
-\frac{\partial_{\widehat{\Theta}} N}{N} \right) \right. \nonumber \\
\left. - \left[\frac{1}{2}\textrm{cosh}(2\overline{\kappa} Z)-\overline{\kappa}^2 Z^2 
- 2\overline{\kappa}^3 Z^2 \textrm{tanh}(\overline{\kappa})\right] M T_0 \right\}. \: \: \: \: \: \: \: \: \: \: \:
\end{eqnarray}

Employing (\ref{Eq:vel_ax1}), with $\partial_Z \Omega_\textrm{c} = 0$ (symmetry), the axial velocity distribution reads
\begin{eqnarray}
\label{Eq:vel_ax_DH} U^{(\textrm{DH})} = -\frac{\partial_X P_0}{2\overline{\eta}}(1-Z^2) + \frac{Ha \overline{\zeta}}{\overline{\eta} \: 
\overline{\kappa}^2_M}\overline{\textrm{E}} 
\left[\frac{\textrm{cosh}(\overline{\kappa} Z)}{\textrm{cosh}(\overline{\kappa})}-1\right] 
- \frac{Ha}{\overline{\eta} \: \overline{\kappa}^2_M}\frac{\overline{\zeta}^2 \partial_X \widehat{\Theta}}{8 \textrm{cosh}^2(\overline{\kappa})} \cdot \nonumber \\
\left\{\left[\frac{\textrm{cosh}(2\overline{\kappa} Z)-\textrm{cosh}(2\overline{\kappa})}{2}+\overline{\kappa}^2(1-Z^2)\right]
\left(\frac{1}{1+\widehat{\Theta}}-\frac{\partial_{\widehat{\Theta}} N}{N}\:-\: M T_0 \right) \nonumber \right. \\
\left. - 2\overline{\kappa}^3(1-Z^2)\textrm{tanh}(\overline{\kappa})\left(\frac{1}{1+\widehat{\Theta}}
\:-\:\frac{\partial_{\widehat{\Theta}} N}{N} + M T_0 \right)\right\}. \: \: \: \: \: \: \: \: \: \: \: \: \:
\end{eqnarray}

Expression (\ref{Eq:vel_ax_DH}) describes the axial velocity profile across the channel if, apart from axial gradients in pressure and an induced 
electric potential, a thermal gradient is present along the channel as well. This gradient may cause contributions to the axial velocity due 
to the temperature dependencies of the electrophoretic mobility and the permittivity as well as due to an axial gradient in salt 
concentration. In the latter case, $\partial_{\widehat{\Theta}} N$ can be expressed by (\ref{Eq:SalinityGrad}) if the concentration 
gradient develops in accordance with the Soret equilibrium. The expressions are applicable to any type of electrokinetic flow 
(electroosmotic pumping or generation of an induced streaming potential by pressure-driven flow) in a slit channel.

For experimental validation, it is commonly more feasible to measure the overall volumetric flow rate  $\dot{V}$ 
in a channel. Within the DH limit, integration of (\ref{Eq:vel_ax_DH}) across the channel width and expressing 
the non-dimensional parameters by their dimensionful counterparts (except for $\overline{\kappa} = \kappa h$) provides
\begin{eqnarray}
\label{Eq:flowrate_DH} \frac{\dot{V}^{(\textrm{DH})}}{2 h \Delta y} = -\frac{h^2}{3 \eta} \partial_x p_0 + \frac{\epsilon \zeta}{\eta} E 
\left[\frac{\textrm{tanh}(\overline{\kappa})}{\overline{\kappa}}-1\right] \: \: \: \: \: \nonumber \\
- \frac{\epsilon \zeta^2}{16 \eta}\partial_x T 
\left\{\left[\frac{\textrm{tanh}(\overline{\kappa})}{\overline{\kappa}}\!-\!1\!-\!\textrm{tanh}^2(\overline{\kappa})\!
+\!\frac{4}{3}\frac{\overline{\kappa}^2}{\textrm{cosh}^2(\overline{\kappa})}\right]
\left(\frac{1}{T}\!-\!\frac{\partial_T n}{n}\!-\!\frac{\epsilon_T}{\epsilon} \right) \right. \: \: \:\:  \: \: \: \: \nonumber \\
\left. \!-\!\frac{8}{3}\frac{\overline{\kappa}^3\textrm{tanh}(\overline{\kappa})}{\textrm{cosh}^2(\overline{\kappa})}\left(\frac{1}{T}
\!-\!\frac{\partial_T n}{n}\!+\!\frac{\epsilon_T}{\epsilon} \right)\right\}, \: \: \: \:
\end{eqnarray}
where $\Delta y$ denotes the extension of the channel in $y$-direction. If the salinity varies with temperature 
according to the Soret equilibrium (\ref{Eq:SalinityGrad}), then 
$\partial_T n/n \equiv \partial_{\widehat{\Theta}} N/(N T_0) =-\overline{S}_0$.

\subsubsection{Thermoosmotic fluid propulsion}\label{subsubsec:ThermoOsmoticFlow}

In isothermal electrokinetic flow through channels, the advective fluid motion described by $U$ (or $U^{(\textrm{DH})}$, respectively) 
is driven either by an externally applied pressure gradient or an externally applied electric field. However, under 
the present assumptions and non-isothermal conditions ($\partial_X \widehat{\Theta} \neq 0$), an axial advection is 
-counterintuitively- present even without a pressure difference ($\partial_X P_0 = 0$) and without an external 
field ($\overline{\textrm{E}} = 0$). 

\begin{figure}
  \centerline{\includegraphics[width=13cm]{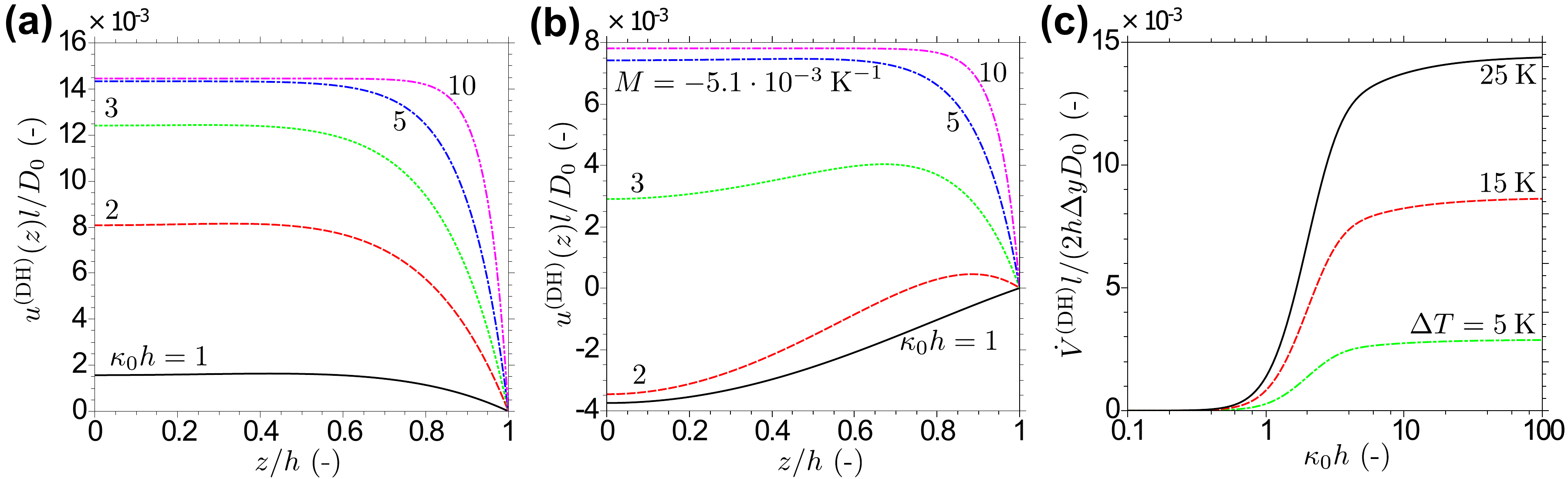}}
  \caption{Illustration of thermoosmotic fluid propulsion. (a) Thermoosmotic axial velocity profile $u^{(\textrm{DH})}(z)$, 
	expressed by (\ref{Eq:vel_ax_DH}) relative to the axial diffusion speed $D_0/l$ (Fickian diffusion coefficient set to 
	$D_0 = 10^{-9}\:\textrm{m}^2 \textrm{s}^{-1}$), for the nominal Debye parameters $\overline{\kappa}_0 = \kappa_0 h = [1, 2, 3, 5, 10]$, 
	$\zeta$ potential $\zeta = -25 \cdot 10^{-3}\:\textrm{V}$ and $\Delta T = 25\:\textrm{K}$. No external pressure gradient 
	or external electric field is applied ($\partial_X P_0 = \overline{\textrm{E}} = 0$). The thermophoretic ion motion is 
	considered by (\ref{Eq:SalinityGrad}) and $\overline{S}_0 = 10^{-3}\:\textrm{K}^{-1}$, while the temperature-dependent 
	dielectric permittivity is included with $M = -5.1 \cdot 10^{-3}\:\textrm{K}^{-1}$. (b) Hypothetical case, where 
	only the temperature dependence of the dielectric permittivity is included ($M = -5.1 \cdot 10^{-3}\:\textrm{K}^{-1}$), 
	whereas thermophoretic ion motion and a temperature-dependent ion mobility are absent. (c) Volumetric flow rate 
	$\dot{V}^{(\textrm{DH})}$ expressed by (\ref{Eq:flowrate_DH}), scaled to $2 h \Delta y D_0/l$, and plotted as a function 
	of $\overline{\kappa}_0$ for a temperature difference of $\Delta T = 5$, $15$, and $25\:\textrm{K}$. All other conditions and 
	parameters are identical to those used in (a).}
\label{Fig:uax}
\end{figure}

Figure \ref{Fig:uax} (a) illustrates the (dimensionful) axial velocity profile $u^{(\textrm{DH})}(z)$ given by (\ref{Eq:vel_ax_DH}) 
(valid within the limits of the DH approximation) for $\Delta T = 25\:\textrm{K}$, while $\partial_X P_0 = \overline{\textrm{E}} = 0$. 
The $\zeta$ potential is set to $\zeta = -25 \cdot 10^{-3}\:\textrm{V}$. The nominal Debye parameter equals one of the following 
values: $\overline{\kappa}_0 = \kappa_0 h = [1, 2, 3, 5, 10]$. Since $u(z)$ is proportional to the inverse of the channel length $l$ 
(not explicitly given herein), in figure \ref{Fig:uax} the ratio $u(z)/v_D$ is shown, where $v_D = D_0/l$ is an axial diffusion 
speed with the Fickian diffusion coefficient set to $D_0 = 10^{-9}\:\textrm{m}^2\:\textrm{s}^{-1}$. Thermophoretic ion motion is 
considered using (\ref{Eq:SalinityGrad}), where $\overline{S}_0 = 10^{-3}\:\textrm{K}^{-1}$, while a temperature-dependent dielectric 
permittivity is included with $M = -5.1 \cdot 10^{-3}\:\textrm{K}^{-1}$. From this plot it is 
apparent that the maximal thermoosmotic velocity is still two to three orders of magnitude smaller than the diffusion speed. For instance, 
for a channel length given by $l = \overline{\kappa}_0/A \kappa_0$ with $\overline{\kappa}_0 = 1$, $A = 0.1$ and 
$\kappa_0^{-1} \approx 10^{-7}\:\textrm{m}$ (i.e. $l \approx 10^{-6}\:\textrm{m}$) one obtains 
$u(z) \lesssim 10^{-5}\:\textrm{m}\:\textrm{s}^{-1}$.

At a casual glance, the velocity profiles plotted in figure \ref{Fig:uax} (a) look similar to typical profiles obtained for 
electroosmotic flow, resembling a plug-like structure for larger values of $\overline{\kappa}_0$. However, despite being not clearly 
visible on the scale of the plot, the largest velocity of thermal origin does not necessarily develop along the channel center plane 
at $z = 0$. Rather than that, for sufficiently small $\overline{\kappa}_0$, one can estimate the location of the axial velocity peak 
to occur at
\begin{equation}
\label{Eq:z_uax_peak} \frac{z}{h} \approx \sqrt{-3 \frac{\textrm{tanh}(\overline{\kappa}_0)}{\overline{\kappa}_0}
\frac{1+M T_0-\partial_{\widehat{\Theta}} N/N}{1-M T_0-\partial_{\widehat{\Theta}} N/N}}.
\end{equation}
Since typically $M < 0$ and $\partial_{\widehat{\Theta}} N/N < 0$, this equation has a real solution $0<z/h<1$ only if 
the temperature dependence of the dielectric permittivity is sufficiently strong. Thus, this dependence causes the development 
of a double velocity maximum in the slit channel. For instance, for the cases shown one finds 
$[\overline{\kappa}_0,z/h] \approx [1,0.43]$, $[2,0.34]$, $[3,0.28]$, 
$[5,0.22]$ and $[10,0.16]$ (accurate solution obtained by a fix-point iteration are $[1,0.42]$, $[2,0.33]$, $[3,0.27]$, 
$[5,0.20]$ and $[10,0.13]$). Hence, for increasing $\overline{\kappa}_0$, the location of the velocity maximum approaches 
the channel center plane.

This effect can be enhanced by (artificially) eliminating the thermophoretic ion motion and the temperature dependence 
of the ion mobility altogether. The velocity profiles for this (hypothetical) case are shown in figure \ref{Fig:uax} (b) 
for the same parameter values as depicted in (a). 
Here, the deviation from the single maximum profile is especially visible for smaller values of $\overline{\kappa}_0$. For 
$\overline{\kappa}_0 = 2$, the axial flow even switches direction as a function of $z$, i.e. it is in one 
direction close to the wall and in the opposite direction close to the channel center plane. For $\overline{\kappa}_0 = 1$, the 
axial flow is opposite to the one observed for larger values of $\overline{\kappa}_0$ throughout the channel 
cross section. The reason for this behavior lies in the multitude of mechanisms to which a temperature-dependent dielectric 
permittivity contributes: Firstly, it adds an extra-source term to the Maxwell stresses 
in the Navier-Stokes equation. Secondly, it alters the electrohydrodynamic pressure of the ions and lastly, it modifies the 
Poisson equation. The latter has direct implications on the specific form of the EDL potential.

In figure \ref{Fig:uax} (c) the volumetric flow rate $\dot{V}$ in the DH limit as given by (\ref{Eq:flowrate_DH}) 
and scaled to $2 h \Delta y D_0/l$ is plotted as a function of $\overline{\kappa}_0$ for a temperature difference of 
$\Delta T = 5$, $15$, and $25\:\textrm{K}$. The other parameters are identical to those of figure \ref{Fig:uax} (a), in particular 
$\partial_X P_0 = \overline{\textrm{E}} = 0$ as well. After a steep increase within $1 \lesssim \overline{\kappa}_0 \lesssim 3$, 
it is found that $\dot{V}^{(\textrm{DH})}$ quickly saturates to
\begin{equation}
\label{Eq:thermoosmotic_flowrate_DH_max} \frac{1}{2 h \Delta y D_0} 
\left(\frac{\dot{V}}{\partial_x T}\right)^{(\textrm{DH})}_{|\overline{\kappa}_0 \rightarrow \infty} 
= \frac{\epsilon \zeta^2}{8 \eta D_0} \left(\frac{1}{T}\!-\!\frac{\partial_T n}{n}\!-\!\frac{\epsilon_T}{\epsilon} \right),
\end{equation}
when $\overline{\kappa}_0 \gtrsim 10$ and for all $\Delta T$ ($\partial_x T = \Delta T/l$). For the parameter set used in figure \ref{Fig:uax}, 
(\ref{Eq:thermoosmotic_flowrate_DH_max}) equals $5.8 \cdot 10^{-4}\: \textrm{K}^{-1}$.

\begin{figure}
	\centerline{\includegraphics[width=6.5cm]{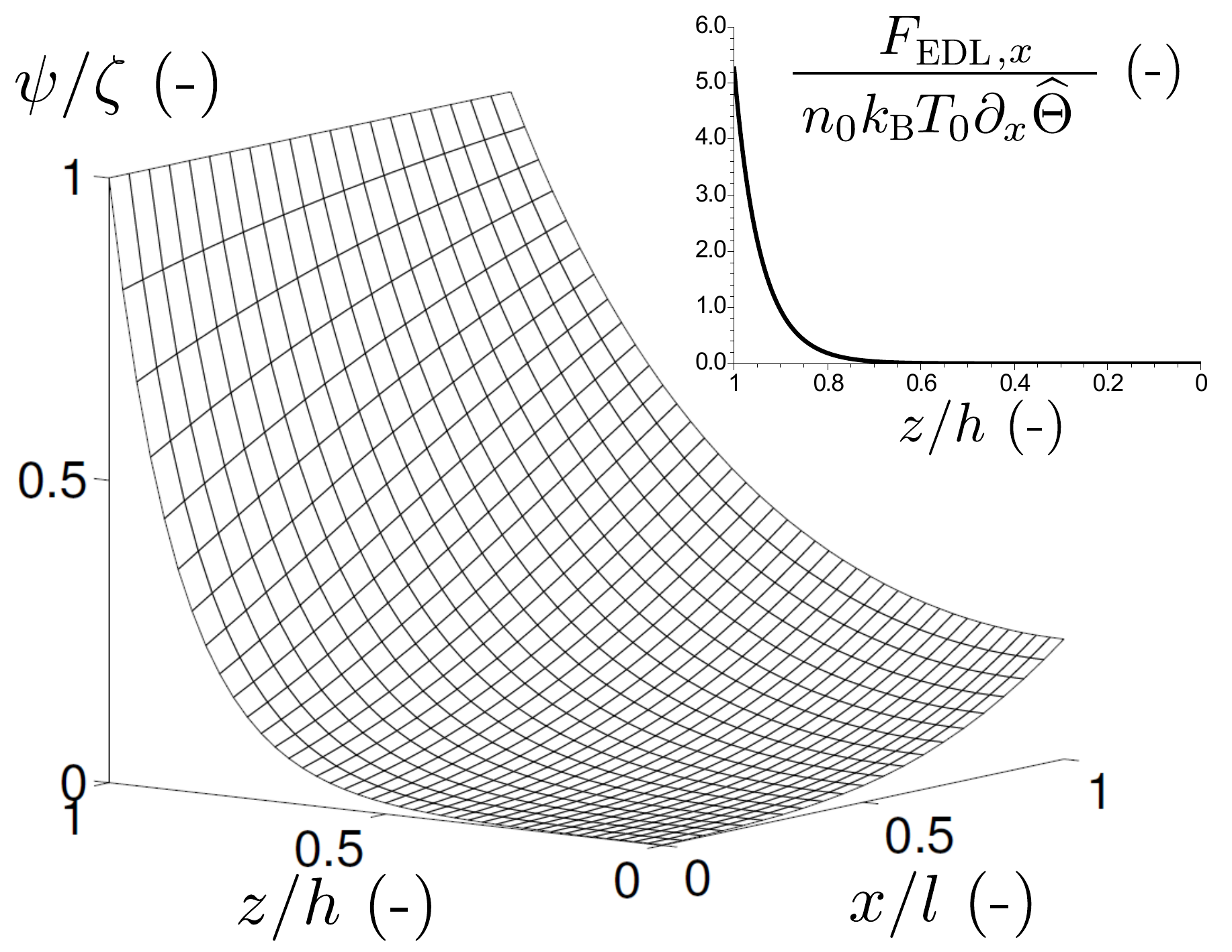}}
	\caption{EDL potential $\psi(x,z)/\zeta$ in the channel for the case of thermophoretic ion motion due 
	to Soret A alone, i.e. both ion species have the same thermophoretic mobility. The nominal Debye parameter 
	is $\overline{\kappa}_0 = 8$ and, for illustrative purposes, $S \Delta T = 2.5$ was chosen (plot is 
	exaggerated). Inset: Effective axial body force density (\ref{Eq:eff_bodyforce_EDLcontrsimple}) due to the interaction 
	of the ion cloud with the charges at the wall boundaries. The externally applied pressure gradient and the 
	induced field are absent ($\partial_X P_0 = \overline{\textrm{E}} = 0$).}
	\label{Fig:EDLprop_schematic}
\end{figure}

The origin of the observed fluid propulsion if only a thermal gradient is present needs some further explanation. 
If $\partial_X \widehat{\Theta} \neq 0$, the EDL thickness varies along the channel, i.e. the EDL potential becomes 
dependent on the axial coordinate. In figure \ref{Fig:EDLprop_schematic} this is shown schematically for the special case 
of identical thermophoretic mobilities of the ion species. In the following, we will refer to this special form of 
thermophoretic ion motion as type \textit{Soret A} ion thermo-diffusion. By contrast, thermally induced ion motion where 
only one ion species moves within a temperature gradient will be referred to as \textit{Soret B} ion thermo-diffusion. 
In figure \ref{Fig:EDLprop_schematic}, for illustrative purposes, $S_0 \Delta T = 2.5$ was chosen (Soret A), 
i.e. the plot is exaggerated. The nominal Debye parameter is set to $\overline{\kappa}_0 = 8$. The combination 
of the (weak) modification of the EDL along the axial coordinate $X$ with the steep lateral gradient of $\Psi$ 
inside the EDL leads to an axial electric field confined to the EDL only, see appendix \ref{sec:app_variaEDLtemp}. 
In turn, this field causes an axial gradient in the electrohydrostatic pressure (being equivalent to the osmotic 
pressure gradient $-k_\textrm{B} T \partial_x n$, where $n = \sum^K_{k=1} n_k$) as well as an axial electromotive 
body force $-\rho_f \partial_x \psi$ (excluding for simplicity the contributions from a temperature-dependent 
dielectric permittivity), which only partially cancel each other. Hence, by contrast to isothermal conditions, the 
ion cloud is not necessarily in mechanical equilibrium. Instead, as shown in appendix \ref{sec:app_streamfct} (expression 
(\ref{Eq:eff_bodyforce_EDLcontr})), a net force density acts on the excess ions in the EDL. Using (\ref{Eq:SalinityGrad}), 
this force density can be described by
\begin{equation}
\label{Eq:eff_bodyforce_EDLcontrsimple} \frac{F_{\textrm{EDL},x}}{k_\textrm{B} T_0 n_0 \partial_x \widehat{\Theta}} 
\approx 2 \left\{[\textrm{cosh}(\Psi^{(\textrm{e})})-1]S_0 T_0 + \Psi^{(\textrm{e})} \textrm{sinh}(\Psi^{(\textrm{e})}) \right\}, 
\end{equation}
The first term in the curly bracket on the RHS expresses the contribution by the thermophoretic ion motion, 
while the second describes the thermo-mechanical effect of a temperature-dependent ion mobility. In this 
linearized form, $F_{\textrm{EDL},x}$ does not vary along the channel center plane and is plotted in the inset of figure 
\ref{Fig:EDLprop_schematic}. As detailed in appendix \ref{sec:app_streamfct}, this force density does not appear to be an 
artifact of the lubrication approximation or the specific scaling used in the present derivation. Being 
proportional to the axial temperature gradient, $F_{\textrm{EDL},x}$ vanishes if the temperature is uniform. Hence, 
under isothermal conditions, the EDL field itself cannot propel the surrounding fluid.

Essentially, regardless of the physical origin, any modification of the EDL causing an imbalance between 
the electromotive and the osmotic pressure should lead to a similar effect. For instance, the essential 
difference of this work compared to the one by \citet{Sasidhar:JCollIntScie1982} is that in the latter an 
axial concentration gradient is applied externally as an independent parameter, while here, 
it is caused by the axial temperature gradient. For large $\overline{\kappa}_0 \rightarrow \infty$, the 
corresponding fluid motion may be regarded as \textit{thermally induced, apparent double layer slip velocity}.

As detailed in the following sections, the thermoosmotic flux driven by a mechanical imbalance of the ion cloud 
in the EDL as described here also contributes to the induced electric field. It should be emphasized that, in this 
study, the incorporation of the Korteweg-Helmholtz electric force in the momentum equation is of crucial importance; 
otherwise the described phenomenon cannot be correctly captured. This distinguishes this study from common 
considerations of isothermal electrokinetics where the force term is not necessary to compute the streaming potential, 
but is merely included to predict the (commonly weak) electroosmotic flow, opposing the pressure-induced flow, and 
to fulfill the Onsager reciprocal condition.

\subsection{Electric currents and induced potential}\label{subsec:currents_potentials}

Besides the EDL potential derived in \S \ref{subsec:EDL}, an induced or externally applied potential may be present, 
which does not affect the local charge density. In studies of pressure driven systems kept at uniform temperature, 
the induced potential is identical to the so-called streaming potential and can be determined by multiplying 
(\ref{Eq:NPE_ax}) by $\overline{\nu}_k$, summing over all ion species, integrating across the channel section 
and setting the result equal to zero. Applying the same procedure to the system considered here results in
\begin{eqnarray} 
\label{Eq:stream_pot_balance}  -\frac{\overline{\textrm{E}}}{1+\widehat{\Theta}} \int^1_0 \sum^K_{k=1} \frac{\overline{\nu}^2_k N_k}{\Pen_k} dZ
+ \frac{\partial_X \widehat{\Theta}}{(1+\widehat{\Theta})^2} \int^1_0 \Psi \sum^K_{k=1} \frac{\overline{\nu}^2_k N_k}{\Pen_k} dZ
+ T_0 \partial_X \widehat{\Theta} \int^1_0 \sum^K_{k=1} \frac{\overline{\nu}_k N_k S_k}{\Pen_k} dZ \nonumber \\
+ \frac{\partial_X N}{N} \int^1_0 \sum^K_{k=1} \frac{\overline{\nu}_k N_k}{\Pen_k} dZ
= \int^1_0 U \sum^K_{k=1} \overline{\nu}_k N_k dZ. \:\:\:\:\:\:\:\:\:\:\:\:\:
\end{eqnarray}

\subsubsection{Streaming current}\label{subsubsec:streaming_current}

When integrated over the extension of the channel in y-direction $\Delta y$, the term on the RHS of (\ref{Eq:stream_pot_balance}) represents 
the total streaming current $I_\textrm{st} = 2 \Delta y \int_0^{h}\rho_f u dz$. According to the leading order terms in (\ref{Eq:Poisson}), the charge density 
$\rho_f/(e \nu n_0) = \sum^{K}_{k=1} \overline{\nu}_k N_k$ is equal to $-2/(\overline{\kappa}^2_M)\partial^2_Z \Psi$. Integrating by parts 
under the given boundary conditions leads to $I_\textrm{st}/(2 \Delta y u_0 e \nu n_0) =2/\overline{\kappa}^2_M \int_0^1 \partial_Z U \partial_Z \Psi dZ$. 
With an expression for $\partial_Z U$ derived from (\ref{Eq:vel_ax1}), this leads to
\begin{equation}
\label{Eq:I_stream} \frac{I_\textrm{st} \nu e}{2 \Delta y \kappa_M u_0 \epsilon k_\textrm{B} T_0} = \overline{I}_{\textrm{st},P} \partial_X P_0 
- \overline{I}_{\textrm{st},\overline{\textrm{E}}} \overline{\textrm{E}} + \overline{I}_{\textrm{st},\Theta} \partial_X \Theta,
\end{equation}
where
\begin{eqnarray}
\label{Eq:I_stream_integrals} \overline{I}_{\textrm{st},P} = \frac{1}{\overline{\kappa}_M \: \overline{\eta}}\left(\overline{\zeta} - \int_0^1{\Psi dZ} \right), \ \ \
\overline{I}_{\textrm{st},\overline{\textrm{E}}} = - \frac{Ha}{\overline{\kappa}^3_M \: \overline{\eta}} \int_0^1{\left(\partial_Z \Psi\right)^2 dZ}, \nonumber \\ 
\overline{I}_{\textrm{st},\Theta} = \frac{Ha}{\overline{\kappa}^3_M \: \overline{\eta}} \frac{1}{\partial_X \Theta} \int_0^1{\partial_Z \Omega \partial_Z \Psi dZ}.
\end{eqnarray}

With (\ref{Eq:Ips_integral_2}) and (\ref{Eq:ddTheta_KP_permit}), the last integral in (\ref{Eq:I_stream_integrals}) can be written as
\begin{eqnarray}
\label{Eq:Integral_ddZOmega_ddZPsi} \int_0^1{\partial_Z \Omega \partial_Z \Psi dZ} = 2 (1+\widehat{\Theta}) \overline{\kappa}^2_M N_0 \partial_X \widehat{\Theta} 
\int_0^1 \! \partial_Z \Psi \left\{ \int [\textrm{sinh}(\widehat{\Psi}) \partial_{\widehat{\Theta}} \widehat{\Psi} 
- \textrm{sinh}(\widehat{\Psi}_\textrm{c})\partial_{\widehat{\Theta}} \widehat{\Psi}_\textrm{c}] dZ \right\} dZ \nonumber \\ 
- \partial_X \widehat{\Theta} \int_0^1 (\partial_Z \Psi)^2 \partial_{\widehat{\Theta}} \Psi dZ + \left(\frac{1}{1+\widehat{\Theta}} 
+ \frac{\partial_{\widehat{\Theta}} N}{N}\right) \partial_X \widehat{\Theta} \int_0^1 \! \partial_Z \Psi \left[ \int\left(\partial_Z \Psi\right)^2 dZ \right]dZ.
\: \: \: \: \: \: \: \: \: \: \: \: \: \: \:
\end{eqnarray}
In general, $\widehat{\Psi}$ is obtained by solving numerically (\ref{Eq:Poisson_1D}), while $\partial_Z \Psi$ is given by (\ref{Eq:ddZPsi2}). 
Furthermore, $\partial_{\widehat{\Theta}} \widehat{\Psi}$ is obtained from (\ref{Eq:ddThetaTildePsi}) by the numerical approach outlined before, while 
$\partial_{\widehat{\Theta}} \Psi$ is computed from (\ref{Eq:ddThetaPsi_2}). Thus, the integrals in (\ref{Eq:I_stream_integrals}) and 
(\ref{Eq:Integral_ddZOmega_ddZPsi}) are fully described and can be evaluated numerically.

Within the DH approximation and after some algebraic manipulations, the (dimensionless) streaming current reads
\begin{eqnarray}
\label{Eq:I_stream_DH} \frac{\left(I_\textrm{st}\right)^{(\textrm{DH})} \nu e}{2 \Delta y \kappa_M u_0 \epsilon k_\textrm{B} T_0} = 
\frac{\overline{\zeta} \partial_X P_0}{\overline{\kappa}_M \overline{\eta}} 
\left[1 - \frac{\textrm{tanh}(\overline{\kappa})}{\overline{\kappa}}\right] \nonumber \\
+ \frac{Ha \overline{\zeta}^2 \overline{\textrm{E}}}{2\overline{\kappa}_M \overline{\eta}} \left(\frac{\overline{\kappa}}{\overline{\kappa}_M}\right)^2
\left[\frac{\textrm{tanh}(\overline{\kappa})}{\overline{\kappa}} - \frac{1}{\textrm{cosh}^2(\overline{\kappa})}\right] 
- \frac{Ha \overline{\zeta}^3 \partial_X \widehat{\Theta}}{2\overline{\kappa}_M \: \overline{\eta}} 
\left(\frac{\overline{\kappa}}{\overline{\kappa}_M}\right)^2 \cdot \nonumber \\
\left\{-\frac{\partial_{\widehat{\Theta}} N}{N} \left[\frac{\overline{\kappa} \textrm{tanh}(\overline{\kappa}) 
- \textrm{tanh}^2(\overline{\kappa})-1/2}{\textrm{cosh}^2(\overline{\kappa})} + \frac{\textrm{tanh}(\overline{\kappa})}{2\overline{\kappa}} 
- \frac{\textrm{tanh}^3(\overline{\kappa})}{3\overline{\kappa}}\right] \right. \nonumber \\
\left. + \frac{1}{1+\widehat{\Theta}}\left[\frac{\overline{\kappa} \textrm{tanh}(\overline{\kappa}) 
- \textrm{tanh}^2(\overline{\kappa})-1/2}{\textrm{cosh}^2(\overline{\kappa})} + \frac{\textrm{tanh}(\overline{\kappa})}{2\overline{\kappa}} 
- \frac{\textrm{tanh}^3(\overline{\kappa})}{3\overline{\kappa}}\right] \right. \nonumber \\
\left. + M T_0 \left[\frac{\overline{\kappa} \textrm{tanh}(\overline{\kappa}) 
- \textrm{tanh}^2(\overline{\kappa})+1/2}{\textrm{cosh}^2(\overline{\kappa})} - \frac{\textrm{tanh}(\overline{\kappa})}{2\overline{\kappa}}  
+ \frac{\textrm{tanh}^3(\overline{\kappa})}{3\overline{\kappa}} \right]\right\},
\end{eqnarray}
where 
\begin{equation}
\label{Eq:KP_ratio} \left(\frac{\overline{\kappa}}{\overline{\kappa}_M}\right)^2 = \frac{N}{1+\widehat{\Theta}}.
\end{equation}
The first term in the curly bracket of (\ref{Eq:I_stream_DH}) denotes the contribution due to a gradient in salt 
concentration, the second quantifies the contribution due the temperature dependence of the electrophoretic ion mobility 
and the last term accounts for the temperature-dependent dielectric permittivity.

\subsubsection{Conduction current}\label{subsubsec:conduction_current}

Going back to (\ref{Eq:stream_pot_balance}), the LHS represents the conduction current $-I_\textrm{cd}$. Since the temperature gradient across the channel is 
negligibly small, in dimensionless form $I_\textrm{cd}$ is given by
\begin{eqnarray}
\label{Eq:I_cd} \frac{I_\textrm{cd} e \nu}{2 \Delta y \kappa_M u_0 \epsilon k_\textrm{B} T_0} = \frac{\overline{\textrm{E}}}{1+\widehat{\Theta}}
\frac{\overline{\kappa}_M}{2\Pen} \int^1_0 \sum^K_{k=1} \overline{\nu}^2_k N_k dZ
- \frac{\partial_X \widehat{\Theta}}{(1+\widehat{\Theta})^2} \frac{\overline{\kappa}_M}{2\Pen} \int^1_0 \Psi \sum^K_{k=1} \overline{\nu}^2_k N_k dZ \nonumber \\
- T_0 \partial_X \widehat{\Theta}\frac{\overline{\kappa}_M}{2\Pen} \int^1_0 \sum^K_{k=1} \overline{\nu}_k N_k S_k dZ 
- \frac{\partial_{\widehat{\Theta}} N \partial_X \widehat{\Theta}}{N} \frac{\overline{\kappa}_M}{2\Pen} \int^1_0 \sum^K_{k=1} 
\overline{\nu}_k N_k dZ,\: \: \: \: \: \: \: \: \:
\end{eqnarray}
where identical (but not necessarily constant) Fickian diffusion coefficients for each ion species (i.e. $D_k \equiv D$) are 
assumed. Note that this is the first time within the derivation where such an assumption is made. For symmetric electrolytes, 
employing (\ref{Eq:ion_distribution}), the sums can be further evaluated to read
\begin{equation}
\label{Eq:I_cd2} \frac{I_\textrm{cd} \nu e}{2\Delta y \kappa_M u_0 \epsilon k_\textrm{B} T_0} = -\overline{I}_{\textrm{cd},\overline{\textrm{E}}} \overline{\textrm{E}} 
+ \overline{I}_{\textrm{cd},\widehat{\Theta}} \partial_X \widehat{\Theta},
\end{equation}
where
\begin{equation}
\label{Eq:I_cd_potential} \overline{I}_{\textrm{cd},\overline{\textrm{E}}} = 
-\frac{N}{1+\widehat{\Theta}} \frac{\overline{\kappa}_M}{\Pen} \int_0^1{\textrm{cosh}(\widehat{\Psi}) dZ},
\end{equation}
and
\begin{eqnarray}
\label{Eq:I_cd_temperature} \overline{I}_{\textrm{cd},\widehat{\Theta}} = -\frac{N}{1+\widehat{\Theta}} \frac{\overline{\kappa}_M}{\Pen}
\int_0^1{\widehat{\Psi}\textrm{cosh}(\widehat{\Psi}) dZ} 
+ S_- T_0 N \frac{\overline{\kappa}_M}{\Pen}\int_0^1{\textrm{sinh}(\widehat{\Psi}) dZ} \nonumber \\
- \frac{1}{2} \Delta S T_0 N \frac{\overline{\kappa}_M}{\Pen}\int_0^1{\textrm{exp}(-\widehat{\Psi}) dZ}
+ \frac{\partial_{\widehat{\Theta}} N}{N} \frac{\left(\partial_Z \Psi \right)_\textrm{s}}{\overline{\kappa}_M \Pen}.
\end{eqnarray}
The intrinsic Soret coefficient of the anion is denoted by $S_-$, while $S_+$ describes the intrinsic Soret coefficient of the cation, and 
$\Delta S = S_+ - S_-$. Note that the conduction current expressed by (\ref{Eq:I_cd2}) still includes the temperature dependencies 
of the electrophoretic mobility and of the (Fickian) diffusion coefficient.

The DH approximation yields
\begin{eqnarray}
\label{Eq:I_cd_DH} \frac{(I_\textrm{cd})^{(\textrm{DH})} \nu e}{2 \Delta y \kappa_M u_0 \epsilon k_\textrm{B} T_0} = 
\frac{\overline{\textrm{E}}}{\Pen} \left(\frac{\overline{\kappa}}{\overline{\kappa}_M}\right) \overline{\kappa} 
\left\{1+\frac{\overline{\zeta}^2}{4(1+\widehat{\Theta})^2} \left[\frac{\textrm{tanh}(\overline{\kappa})}{\overline{\kappa}} 
+ \frac{1}{\textrm{cosh}^2(\overline{\kappa})} \right]\right\} \nonumber \\
-\frac{\partial_X \widehat{\Theta}}{\Pen}\left(\frac{\overline{\kappa}}{\overline{\kappa}_M}\right)\overline{\kappa}\left\{
\frac{\overline{\zeta}}{1+\widehat{\Theta}} \frac{\textrm{tanh}(\overline{\kappa})}{\overline{\kappa}} 
+ \frac{\overline{\zeta}^3}{2(1+\widehat{\Theta})^3}\frac{\textrm{tanh}(\overline{\kappa})}{\overline{\kappa}}\left[\frac{\textrm{tanh}^2(\overline{\kappa})}{3}
+ \frac{1}{\textrm{cosh}^2(\overline{\kappa})} \right] \right. \nonumber \\
\left. - \overline{\zeta} \left[\frac{1}{2}(2S_- + \Delta S) T_0 + \frac{\partial_{\widehat{\Theta}} N}{N}\right] \textrm{tanh}(\overline{\kappa}) 
+ \frac{1}{2}\Delta S T_0 (1+\widehat{\Theta}) \overline{\kappa} \right\}.\:\:\:\:\:\:\:\:\:\:\:\:\:
\end{eqnarray}

\subsubsection{Induced electric field dominated by external pressure difference}\label{subsubsec:general_streaming_potential}

The induced electric field must fulfill (\ref{Eq:stream_pot_balance}), i.e. $I_\textrm{st} + I_\textrm{cd} = 0$. From (\ref{Eq:I_stream}) 
and (\ref{Eq:I_cd2}), one finds for the local (dimensionless)
streaming field
\begin{equation}
\label{Eq:induced_pot} \overline{\textrm{E}} = \frac{\overline{I}_{\textrm{st},P} \partial_X P_0 + \left(\overline{I}_{\textrm{st},\widehat{\Theta}} 
+ \overline{I}_{\textrm{cd},\widehat{\Theta}}\right) \partial_X \widehat{\Theta}}{\overline{I}_{\textrm{st},\overline{\textrm{E}}} 
+ \overline{I}_{\textrm{cd},\overline{\textrm{E}}}}.
\end{equation}

At first, we want to consider slit channels where the induced field is mainly caused by the 
externally applied pressure gradient $\partial_x p_0$. The latter was introduced as an integration constant 
in the course of the derivation of (\ref{Eq:momentum_x2}), which implies that $p_0$ is not necessarily 
identical to the total fluid pressure. The induced field one obtains corresponds to the conventional streaming 
field known for isothermal systems. Hence, if the dimensionless externally applied pressure gradient is much larger 
than the dimensionless thermal gradient, we well refer to the induced field as the streaming field also under 
non-isothermal conditions. After reinserting the definitions of the dimensionless parameters to obtain dimensional 
values, the local streaming field per applied pressure gradient is given by
\begin{equation}
\label{Eq:stream_pot} \frac{E_p}{\partial_x p_0} = 
\frac{k_\textrm{B} T_0}{e \nu} \frac{A \overline{\kappa}_0}{\kappa_0 u_0 \eta_0}\frac{\overline{\textrm{E}}}{\partial_X P_0},
\end{equation}
where $\overline{\textrm{E}}$ is given by (\ref{Eq:induced_pot}) and the subscript $p$ added to $E$ marks the streaming 
field induced by an external pressure difference. Note that while the characteristic parameters $A$ and $u_0$ need to stay 
in a certain range so that the simplified governing equations remain valid, the value of the local streaming field expressed by
(\ref{Eq:stream_pot}) does not depend on the specific choice of these parameters. This is because the dimensionless 
functions $\overline{I}_{\textrm{st},\varphi}$ ($\varphi=P_0,\overline{\textrm{E}},\widehat{\Theta}$) and 
$\overline{I}_{\textrm{cd},\varrho}$ ($\varrho=\overline{\textrm{E}},\widehat{\Theta}$) depend on these parameters in a reciprocal 
fashion (in comparison with the pre-factor) so that the specific values cancel out. This can best be verified by considering 
the local streaming field in the DH limit, which reads 
\begin{eqnarray}
\label{Eq:stream_pot_DH} -\left(\frac{E_p}{\partial_x p_0}\right)^{(\textrm{DH})} = 
\left(F_\textrm{CS}\:+\:\frac{\zeta^2}{2} \frac{\epsilon}{D \eta} F_\alpha \right)^{-1}
\left\{\frac{\zeta}{\kappa^2 D \eta} \left[1\:-\:\frac{\textrm{tanh}(\overline{\kappa})}{\overline{\kappa}}\right] \right. \nonumber \\
\left. + \frac{\zeta^3}{2}\frac{\epsilon}{D \eta} \frac{\Delta T/T_0}{\Delta p_0} \frac{\partial_{\widehat{\Theta}}N}{N}
\left[F_{\beta}\:-\:\frac{\textrm{tanh}(\overline{\kappa})}{\overline{\kappa}}\left(\frac{\textrm{tanh}^2(\overline{\kappa})}{3}\:
-\:\frac{1}{2}\right)\right] \right. \nonumber \\
\left. - \frac{\zeta^3}{2}\frac{\epsilon}{D \eta} \frac{\Delta T/T_0}{\Delta p_0}\frac{T_0}{T}
\left[F_{\beta}\:-\:\frac{\textrm{tanh}(\overline{\kappa})}{\overline{\kappa}}\left(\frac{\textrm{tanh}^2(\overline{\kappa})}{3}\:
-\:\frac{1}{2}\right) \right] \right. \nonumber \\
\left. - \frac{\zeta^3}{2}\frac{\epsilon}{D \eta} \frac{M \Delta T}{\Delta p_0}
\left[F_{\beta}\:+\frac{1}{\textrm{cosh}^2(\overline{\kappa})}\:
+\:\frac{\textrm{tanh}(\overline{\kappa})}{\overline{\kappa}}\left(\frac{\textrm{tanh}^2(\overline{\kappa})}{3}\:-\:\frac{1}{2}\right)\right] \right. \nonumber \\
\left. - \zeta \frac{\Delta T/T_0}{\Delta p_0}\frac{T_0}{T}\frac{\textrm{tanh}(\overline{\kappa})}{\overline{\kappa}}
\left[1+\:\frac{\overline{\zeta}^2}{2}\left(\frac{T_0}{T}\right)^2\left[\frac{\textrm{tanh}^2(\overline{\kappa})}{3}\:
+\:\frac{1}{\textrm{cosh}^2(\overline{\kappa})}\right] \right] \right. \nonumber \\
\left. + \zeta \frac{\Delta T/T_0}{\Delta p_0}\frac{\textrm{tanh}(\overline{\kappa})}{\overline{\kappa}}
\left[\frac{1}{2}\left(2S_-\:+\:\Delta S\right)T_0\:+\:\frac{\partial_{\widehat{\Theta}}N}{N}\right]\:
-\:\frac{1}{2}\frac{\Delta S \Delta T}{\Delta p_0} \frac{k_\textrm{B} T}{e \nu} \right\}, \ \ \ \ \ \ \
\end{eqnarray}
where $\Delta T/\Delta p_0 \equiv \partial_x T/\partial_x p_0$ (the -externally applied- temperature and 
pressure gradients are constant herein) and $T = T_0 + \Delta T$ is the temperature at position $x=l$. Furthermore,
\begin{equation}
\label{Eq:F_CS} F_\textrm{CS} = \int_0^1 \textrm{cosh}(\widehat{\Psi}) dZ \approx 1\:
+\:\left(\frac{e \nu \zeta}{2 k_\textrm{B} T_0}\right)^2 \left(\frac{T_0}{T}\right)^2
\left[\frac{\textrm{tanh}(\overline{\kappa})}{\overline{\kappa}} + \frac{1}{\textrm{cosh}^2(\overline{\kappa})}\right]
\end{equation}
and
\begin{equation}
\label{Eq:F_alpha2_F_beta} F_\alpha = \frac{\textrm{tanh}(\overline{\kappa})}{\overline{\kappa}} - \frac{1}{\textrm{cosh}^2(\overline{\kappa})}, \ \ \
F_\beta = \frac{\overline{\kappa} \textrm{tanh}(\overline{\kappa}) - \textrm{tanh}^2(\overline{\kappa}) - 1/2}{\textrm{cosh}^2(\overline{\kappa})}.
\end{equation}
In (\ref{Eq:stream_pot_DH}), since the hydrodynamic radii of the dissolved ions are practically unaffected by temperature, 
the identity $D \eta /T = D_0 \eta_0 /T_0 = const$ holds as a good approximation and can be incorporated if $\kappa^2 D \eta$ is 
replaced by $\kappa^2_0 D_0 \eta_0 N \epsilon/\epsilon_0$. In other words, in the term describing the contribution by pressure-induced 
streaming (first term on the RHS of (\ref{Eq:stream_pot_DH})), the temperature dependencies of the Fickian diffusion coefficient, 
viscosity and of the ionic mobility exactly cancel. This is accurate in the limit of identical Fickian diffusion coefficients for each 
ion species but should still be a reasonably good approximation in the more realistic case where they differ slightly from each other.
As mentioned, equation (\ref{Eq:stream_pot_DH}) indeed does not depend on $A$ or $u_0$. It is a first order approximation in $A$ and requires 
that $A^2\ll 1$, $\Pen_k\leq {\cal O}(1)$, $\Delta T/T < 1$, $S_k \Delta T \leq {\cal O}(1)$, 
$M \Delta T \leq {\cal O}(1)$, $\Rey \leq {\cal O}(A)$, $Pe_T \leq {\cal O}(A)$ and $Ha/\overline{\kappa}^2_0 \leq {\cal O}(1)$.

For a non-zero temperature difference, since $\overline{\kappa}$, $\epsilon$, $\eta$ and $D$ depend on the local 
temperature, the streaming field according to (\ref{Eq:stream_pot}), respectively (\ref{Eq:stream_pot_DH}), is not a constant but a (weak) 
function of the axial coordinate $x$. The total streaming potential difference per applied pressure difference along the length of the 
channel, $\Delta \phi_{\textrm{st},p}/\Delta p_0$, has to be computed by numerical integration of (\ref{Eq:stream_pot}), respectively of 
(\ref{Eq:stream_pot_DH}), from $x = 0$ to $x = l$.

\section{Analysis of specific cases}\label{sec:results}

\subsection{Validation}\label{subsec:verification}

For a vanishing temperature difference, (\ref{Eq:stream_pot_DH}) agrees with the well-known expression of the streaming potential for 
slit channels with small $\zeta$ potentials discussed in the literature \citep{Masliyah:JohnWiley2009}. Furthermore, for channels 
subjected to an axial temperature gradient but without any surface charge ($\zeta \equiv 0$), one finds
\begin{equation}
\label{Eq:stream_pot_zero_surface_charge} \left(\frac{E_T}{\partial_x T}\right)^{(\textrm{DH})}_{|\zeta \rightarrow 0} = 
\frac{1}{2}\Delta S \frac{k_\textrm{B} T}{e \nu}.
\end{equation}
This represents the conventional Soret equilibrium for a symmetric electrolyte expressed by (\ref{Eq:thermoelectric_pot_app}). 
Finally, for the limiting case of infinitely thin EDLs compared to $h$, the nominal Debye parameter 
becomes very large ($\overline{\kappa}_0 \rightarrow \infty$ and thus also $\overline{\kappa} \rightarrow \infty$) 
and one finds
\begin{equation}
\label{Eq:stream_pot_DH_verylargeKP} -\left(\frac{E_p}{\partial_x p_0}\right)^{(\textrm{DH})}_{|\overline{\kappa}_0 \rightarrow \infty} = 
\frac{\zeta}{\kappa^2 D \eta} - \frac{1}{2}\frac{\Delta S \Delta T}{\Delta p_0} \frac{k_\textrm{B} T}{e \nu}, 
\end{equation}
which is simply the linear superposition of the (temperature-dependent) local value of the classical Smoluchowski limit 
and the contribution of the Soret equilibrium.

\subsection{Induced electric field with axial salt redistribution according to Soret equilibrium}\label{subsec:Streampot_Soret}

If the axial salt redistribution within the temperature gradient can be described by (\ref{Eq:SalinityGrad}) (case 'S': 
$\partial_{\widehat{\Theta}} N/N = -\overline{S} T_0$, where $\overline{S} = (S_- + S_+)/2$), the corresponding terms in the 
last row of (\ref{Eq:stream_pot_DH}) exactly cancel each other. Considering temperature-independent intrinsic Soret coefficients, 
one has in addition $\overline{S} \equiv \overline{S}_0$. Thus
\begin{eqnarray}
\label{Eq:stream_pot_DH_Soret} -\left(\frac{E_p}{\partial_x p_0}\right)^{(\textrm{S,DH})} = 
\left(F_\textrm{CS}\:+\:\frac{\zeta^2}{2} \frac{\epsilon}{D \eta} F_\alpha\right)^{-1}
\left\{\frac{\zeta}{\kappa^2 D \eta} \left[1\:-\:\frac{\textrm{tanh}(\overline{\kappa})}{\overline{\kappa}}\right] \right. \nonumber \\
\left. - \frac{\zeta^3}{2}\frac{\epsilon}{D \eta} \frac{\overline{S}_0 \Delta T}{\Delta p_0}
\left[F_{\beta}\:-\:\frac{\textrm{tanh}(\overline{\kappa})}{\overline{\kappa}}\left(\frac{\textrm{tanh}^2(\overline{\kappa})}{3}\:
-\:\frac{1}{2}\right)\right] \right. \nonumber \\
\left. - \frac{\zeta^3}{2}\frac{\epsilon}{D \eta} \frac{\Delta T/T_0}{\Delta p_0}\frac{T_0}{T}
\left[F_{\beta}\:-\:\frac{\textrm{tanh}(\overline{\kappa})}{\overline{\kappa}}\left(\frac{\textrm{tanh}^2(\overline{\kappa})}{3}\:
-\:\frac{1}{2}\right) \right] \right. \nonumber \\
\left. - \frac{\zeta^3}{2}\frac{\epsilon}{D \eta} \frac{M \Delta T}{\Delta p_0}
\left[F_{\beta}\:+\frac{1}{\textrm{cosh}^2(\overline{\kappa})}\:
+\:\frac{\textrm{tanh}(\overline{\kappa})}{\overline{\kappa}}\left(\frac{\textrm{tanh}^2(\overline{\kappa})}{3}\:-\:\frac{1}{2}\right)\right] \right. \nonumber \\
\left. - \zeta \frac{\Delta T/T_0}{\Delta p_0}\frac{T_0}{T}\frac{\textrm{tanh}(\overline{\kappa})}{\overline{\kappa}}
\left[1\:+\:\frac{\overline{\zeta}^2}{2}\left(\frac{T_0}{T}\right)^2\left[\frac{\textrm{tanh}^2(\overline{\kappa})}{3}\:
+\:\frac{1}{\textrm{cosh}^2(\overline{\kappa})}\right] \right] \right. \nonumber \\
\left. - \frac{1}{2}\frac{\Delta S \Delta T}{\Delta p_0} \frac{k_\textrm{B} T}{e \nu} \right\}. \ \
\end{eqnarray}
As apparent from the latter equation, the (conventional) thermo-electric field caused by a difference of the intrinsic Soret 
coefficients of the ion species (i.e. $\Delta S \neq 0$) can simply be added to the electric field induced by effects related 
to the confinement of the electrolyte between walls carrying a surface charge. To focus on these more unconventional effects and 
along with setting $\partial_{\widehat{\Theta}} N/N = -\overline{S}_0 T_0$, in the following the term proportional to $\Delta S$ 
is ignored. Essentially, this corresponds to $\Delta S = 0$ and $\overline{S}_0 = S_- = S_+ = S_0$ (Soret A), i.e. the ion species 
have the same thermophoretic mobility in a temperature gradient. Relating the non-isothermal streaming field to the 
isothermal Smoluchowski limit $-\zeta/(\kappa^2_0 D_0 \eta_0)$ and using $D \eta/T = D_0 \eta_0/T_0$ for ions with hydrodynamic 
radii remaining unaffected by temperature variations, one finds
\begin{eqnarray}
\label{Eq:stream_pot_DH_SoretA} \left(\frac{E_p}{\partial_x p_0}\right)^{(\textrm{A,DH})} \frac{\kappa^2_0 D_0 \eta_0}{\zeta} = 
\left(F_\textrm{CS}\:+\:\overline{\zeta}^2\frac{\varsigma_0}{2}\frac{\epsilon}{\epsilon_0} \frac{T_0}{T} F_\alpha\right)^{-1} 
\left\{\frac{\epsilon}{\epsilon_0 N} \left[\frac{\textrm{tanh}(\overline{\kappa})}{\overline{\kappa}}\:-\:1\right] \right. \nonumber \\
\left. +\: \overline{\zeta}^2 S_0 T_0 \frac{\epsilon}{\epsilon_0} \frac{p_{\textrm{osm},0}}{\Delta p_0} \frac{\Delta T}{T}
\left[F_{\beta}\:-\:\frac{\textrm{tanh}(\overline{\kappa})}{\overline{\kappa}}\left(\frac{\textrm{tanh}^2(\overline{\kappa})}{3}\:
-\:\frac{1}{2}\right)\right] \right. \nonumber \\
\left. + \overline{\zeta}^2 \frac{\epsilon}{\epsilon_0} \frac{p_{\textrm{osm},0}}{\Delta p_0} \frac{\Delta T}{T}
\left[F_{\beta}\:-\:\frac{\textrm{tanh}(\overline{\kappa})}{\overline{\kappa}}\left(\frac{\textrm{tanh}^2(\overline{\kappa})}{3}\:
-\:\frac{1}{2}\right) \right] \right. \nonumber \\
\left. + \overline{\zeta}^2 M_0 T_0 \frac{\epsilon}{\epsilon_0} \frac{p_{\textrm{osm},0}}{\Delta p_0} \frac{\Delta T}{T}
\left[F_{\beta}\:+\frac{1}{\textrm{cosh}^2(\overline{\kappa})}\:
+\:\frac{\textrm{tanh}(\overline{\kappa})}{\overline{\kappa}}\left(\frac{\textrm{tanh}^2(\overline{\kappa})}{3}\:-\:\frac{1}{2}\right)\right] \right. \nonumber \\
\left. + \frac{\kappa^2_0 D_0 \eta_0}{\Delta p_0} \frac{\Delta T}{T}\frac{\textrm{tanh}(\overline{\kappa})}{\overline{\kappa}}
\left[1\:+\:\frac{\overline{\zeta}^2}{2}\left(\frac{T_0}{T}\right)^2
\left[\frac{\textrm{tanh}^2(\overline{\kappa})}{3}\:+\:\frac{1}{\textrm{cosh}^2(\overline{\kappa})}\right] \right] \right\}, \ \ \ \ \ \ \ \ \ \ \
\end{eqnarray}
where
\begin{equation}
\label{Eq:intrinsic_Peclet} \varsigma_0 = \frac{\epsilon_0}{D_0 \eta_0} \left(\frac{k_\textrm{B} T_0}{e \nu}\right)^2 
= 2\frac{n_0 k_\textrm{B} T_0}{\kappa^2_0 D_0 \eta_0}, 
\end{equation}
denotes the intrinsic P\'eclet number and $p_{\textrm{osm},0} = n_0 k_\textrm{B} T_0$ expresses the osmotic reference pressure of the ion cloud. 
In (\ref{Eq:stream_pot_DH_SoretA}), for simplicity and consistency (the intrinsic Soret coefficient was already linearized in temperature), 
the temperature dependence of the dielectric permittivity and of the ion electromobility were neglected, so that $M \approx M_0$ and $1/T \approx 1/T_0$.

By examining the pre-factors in front of each of the square brackets on the RHS of (\ref{Eq:stream_pot_DH_SoretA}) one can deduce that 
$[E_p/(\partial_x p_0)]^{(\textrm{A,DH})}$ is caused by three qualitatively different physical effects. The first term within the 
curly bracket on the RHS simply denotes the induced Smoluchowski field, corrected with respect to confinement and temperature. 
As mentioned before, the latter correction is mainly due to the local values of the temperature-dependent dielectric permittivity and 
salt concentration, whereas the temperature dependence of the Fickian diffusion coefficient and viscosity does not play a role. The 
electric field induced by this term is proportional to the externally applied pressure gradient, i.e. it vanishes for 
$\partial_x p_0 = 0$. By contrast, the induced electric fields affiliated with the other four terms in the curly bracket do not necessarily 
vanish if the external pressure gradient is absent. Rather than that, they are proportional to the temperature difference. Among these four terms, 
those three multiplied by $p_{\textrm{osm},0}$ are contributions due to a thermoosmotically propelled liquid flow, i.e. they go along 
with fluid advection. By contrast, the last term in the curly bracket captures a non-advective effect, i.e. it can be derived 
by solving the Nernst-Planck and Poisson equation alone, without relying on the momentum equation. As discussed in a recent 
publication and as will be revisited later on, the corresponding field is caused by selective ion diffusion within the EDL, which 
depends on the polarity of the respective ion and is related to the temperature-dependent electrophoretic mobility of the 
ions \citep{Dietzel:PRL2016}. 

In order to estimate the magnitudes of the contributing effects relative to each other, in the next section typical values 
of the characteristic parameters are reviewed.

\subsection{Typical values of characteristic parameters}\label{subsec:estimates}

In the first two columns of table \ref{Tbl:thermoprops_ratios}, the electrolyte properties used in this study are summarized, 
where the solvent properties are based on pure water and given in the first three rows. The electric and transport properties 
of the solute, listed in the fourth to seventh row of the first two columns, refer to a 0.01\:M NaCl electrolyte solution. 
The corresponding values for a KCl electrolyte are of the same order of magnitude. 
If not stated otherwise, all values were determined at $25\:^{o}\textrm{C}$ ($T_0 = 298\:\textrm{K}$). In this study, 
$\Delta T$, $\Delta p_0$, $\zeta$ and $\overline{\kappa}_0$ were varied. The ranges of values taken by 
these parameters are summarized in the third and fourth column. The fifth and sixth column provide a selection of 
calculated parameters relevant for the verification of the scaling used in the derivation. In this context, $A=0.1$ and 
$n_0 = 0.01\:\textrm{M}$ \citep{Mansouri:JPhysChemB2007} were fixed values so that the nominal reference EDL thickness is 
calculated to be $\kappa^{-1}_0 \approx 10^{-7} \textrm{m}$. The nominal channel height is determined according to 
$h = \overline{\kappa}_0/\kappa_0$, while the channel length is determined by $l = \overline{\kappa}_0/(A \kappa_0)$. 
A conventional parabolic velocity profile in a slit channel was assumed to estimate the characteristic flow velocity, i.e. 
$u_0=\Delta p_0 A \overline{\kappa}_0/(3\kappa_0 \eta_0)$. Then, the Reynolds number is given by 
$\Rey = A \rho \Delta p_0 \overline{\kappa}^2_0/(3\kappa^2_0 \eta^2_0)$, the thermal P\'eclet number by 
$\Pen_T = A \Delta p_0 \overline{\kappa}^2_0 /(3\kappa^2_0 \alpha_0 \eta_0)$, and the Hartmann number reads 
$Ha = 6 n_0 k_\textrm{B} T_0/\Delta p_0$. The ionic P\'eclet number, $\Pen_k = l u_0/D_k$, is estimated by 
$\Pen_k = \Delta p_0 \overline{\kappa}^2_0/(3 \kappa^2_0 D_0 \eta_0)$. The selected pressure differences are in the range 
of $1\:\textrm{Pa} \leq \Delta p_0 \leq 10^2\: \textrm{Pa}$, which correspond to pressure gradients of 
$10^4\:\textrm{Pa}\:\textrm{m}^{-1} \leq |\partial_x p_0| \leq 10^8\: \textrm{Pa}\:\textrm{m}^{-1}$. 
For comparison, pressure gradients typically applied in studies concerned with electrokinetic streaming are of the order of 
$|\partial_x p_0| = {\cal O}(10^6)-{\cal O}(10^9)\:\textrm{Pa}\:\:\textrm{m}^{-1}$ 
\citep{Yang:JMicroMechEng2003,Mansouri:JPhysChemB2007,vdHeyden:PRL2005}. As mentioned in \S \ref{sec:model_eqn}, the 
intrinsic Soret coefficients (or equivalently, the ionic heats of transport of each ion species) can be determined experimentally 
only relative to a reference ion \citep{Agar:ProcRsSocLondA1960}, whose thermophoretic mobility is arbitrarily 
set to zero. Since we focus here on the case that $\Delta S = 0$, it is specifically assumed that all the $S_k$ 
(respectively the $Q_k$) take the same value for each ion species $k$, i.e. one cannot specify a reference ion. Hence, 
one may legitimately argue that it is not readily possible to determine $S_0$ experimentally. For this work, this 
circumstance is ignored and the order-of-magnitude of the coefficients provided in the literature is taken as a rough estimate.
\begin{table}

	\begin{center}			
		\begin{tabular}{c|c||c|c||c|c}
			\multicolumn{2}{c||}{Fluid properties} & \multicolumn{2}{c||}{Operation parameters} & \multicolumn{2}{c}{Scaling parameters}  \\ \hline
			$\rho_0\:(\textrm{kg}\:\textrm{m}^{-3})$ & $997$$^{[1]}$ & $\Delta T\:(\textrm{K})$ & $5$-$25$ & $u_0\:(\textrm{m}\:\textrm{s}^{-1})$ & $10^{-6}$-$10^{-2}$ \\
			$\eta_0\:(10^{-3}\:\textrm{Pa s})$ & $0.89$$^{[1]}$ & $|\Delta p_0|\:(\textrm{Pa})$ & $10^0$-$10^2$ & $\Pen_T$ (-) & $10^{-6}$-$10^{0}$ \\
			$\alpha_0\:(\textrm{m}^2\:\textrm{s}^{-1})$ & $1.45 \cdot 10^{-7}$$^{[1]}$ & $\zeta\:(10^{-3}\:\textrm{V})$ & $5$-$125$ & $\Rey$ (-) & $10^{-7}$-$10^{-1}$ \\
			$D_0\:(\textrm{m}^2\:\textrm{s}^{-1})$ & $10^{-9}$$^{[3]}$ & $\overline{\kappa}_0$\:(-) & $10^{-1}$-$10^2$ & $\Pen_k$ (-) & $10^{-3}$-$10^{3}$ \\
			$S_0 T_0\:(-)$ & $10^{-1}$-$10^{0}$$^{[4]}$ & & & $Ha/\overline{\kappa}^2_0$ (-) & $10^{-4}$-$10^{2}$ \\
			$\epsilon/\varepsilon_0$\:(-) & $78.14$$^{[2]}$ & & & & \\
			$M_0 T_0\:(-)$ & $-1.52$$^{[2]}$ & & & & \\
		\end{tabular}
	\end{center}
\caption{Thermophysical properties and characteristic numbers. The electrolyte properties are listed in the first two columns, where 
the solvent properties are based on pure water and listed in the first three rows. The electric and transport properties of the solute, 
listed in the fourth to seventh row, refer to a 0.01\: M NaCl electrolyte solution. The corresponding values for a KCl electrolyte are 
of the same order of magnitude. If not stated otherwise, all values were determined at $25\:^{o}\textrm{C}$ ($T_0 = 298\:\textrm{K}$). 
The third and fourth column list the range taken by the parameters varied in this study. The fifth and sixth column provide a selection 
of calculated parameters, which are relevant for the verification of the scaling used in the derivation. In this context, $A=0.1$ and 
$n_0 = 0.01\: \textrm{M}$ are fixed values, leading to a nominal reference EDL thickness of $\kappa^{-1}_0 \approx 10^{-7}\: \textrm{m}$. The 
nominal channel height is given by $h = \overline{\kappa}_0/\kappa_0$, while its length is determined by 
$l = \overline{\kappa}_0/(A \kappa_0)$.
[1] - \citet{Lide:CRCPubComp2009}, [2] - \citet{Buchner:PhysChemA1999}, [3] - \citet{Takeyama:JPhysSocJap1983}, 
[4] - \citet{Takeyama:JPhysSocJap1983}, \citet{Agar:ProcRsSocLondA1960}, \citet{Leaist:JSolChem1990}}
\label{Tbl:thermoprops_ratios}
\end{table}

From table \ref{Tbl:thermoprops_ratios} it becomes apparent that the assumptions about the magnitude of specific 
parameters underlying the present derivation are fulfilled, with the exception of $Ha/\overline{\kappa}^2_0$ and $\Pen_k$. 
The latter is significantly larger than unity for the upper limiting case of $\Delta p_0 = 100\:\textrm{Pa}$ and 
$\overline{\kappa}_0 = 100$ ($h \approx 10 \mu \textrm{m}$). For this case, the separate treatment of advective and diffusive effects 
used to derive the EDL potential might be questionable, although this is commonly neglected in most studies of the isothermal case, 
even for larger $h$ and $\Delta p_0$. This is legitimate since, by contrast to the flow around a charged colloidal particle, advectively 
driven charge in- and ejection into and from the EDL is of little importance in fully developed flow as treated herein. For values of 
$\Delta p_0 \leq {\cal O}(10^1)$ and $\overline{\kappa}_0 \leq {\cal O}(10^1)$, $\Pen_k$ is indeed $\leq {\cal O}(1)$. 
Furthermore, $\Pen_k > 1$ refers to a regime where the Soret effect as well as the temperature dependencies of the 
electromobility and of the permittivity have little effect. Therefore, corresponding limitations are less important for the main 
conclusions drawn in this study. Given the definition of $Ha$, it becomes unbounded if $\Delta p_0 \rightarrow 0$. In this case, 
instead of being the externally applied pressure difference, $\Delta p_0$ needs to be replaced by the osmotic ion pressure in the EDL 
so that $Ha \equiv 1$. The rescaling would not lead to the inclusion of the (now neglected) electro-migration terms on the LHS of 
(\ref{Eq:momentum_x}) and (\ref{Eq:momentum_z}), since they would still be ${\cal O}(A^2)$ smaller than the corresponding terms on the 
RHS. From this discussion it follows that values of $Ha/\overline{\kappa}^2_0$ larger than unity are acceptable.

\subsection{Relative magnitude of contributing effects}\label{subsec:magntitudes}

By choice of the employed scaling, the first term in the curly bracket of (\ref{Eq:stream_pot_DH_SoretA}) is of ${\cal O}(1)$, 
whereas the pre-factors in front of the other square brackets (whose contents are of ${\cal O}(1)$ as well) determine their 
magnitude relative to the first term. According to table \ref{Tbl:thermoprops_ratios}, not only $\epsilon/\epsilon_0$ but also 
$S_0 T_0$ and $|M_0 T_0|$ are of ${\cal O}(1)$, so that those terms in (\ref{Eq:stream_pot_DH_SoretA}) multiplied by 
$p_{\textrm{osm},0}$ are all of the same order of magnitude. With the reference condition defined in the caption of 
table \ref{Tbl:thermoprops_ratios}, one has $p_{\textrm{osm},0} = 24.7\:\textrm{Pa}$, i.e. the osmotic pressure is of similar 
magnitude as the pressure difference typically applied to drive the flow in micro- and nanochannels (see table 
\ref{Tbl:thermoprops_ratios}). Consequently, for $\Delta T = 25\:\textrm{K}$, one finds that the thermoosmotically induced fields 
contribute less than $10\:\%$ to the overall field as long as $\overline{\zeta} \lesssim 1$ and $\Delta p_0 \gtrsim p_{\textrm{osm},0}$. 
For highly charged channels (e.g. $\overline{\zeta} \approx 5$) and $\Delta p_0 \lesssim p_{\textrm{osm},0}$, the thermoosmotic 
contribution can be more significant. With the parametric values provided in table \ref{Tbl:thermoprops_ratios} one finds that 
$\kappa^2_0 D_0 \eta_0 \approx 10^2\:\textrm{Pa} > p_{\textrm{osm},0}$, i.e. the contribution to the induced field by the last 
term on the RHS of (\ref{Eq:stream_pot_DH_SoretA}) might be more important than the thermoosmotic contributions, at least at 
low $\zeta$ potentials.

\subsection{Induced field for very large and very small $\overline{\kappa}$  (Soret A)}\label{subsec:streaming}

On the one hand, for $\overline{\kappa}_0 \rightarrow \infty$ (and therefore also $\overline{\kappa} \rightarrow \infty$) 
$\textrm{tanh}(\overline{\kappa})/\overline{\kappa} \rightarrow 0$, so that apart from $F_\beta$ 
also $F_\alpha$ approaches zero, while the term in the first round bracket on the RHS of (\ref{Eq:stream_pot_DH_SoretA}) (i.e. 
the denominator ) goes to unity. Consequently, the ratio between the induced (streaming) fields at non-isothermal 
and isothermal conditions [the latter being equal to $\zeta/(\kappa^2_0 D_0 \eta_0)$] becomes
\begin{equation}
\label{Eq:stream_pot_DH_verylargeKP} \left[\frac{E_p}{(E_p)_\textrm{isoth}}\right]^{(\textrm{A,DH})}_{|{\overline{\kappa}_0 \rightarrow \infty}} = 
\frac{\epsilon}{\epsilon_0 N} = (1+M_0 \Delta T \Theta) \textrm{exp}(S_0 \Delta T \Theta).
\end{equation}
Subsequent integration in axial direction with a linear temperature profile ($\Theta = X$) leads to
\begin{equation}
\label{Eq:stream_pot_DH_verylargeKP_integrated} \left[\frac{\Delta \phi_{0,p}}{(\Delta \phi_{0,p})_\textrm{isoth}}\right]^{(\textrm{A,DH})}_{|{\overline{\kappa}_0 \rightarrow \infty}}
\approx 1 + \frac{1}{2}(S_0 + M_0) \Delta T.
\end{equation}
Hence, in this limit, the variation of the streaming potential with temperature is (approximately) proportional to the 
temperature difference as well as to the sum of the intrinsic Soret coefficient and the relative change of the dielectric 
permittivity with temperature. On the contrary, the temperature dependencies of the electrophoretic mobility, viscosity 
and of the Fickian diffusivity have no effect. The observed increase in streaming potential is due to the average 
increase of the EDL thickness with temperature by the factor of $\sqrt{\epsilon/(\epsilon_0 N)}$. Therefore, for very 
large $\overline{\kappa}_0$, it is sufficient to account for non-isothermal effects by using an average of the increased 
EDL thickness, which is given by $(\kappa^{-1}_0)^* = \kappa^{-1}_0 \sqrt{1 + (S_0 + M_0) \Delta T/2}$.

On the other hand, in the limit of $\overline{\kappa}_0 \rightarrow 0$, one has 
$\textrm{tanh}(\overline{\kappa})/\overline{\kappa} \rightarrow 1$, 
$F_\alpha \rightarrow 0$ and $F_{\beta} \rightarrow -1/2$. In this case, the induced field is independent of the externally 
applied pressure gradient since the first term in the curly brackets on the RHS of (\ref{Eq:stream_pot_DH_SoretA}) vanishes. 
Multiplication by $\partial_x p_0/\partial_x T$ leads to
\begin{eqnarray}
\label{Eq:stream_pot_DH_smallKP_local} \left(\frac{E_T}{\partial_x T}\right)^{(\textrm{A,DH})}_{|\overline{\kappa}_0 \rightarrow 0} = 
\frac{\zeta}{T}, \ \ \ \ \ \
\end{eqnarray}
while integration in axial direction results in
\begin{eqnarray}
\label{Eq:stream_pot_DH_smallKP} \left(\frac{\Delta \phi_{0,T}}{\Delta T}\right)^{(\textrm{A,DH})}_{|\overline{\kappa}_0 \rightarrow 0} = 
-\frac{\zeta}{\Delta T} \textrm{ln}\left(1+\frac{\Delta T}{T_0}\right) \approx -\frac{\zeta}{T_0}. \ \ \ \ \ \
\end{eqnarray}
This induced potential difference is present for any value of the external pressure gradient as long as $\Delta T$ is non-vanishing. 
This behavior differs from the one obtained under isothermal conditions, where the induced field approaches zero for 
$\overline{\kappa}_0 \rightarrow 0$. For vanishing $\overline{\kappa}_0$, the contribution of the streaming current to the induced 
field [first four terms on the RHS of (\ref{Eq:stream_pot_DH_SoretA})] vanishes regardless of the temperature distribution. This is 
the result of two counteracting effects: decreasing $\overline{\kappa}_0$ leads to a uniform charge density $\rho_\textrm{f}$ across 
the channel, potentially increasing the streaming current (which is proportional to $\rho_\textrm{f} u_0$). Yet, 
$\overline{\kappa}_0 \rightarrow 0$ implies either $h \rightarrow 0$ or $\kappa_0 \rightarrow 0$, where the latter commonly corresponds 
to $n_0 \rightarrow 0$. Since the area-averaged velocity behaves approximately according to $u_0=\Delta p_0 A h/(3 \eta_0)$, it goes 
to zero for vanishing $h$. Alternatively, for vanishing $n_0$, the charge density goes to zero so that the net result of 
$\overline{\kappa}_0 \rightarrow 0$ is a vanishing streaming current. At constant temperature, the streaming current is the only 
mechanism present to separate charges. This implies that in the absence of any streaming current and at uniform temperature, no induced 
field can be generated. However, under non-isothermal conditions, charges are also separated if one ion species moves differently in 
a thermal gradient than the respective counter ion. This can be accomplished by (at least) two mechanisms: the first is simply based 
on different thermophoretic mobilities within the thermal gradient ($\Delta S \neq 0$, Soret B), which leads to the well-known Soret 
equilibrium in bulk electrolytes. The second uses a combination of a temperature-dependent electrophoretic mobility of the ions and 
the selection of one ion species over the other by means of interaction with the surface charge along the wall. The latter mechanism 
induces the finite potential expressed by (\ref{Eq:stream_pot_DH_smallKP}) and is explained in detail in \citet{Dietzel:PRL2016}. It can be 
understood by referring to the Boltzmann distribution (\ref{Eq:Boltzmann}): for a given value of the potential $\Psi$, a temperature difference 
causes a difference in ion concentration, which is dependent on the polarity of the specific ion. Subsequently, the concentration 
differences trigger diffusive ion fluxes and a transport of a net charge in regions with a non-vanishing space charge. Note that all 
terms in (\ref{Eq:stream_pot_DH_SoretA}) proportional $p_{\textrm{osm},0}$ vanish for both very small or very large values of 
$\overline{\kappa}_0$. Instead and as will be detailed later, corresponding terms are maximal at $\overline{\kappa}_0 \approx 1$.

Since all contributions in the curly bracket on the RHS of (\ref{Eq:stream_pot_DH_SoretA}) are additive, while the magnitude of the first 
term is little affected by temperature and its characteristic behavior with respect to confinement is well-known, in the following 
the first term will be disregarded. This is equivalent to considering a non-isothermal channel without exposing it to an externally 
applied pressure difference.

\subsection{Thermoosmotic streaming field (Soret A)}

From (\ref{Eq:stream_pot}), maintaining the thermal gradient while setting the externally applied pressure difference, 
$\Delta p_0$ (defined in the course of the derivation of (\ref{Eq:momentum_x2})), to zero, one obtains the induced thermoelectric 
field of a confined symmetric electrolyte, which is given by
\begin{equation}
\label{Eq:thermo_pot} \frac{E_T}{\partial_x T} = \frac{k_\textrm{B}}{e \nu} 
\frac{\overline{I}_{\textrm{st},\widehat{\Theta}} + \overline{I}_{\textrm{cd},\widehat{\Theta}}}{\overline{I}_{\textrm{st},\Phi} + \overline{I}_{\textrm{cd},\Phi}},
\end{equation}
where the subscript $T$ added to $E$ indicates an induced field solely caused by a temperature difference. 

Within the DH approximation, from (\ref{Eq:stream_pot_DH_SoretA}) (Soret A) one finds
\begin{eqnarray}
\label{Eq:stream_pot_DH_SoretA_nopress} \left(\frac{E_T}{\partial_x T}\right)^{(\textrm{A,DH})} \frac{T}{\zeta}= 
\left(F_\textrm{CS}\:+\:\overline{\zeta}^2\frac{\varsigma_0}{2}\frac{\epsilon}{\epsilon_0} \frac{T_0}{T} F_\alpha\right)^{-1} \nonumber \\
\left\{\overline{\zeta}^2 \frac{\varsigma_0}{2} \frac{\epsilon}{\epsilon_0} S_0 T_0
\left[F_{\beta}\:-\:\frac{\textrm{tanh}(\overline{\kappa})}{\overline{\kappa}}\left(\frac{\textrm{tanh}^2(\overline{\kappa})}{3}\:
-\:\frac{1}{2}\right)\right] \right. \nonumber \\
\left. + \overline{\zeta}^2 \frac{\varsigma_0}{2} \frac{\epsilon}{\epsilon_0} 
\left[F_{\beta}\:-\:\frac{\textrm{tanh}(\overline{\kappa})}{\overline{\kappa}}\left(\frac{\textrm{tanh}^2(\overline{\kappa})}{3}\:
-\:\frac{1}{2}\right) \right] \right. \nonumber \\
\left. + \overline{\zeta}^2 \frac{\varsigma_0}{2} \frac{\epsilon}{\epsilon_0} M_0 T_0
\left[F_{\beta}\:+\frac{1}{\textrm{cosh}^2(\overline{\kappa})}\:
+\:\frac{\textrm{tanh}(\overline{\kappa})}{\overline{\kappa}}\left(\frac{\textrm{tanh}^2(\overline{\kappa})}{3}\:-\:\frac{1}{2}\right)\right] \right. \nonumber \\
\left. + \frac{\textrm{tanh}(\overline{\kappa})}{\overline{\kappa}}
\left[1\:+\:\frac{\overline{\zeta}^2}{2}\left(\frac{T_0}{T}\right)^2
\left[\frac{\textrm{tanh}^2(\overline{\kappa})}{3}\:+\:\frac{1}{\textrm{cosh}^2(\overline{\kappa})}\right] \right] \right\}. \ \ \ \
\end{eqnarray}
This thermoelectric field is a confinement effect since it vanishes for $\overline{\kappa}_0 \rightarrow \infty$ (i.e. 
also $\overline{\kappa} \rightarrow \infty$). At $\overline{\kappa}_0 \rightarrow 0$, the local field is given by 
(\ref{Eq:stream_pot_DH_smallKP_local}).

\begin{figure}
	\centerline{\includegraphics[width=13cm]{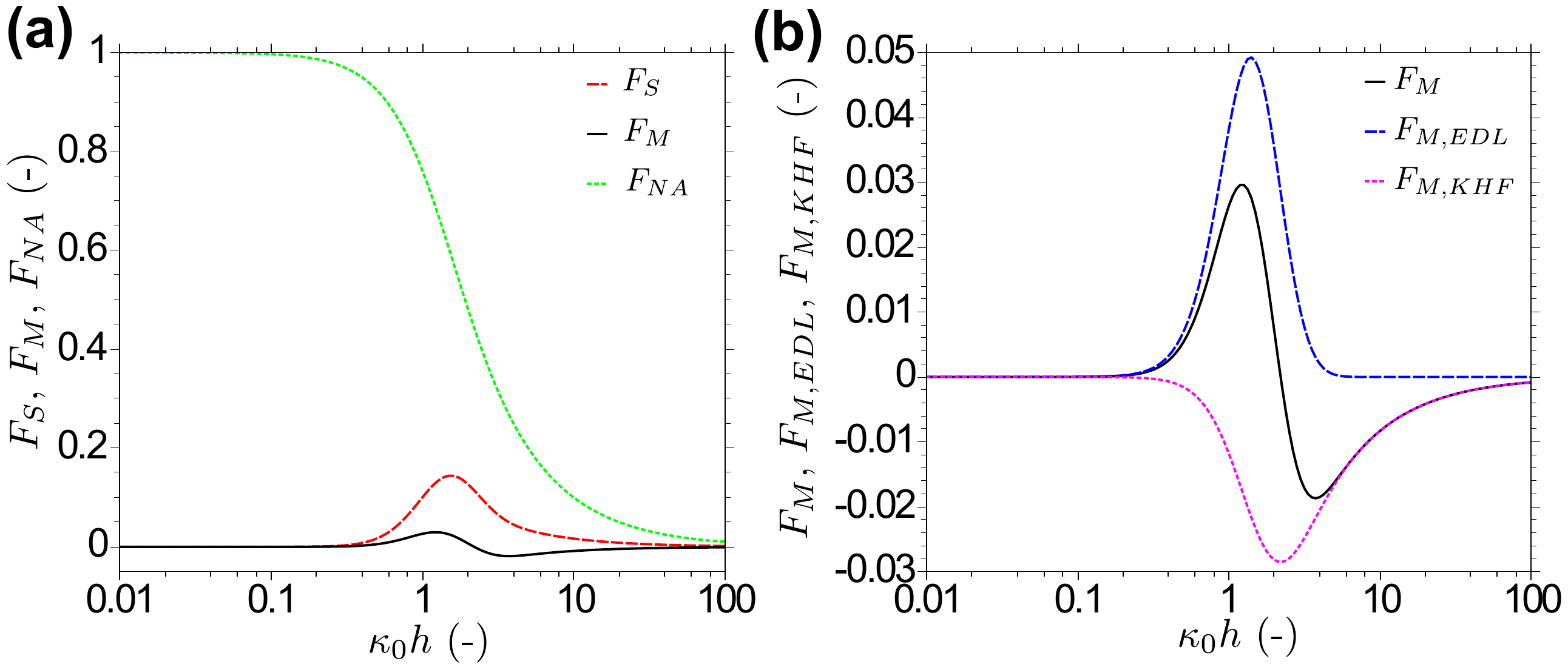}}
	\caption{Dependence of the individual terms contributing to (\ref{Eq:stream_pot_DH_SoretA_nopress}) on 
	$\overline{\kappa} \approx \overline{\kappa}_0 = \kappa_0 h$. (a) Comparison of $F_S$, $F_M$ and $F_\textrm{NA}$ as 
	a function of $\overline{\kappa}_0$. (b) Comparison of $F_M$, $F_{M,\textrm{EDL}}$ and $F_{M,\textrm{KHF}}$ as a function of 
	$\overline{\kappa}_0$.}	\label{Fig:Comp_Soret_Emob_Permit_DHlimit_VS_KP}
\end{figure}

The first, second and third term in the curly bracket of (\ref{Eq:stream_pot_DH_SoretA_nopress}) represent the 
contributions to the induced field due to thermoosmotic fluid propulsion either caused by the temperature-dependent local salt 
concentration, the temperature-dependent electrophoretic ion mobility or by the temperature-dependent dielectric permittivity, 
respectively. The value of the ionic P\'eclet number $\varsigma_0$ is typically about $0.5$ [see \S \ref{sec:axial_flow} 
or as computed by (\ref{Eq:intrinsic_Peclet})], while $\epsilon/\epsilon_0$, $S_0 T_0$ and $|M_0 T_0|$ are all of {\cal O}(1). 
Consequently, for $\overline{\zeta} \lesssim 1$, the corresponding terms in the curly bracket of (\ref{Eq:stream_pot_DH_SoretA_nopress}) 
are at most about one fourth of the last term. Nevertheless, at higher $\zeta$ potentials all terms might be of the same order of 
magnitude. To identify the most important terms for the three characteristic regimes of either $\overline{\kappa}_0 \ll 1$, 
$\overline{\kappa}_0 \approx 1$ and $\overline{\kappa}_0 \gg 1$, it is useful to compare the dependencies of these terms on 
$\overline{\kappa}_0$ alone. In this context, in figure \ref{Fig:Comp_Soret_Emob_Permit_DHlimit_VS_KP} (a), the 
dependencies of the content of the first 
three square brackets of (\ref{Eq:stream_pot_DH_SoretA_nopress}) on $\overline{\kappa} \approx \overline{\kappa}_0$ are plotted and 
compared to $F_\textrm{NA} = \textrm{tanh}(\overline{\kappa}_0)/\overline{\kappa}_0$, representing the fourth term in the curly bracket of 
(\ref{Eq:stream_pot_DH_SoretA_nopress}) (for a first orientation and simplicity, its dependence on $\overline{\zeta}$ is neglected). 
The latter term is maximal at $\overline{\kappa}_0 \rightarrow 0$ and vanishing rapidly for larger values of $\overline{\kappa}_0$. 
The content of the first and the second square brackets, respectively, of (\ref{Eq:stream_pot_DH_SoretA_nopress}) is identical and 
abbreviated in the following by $F_S$. It is maximal at $\overline{\kappa}_0 \approx 1$, whereas it is vanishingly small for values 
of $\overline{\kappa}_0$ much larger or much smaller than unity. Lastly, the term representing the effect of a temperature-dependent 
permittivity [i.e. the content of the third square bracket of (\ref{Eq:stream_pot_DH_SoretA_nopress}), in the following abbreviated 
by $F_M$] changes sign at about $\overline{\kappa}_0 \approx 2$; for significantly smaller or larger values of $\overline{\kappa}_0$ 
it is vanishingly small as well. While a thermophoretic ion motion or a temperature-dependent electrophoretic ion mobility 
induces a thermoelectric field via thermoosmosis solely by means of the circumstance that the ion cloud in the 
non-isothermal EDL is not in mechanical equilibrium, thermoelectricity driven by a temperature-dependent permittivity is 
in addition caused by a thermoosmotic effect affiliated with the expression $(\bnabla \phi)^2 \bnabla \epsilon/2$, being part 
of the Korteweg-Helmholtz force in the momentum equations. As derived by \citet{Derjaguin:NewYork1987}, the electric field induced 
by the flow driven by this force alone (superscript 'Dn') is given by
\begin{equation} 
\label{Eq:thermoelctric_Derjaguin} \left(\frac{E_T}{\partial_x T}\right)^{(\textrm{Dn,DH})} \frac{T}{\zeta} = 
-\frac{\epsilon^2 \gamma \zeta^2 \kappa^2}{4 \eta \sigma^{(\infty)}}
\left[\frac{\textrm{tanh}(\overline{\kappa})}{\overline{\kappa}} + \textrm{tanh}^2(\overline{\kappa}) - 1 
- \frac{2}{3}\frac{\textrm{tanh}^3(\overline{\kappa})}{\overline{\kappa}}\right],
\end{equation}
where $\gamma = T d_T \epsilon/\epsilon = M T$, and $\sigma^{(\infty)} = \epsilon \kappa^2 D$ denotes the electric 
conductivity of the bulk. In the original expression of Derjaguin and co-workers, besides carrying a sign error in front of 
the last term on the RHS of (\ref{Eq:thermoelctric_Derjaguin}), $\gamma$ was replaced by $\gamma^* = 1+\gamma$. Nevertheless, as 
explained in detail in the Supplemental Material of \citet{Dietzel:PRL2016}, for a channel whose ends are assumed to be in 
electrochemical equilibrium with the (non-isothermal) reservoir (as it is the case in the scenario considered herein), the correct 
form is given by (\ref{Eq:thermoelctric_Derjaguin}). Furthermore, per convection, in the original derivation $I_\textrm{st}=I_\textrm{cd}$ was set 
to compute the induced electric field, while in the present work $I_\textrm{st}+I_\textrm{cd}=0$ was used. Thus, to match the 
present sign definition, the sign of the original expression of \citet{Derjaguin:NewYork1987} was switched. Finally, the original 
derivation was performed using the Gaussian CGS unit system, whereas (\ref{Eq:thermoelctric_Derjaguin}) is written in the SI unit 
system. This implies that a factor of $1/(4 \pi)^2$ is omitted.

According to (\ref{Eq:stream_pot_DH_SoretA_nopress}) and replacing $1/\textrm{cosh}^2(\overline{\kappa})$ by 
$1-\textrm{tanh}^2(\overline{\kappa})$, the thermoosmotic contribution to the induced field by means of a temperature-dependent 
dielectric permittivity, $E^{(\textrm{M,DH})}_T$, can be rewritten to read
\begin{eqnarray}
\label{Eq:stream_pot_DH_permitonly} \left(\frac{E_T}{\partial_x T}\right)^{(\textrm{M,DH})} \frac{T}{\zeta}= 
\overline{\zeta}^2 \varsigma_0 \frac{\epsilon}{\epsilon_0} M_0 T_0 \frac{F_{M,\textrm{EDL}}+F_{M,\textrm{KHF}}}
{F_\textrm{CS}\:+\:\frac{\overline{\zeta}^2}{2} \varsigma_0 \frac{\epsilon}{\epsilon_0} \frac{T_0}{T} F_\alpha}, \ \ \ \
\end{eqnarray}
with
\begin{equation}
\label{Eq:F_MEDL_F_MKHF} F_{M,\textrm{EDL}} = \frac{\overline{\kappa} \textrm{tanh}(\overline{\kappa})
-\textrm{tanh}^2(\overline{\kappa})}{2\textrm{cosh}^2(\overline{\kappa})}, \\
F_{M,\textrm{KHF}} = -\frac{1}{4}\left[\frac{\textrm{tanh}(\overline{\kappa})}{\overline{\kappa}}+\textrm{tanh}^2(\overline{\kappa})-1
-\frac{2}{3}\frac{\textrm{tanh}^3(\overline{\kappa})}{\overline{\kappa}}\right]. \nonumber
\end{equation}
Employing $T_0/(D_0 \eta_0) = T/(D \eta)$ and reversing the linearization of $M$ [i.e. $M$ instead of $M_0$ is used 
in (\ref{Eq:stream_pot_DH_permitonly}) to derive (\ref{Eq:Permit_prefactor})], one can readily show that
\begin{equation}
\label{Eq:Permit_prefactor} \overline{\zeta}^2 \varsigma_0 \frac{\epsilon}{\epsilon_0} M T_0 =
 \frac{\epsilon^2 \gamma \zeta^2 \kappa^2}{\sigma^{(\infty)} \eta}.
\end{equation}
Thus, $F_{M,\textrm{KHF}}$ multiplied by (\ref{Eq:Permit_prefactor}) corresponds to the induced thermoelectric 
field derived by \citet{Derjaguin:NewYork1987}. As mentioned, the corresponding field contribution is 
directly related to the expression $(\bnabla \phi)^2 \bnabla \epsilon/2$ of the Korteweg-Helmholtz force, so that 
corresponding effects will be marked by the subscript 'KHF'. Nevertheless, a temperature-dependent permittivity 
induces an additional thermoosmotic flow, which is caused by the variation of the EDL potential along the channel and 
the mechanical imbalance of the ion cloud in the EDL. In turn, this leads to an induced thermoelectric field as well, whose 
contribution is represented by $F_{M,\textrm{EDL}}$. In the original derivation of Derjaguin and co-workers this 
part is missing, as in their work it was implicitly assumed that the EDL potential does not vary along the channel. 
In the present work, it was verified that only the term proportional to $F_{M,\textrm{EDL}}$ is obtained if the expression 
$(\bnabla \phi)^2 \bnabla \epsilon/2$ is omitted from the Korteweg-Helmholtz force. Since $F_{M,\textrm{EDL}}$ 
arises from the circumstance that the ion cloud in a non-isothermal EDL is not necessarily in mechanical equilibrium, 
in the following, corresponding effects will be referred to by the subscript 'EDL'.

In figure \ref{Fig:Comp_Soret_Emob_Permit_DHlimit_VS_KP} (b), to illustrate the contribution to the induced field by 
$F_{M,\textrm{KHF}}$ in comparison to $F_{M,\textrm{EDL}}$, both expressions are plotted as a function of 
$\overline{\kappa}_0 \approx \overline{\kappa}$, with $F_M = F_{M,\textrm{EDL}}+F_{M,\textrm{KHF}}$. The absolute values 
of both expressions are of the same order of magnitude, while $F_{M,\textrm{EDL}}$ is more localized around 
$\overline{\kappa} = 1$. In comparison, $F_{M,\textrm{KHF}}$ is non-vanishing in a wider range of $\overline{\kappa}$ values. 
This characteristics and the opposite sign of the latter in comparison 
to $F_{M,\textrm{EDL}}$ is the reason for the sign change of $F_M$ displayed in figure 
\ref{Fig:Comp_Soret_Emob_Permit_DHlimit_VS_KP} (a) and (b), which is a direct consequence of the corresponding change 
in the thermoosmotic velocity profile discussed along with figure \ref{Fig:uax} (b). Hence, the contribution to the induced 
thermoelectric field caused by the KHF term has the opposite sign of the field induced thermoosmotically by the EDL term, 
affiliated with the mechanical imbalance of the ion cloud in the EDL due to a temperature-dependent permittivity.

While under the given assumptions the total induced thermoelectric field is given by (\ref{Eq:stream_pot_DH_SoretA_nopress}), 
the last term in the curly bracket of (\ref{Eq:stream_pot_DH_SoretA_nopress}) was analyzed in depth in a recent publication 
\citep{Dietzel:PRL2016}, as mentioned. In that work it was demonstrated that the corresponding contribution to the induced 
field is due to a non-advective effect, which is affiliated with the selective electro-migration of the ions caused by a 
temperature-dependent electromobility. As illustrated in figure \ref{Fig:Comp_Soret_Emob_Permit_DHlimit_VS_KP} (a), 
this contribution is indeed the dominant one. It is particularly pronounced under extreme confinement, i.e. for 
$\overline{\kappa}_0 \rightarrow 0$. In the following, to focus on advective effects induced by a non-uniform temperature, 
the corresponding term representing the non-advective contribution will be disregarded. This is particular 
relevant if, apart from high values of $|\zeta|$, an electrolyte is used which exhibits higher values of the intrinsic P\'eclet 
number (e.g. due to a very small Fickian ion diffusivity). Given that the remaining electric field is caused by a mechanical 
propulsion of the fluid due to a temperature gradient alone, it will be referred to as thermoosmotic streaming field (TOSF). 
The TOSF is to be understood as an electric field per temperature gradient, i.e. it has the same physical units as the conventional 
Seebeck coefficient.

\begin{figure}
	\centerline{\includegraphics[width=13cm]{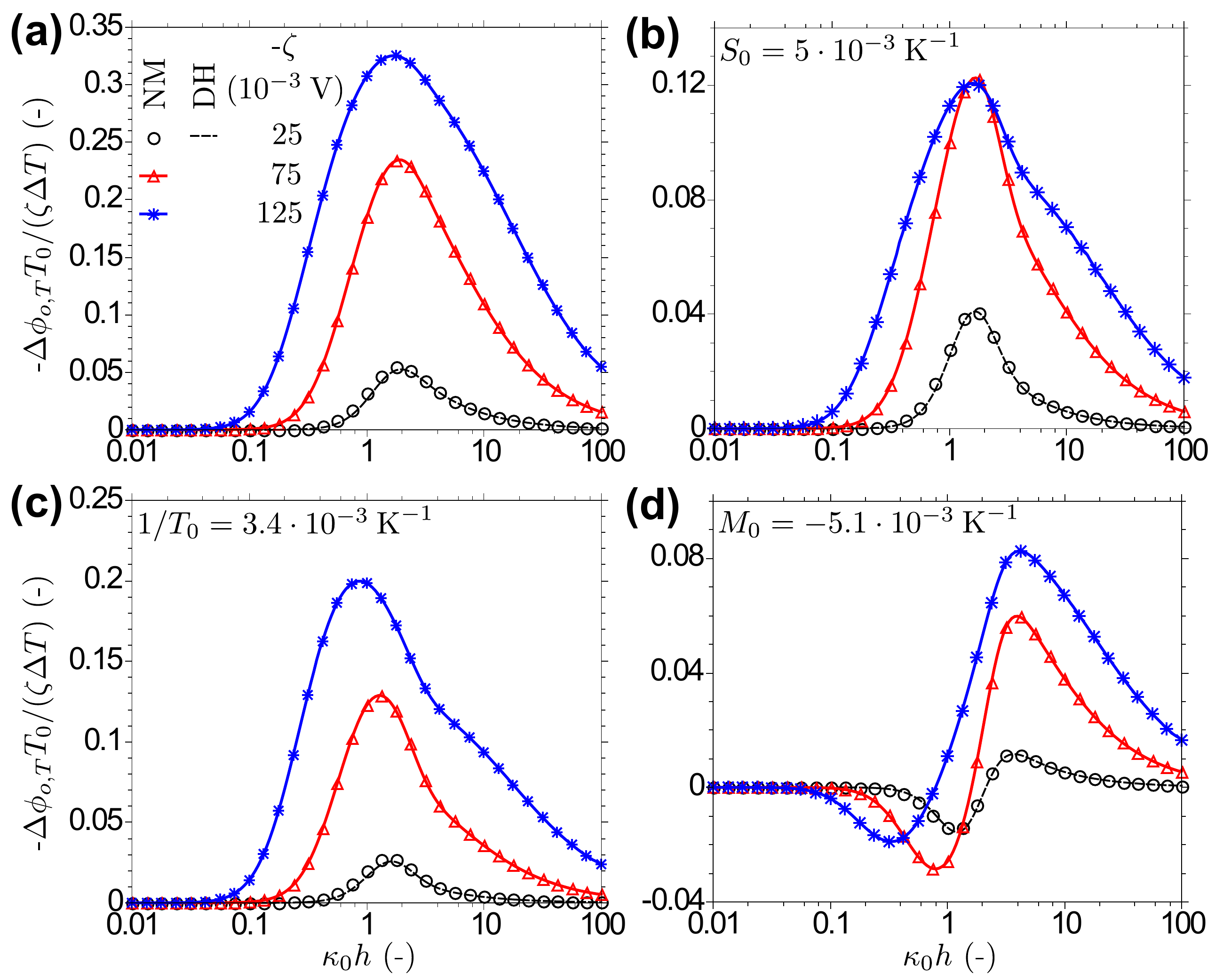}}
	\caption{Induced electric potential $\Delta \phi_{0,T}/\Delta T$ relative to $-\zeta/T_0$ representing the thermoosmotic 
	streaming field (TOSF) of a confined symmetric electrolyte as a function of $\overline{\kappa}_0 = \kappa_0 h$ and for 
	$\zeta = [-25, -75, -125] \cdot 10^{-3}\:\textrm{V}$). For the smallest value of the $\zeta$ potential, the numerical solutions 
	('NM') are compared to those obtained within the Debye-H\"uckel approximation ('DH'). (a) Complete solution, where 
	a thermophoretic ion motion with equal thermophoretic ion mobilities of each ion species 
	(Soret A, $S_0 = 5 \cdot 10^{-3}\:\textrm{K}^{-1}$) is present next to temperature dependencies of the ion mobility 
	and the dielectric permittivity ($M_0 = -5.1 \cdot 10^{-3}\:\textrm{K}^{-1}$). (b) Isolated effect of the thermophoretic 
	ion motion, while the temperature dependencies of the ion mobilities and of the dielectric permittivity are excluded 
	($S_0 = 5 \cdot 10^{-3}\:\textrm{K}^{-1}$, $M_0 = 0$). (c) Isolated effect of the temperature-dependent electrophoretic 
	ion mobility, while the temperature dependence of the permittivity and thermophoretic ion motion are excluded 
	($S_0 = M_0 = 0$) (d) Isolated effect of the temperature-dependent dielectric permittivity, while the temperature 
	dependence of the ion mobility and the thermophoretic ion motion are excluded 
	($M_0 = -5.1 \cdot 10^{-3}\:\textrm{K}^{-1}$, $S_0 = 0$). The legend shown in (a) is also valid in (b)-(d).}
	\label{Fig:thermoelectric_pot_VS_kp_zetavary}
\end{figure}

In figure \ref{Fig:thermoelectric_pot_VS_kp_zetavary} (a), the overall TOSF is shown, which includes the field contributions 
caused by thermoosmotic effects affiliated with the thermophoretic ion motion, the temperature-dependent electrophoretic ion 
mobility and the temperature-dependent permittivity, while the contribution due to non-advective effects, as described by 
the last term in the curly bracket of (\ref{Eq:stream_pot_DH_SoretA_nopress}), is omitted. The local TOSF is integrated 
(numerically) along the channel and plotted as a function of $\overline{\kappa}_0 = \kappa_0 h$ in form of the potential difference 
$-\Delta \phi_{0,T} / \Delta T$, scaled to $\zeta/T_0$. The $\zeta$ potential is used as a parameter and set equal to 
$\zeta = -25$, $-75$, or to $-125 \cdot 10^{-3}\:\textrm{V}$, respectively. With respect to the full solution (not relying 
on the Debye H\"uckel approximation), $\Delta \phi_{0,T} / \Delta T$ was calculated by integrating (\ref{Eq:thermo_pot}) 
numerically along the channel. The thermophoretic ion motion according to 
Soret A (i.e. thermophoretic ion mobilities are identical for each ion species, $S_0 = 5 \cdot 10^{-3}\:\textrm{K}^{-1}$) as well 
as the temperature dependencies of the ion mobility and of the dielectric permittivity ($M_0 = -5.1 \cdot 10^{-3}\:\textrm{K}^{-1}$) 
are included. The temperature difference is set to $\Delta T = 25\:\textrm{K}$. For the lowest $\zeta$ potential, the numerical 
result is compared to the DH approximation given by the (numerically integrated) expression (\ref{Eq:stream_pot_DH_SoretA_nopress}), 
demonstrating very good agreement. Figure \ref{Fig:thermoelectric_pot_VS_kp_zetavary} (a) indicates that also at higher values of 
$|\zeta|$ the TOSF is particularly present if half of the channel height is of the same order of magnitude as the EDL thickness, i.e. for 
$\overline{\kappa}_0 \approx 1$. By contrast, it vanishes for very small or very large values of $\overline{\kappa}_0$. This 
confirms earlier findings obtained from figure \ref{Fig:Comp_Soret_Emob_Permit_DHlimit_VS_KP} (a). This characteristic behavior can 
be understood as follows: the varying thickness of the EDL with temperature leads to an EDL potential depending also on the axial 
coordinate $x$. The driving forces remain constrained to the EDL and propel fluid by electrohydrostatic and electroosmotic action. 
Subsequently, the corresponding advective charge transport leads to a finite TOSF. 
Increasing $h$ beyond $\kappa^{-1}_0$ (i.e. $\overline{\kappa}_0 \rightarrow \infty$) does not enhance the thermo-electroosmotic 
fluid propulsion in the EDL. However, the conduction current increases with increasing cross section, leading to a decrease of 
the TOSF. For the opposite limit $h \ll \kappa^{-1}_0$ (i.e. $\overline{\kappa}_0 \rightarrow 0$), 
the excess ion distribution across the slit channel becomes uniform and the local modification of the EDL thickness is of no 
importance, so that the thermo-electroosmotic fluid propulsion vanishes. Notable is the relatively sharp peak in the TOSF and 
its limitation to a comparatively narrow $\overline{\kappa}_0$ range: for $|\zeta| = 25 \cdot 10^{-3}\:\textrm{V}$, $\Delta \phi_{0,T}$ 
at $\overline{\kappa}_0 = 10$ is already only a little more than $10\%$ of the maximum value obtained at $\overline{\kappa}_0 = 2$. 
For higher $\zeta$ potentials the TOSF is present within a much broader range of $\overline{\kappa}_0$, reaching peak values of about $35\:\%$ of $-\zeta/T_0$. 
The qualitative behavior of the TOSF with respect to $\overline{\kappa}_0$ stands in sharp contrast not only to the conventional 
streaming field induced by an externally applied pressure difference, which is maximal for  $\overline{\kappa}_0 \rightarrow \infty$, 
but also to the non-advectively induced thermoelectric field captured by the last term in the curly bracket of 
(\ref{Eq:stream_pot_DH_SoretA_nopress}). As illustrated in figure \ref{Fig:Comp_Soret_Emob_Permit_DHlimit_VS_KP} (a) 
the latter is maximal in the limit of $\overline{\kappa}_0 \rightarrow 0$ and goes rapidly to zero for larger values of 
$\overline{\kappa}_0$.

To analyze the significance of either the thermophoretic ion motion, the temperature-dependent ion mobility, or the 
temperature dependence of the dielectric permittivity with respect to the TOSF, in figure \ref{Fig:thermoelectric_pot_VS_kp_zetavary} 
(b)-(d) the individual contributions to the TOSF are shown. In (b), the (hypothetical) case is considered that only the thermophoretic 
ion motion is present as a non-isothermal effect, where $S_0 = 5 \cdot 10^{-3}\:\textrm{K}^{-1}$ and $\Delta T = 25\:\textrm{K}$ 
is used. The temperature dependencies of the ion electromobility and of the dielectric permittivity are excluded. The $\zeta$ potential 
in (b)-(d) takes on the same values as used in (a), and the legend depicted in (a) is valid for (b)-(d) as well. The TOSF induced 
by the thermophoretic ion motion alone follows qualitatively the behavior of the overall TOSF. However, beyond $\zeta = -75 \cdot 10^{-3}\:\textrm{V}$ 
the peak value of the TOSF levels off at about $12\:\%$ of $-\zeta/T_0$. As verified by artificially deactivating the corresponding 
term in the computation, this is caused by the electroosmotic counterflow generated by the induced thermoelectric field itself, 
an effect captured by the second term in the first round bracket of (\ref{Eq:stream_pot_DH_SoretA_nopress}) (proportional to $F_\alpha$). 
The effect is more pronounced in the full, numerically solved model than within the DH approximation. In addition and more peculiar, 
this saturation effect is not observed in the overall TOSF. The reason for this will be explained in the following.

In figure \ref{Fig:thermoelectric_pot_VS_kp_zetavary} (c), the TOSF is plotted as a function of $\overline{\kappa}_0$ 
for the same values of the $\zeta$ potential as for the other plots, while the only thermoosmotic effect considered is the 
one caused by a temperature-dependent ion electromobility. The TOSF induced by this isolated effect follows qualitatively
the one shown in (a) and (b), i.e. it is maximal at $\overline{\kappa}_0 \approx 1$ and vanishingly small for $\overline{\kappa}_0 \rightarrow 0$ and 
$\overline{\kappa}_0 \rightarrow \infty$. Notably, within the DH approximation given by (\ref{Eq:stream_pot_DH_SoretA_nopress}), 
the terms describing the dependence of the TOSF on $\overline{\kappa}$ are absolutely identical (and equal to $F_S$) for both the 
contribution by thermodiffusion and by a temperature-dependent electromobility of the ions. Thus, for the given value of $S_0$ and for every 
$\overline{\kappa}_0$, the TOSF caused by a temperature-dependent ion electromobility should be about two thirds of the TOSF 
induced by thermodiffusion. This holds for small $\zeta$ potentials but not for higher ones. Particularly, the saturation 
of the TOSF observed at $\zeta = 125\cdot 10^{-3}\:\textrm{V}$ for the thermodiffusion case, displayed in figure 
\ref{Fig:thermoelectric_pot_VS_kp_zetavary} (b), cannot be observed in (c). This can be explained by considering the pre-factor 
of the second term in the first round bracket of (\ref{Eq:stream_pot_DH_SoretA_nopress}), describing the effect of the electroosmotic 
backflow. In case of a temperature-dependent electrophoretic ion mobility, this pre-factor is temperature-dependent and becomes 
smaller at higher temperature. This weakens the effect of the electroosmotic backflow. By contrast, in case of thermodiffusion being 
the sole non-isothermal effect, this pre-factor is not temperature-dependent, rendering the affiliated electroosmotic backflow more 
significant, especially at higher values of $|\zeta|$. Despite these differences and in light of the qualitative and quantitative similarity 
between the contributions to the TOSF either due to thermodiffusion or due to a temperature-dependent ion electromobility (especially at 
lower $\zeta$ potentials), deciding in practice whether a TOSF is primarily induced by either of these two mechanisms appears 
to be a formidable challenge. This is corroborated by considering that accurate values of $S_0$ are not available.

Finally, for the (hypothetical) case that only the temperature-dependent permittivity is present as a non-isothermal effect, 
figure \ref{Fig:thermoelectric_pot_VS_kp_zetavary} (d) depicts the ratio $(-\Delta \phi_{0,T}/\Delta T)/(\zeta/T_0)$ as a function 
of $\overline{\kappa}_0$, where $M_0 = -5.1 \cdot 10^{-3}\:\textrm{K}^{-1}$ and $\Delta T = 25\:\textrm{K}$. The temperature 
dependence of the ion electromobility and the thermophoretic ion motion are excluded. As the main difference to the cases treated in 
(b) and (c), the TOSF found in this case changes sign with increasing $\overline{\kappa}_0$. As mentioned in the context of figure 
\ref{Fig:Comp_Soret_Emob_Permit_DHlimit_VS_KP}, this is a direct consequence of the corresponding change in the thermoosmotic velocity 
profile discussed along with figure \ref{Fig:uax} (b). Consequently, the TOSF induced by a temperature-dependent permittivity by means of 
a mechanical imbalance of the ion cloud in the EDL has the opposite sign of the TOSF induced by a temperature-dependent permittivity by means 
of the $(\bnabla \phi)^2 \bnabla \epsilon/2$ term of the Korteweg-Helmholtz force. The saturation effect observed in case of thermodiffusion 
is not observed for a temperature-dependent permittivity. Similar to the case of a temperature-dependent ion electromobility described 
in (c), the pre-factor of the the second term in the round bracket of (\ref{Eq:stream_pot_DH_SoretA_nopress}) is temperature-dependent, 
which is found to weaken the effect of the electroosmotic backflow.

In summary, depending on the actuation mechanism, the thermoosmotically induced electric fields per temperature gradient (i.e. the effective 
Seebeck coefficients) are approximately $5$-$35\%$ of $\zeta/T_0$. For $\zeta = 25$-$125\cdot 10^{-3}\:\textrm{V}$ this amounts to 
$4$-$147\cdot 10^{-6}\:\textrm{V}\:\textrm{K}^{-1}$. This is about an order of magnitude lower than the Seebeck coefficient of 
semiconductors but higher than that of many metals.

\section{Conclusions}
 
A semi-analytical model for non-isothermal electrokinetic transport of a symmetric dilute electrolyte 
in a slit channel subjected to axial gradients in pressure and temperature was developed. The 
derivation relies on a perturbation expansion in a small parameter, the height-to-length ratio of the 
channel. The model takes into account the temperature dependencies of the electrophoretic ion mobility 
and of the dielectric permittivity of the solvent, next to thermophoretic ion motion (intrinsic Soret effect), 
whereas effects caused by the temperature dependencies of density, heat capacity and heat conductivity 
of the solvent are shown to be negligible to leading order in the temperature variation and in the 
expansion coefficient. 

In the framework of the Debye-H\"uckel (DH) approximation, analytical expressions for the electric 
potential inside the electric double layer (EDL) were determined and compared to numerical results, 
where the wall $\zeta$ potential was increased up to $125 \cdot 10^{-3}\:\textrm{V}$. For diffusion-dominated 
ion transport and constant $\zeta$ potential along the channel walls, it was found that a thermophoretic 
ion motion increases the EDL thickness exponentially with temperature, while a temperature-dependent 
ion mobility increases the EDL thickness according to the inverse square root of the temperature. By contrast, 
given the typical decrease of permittivity with temperature, this dependency is shown to 
shrink the EDL with increasing temperature. The thermally induced expansion/shrinkage of the EDL 
(normal to the main flow direction) varies in axial direction along the channel and leads, together 
with the strong variation of the potential within the EDL, to a large axial field within the EDL. 

Based on this EDL potential, expressions for the flow field and the induced electric field under the combined 
effects of axial pressure and temperature were derived. The electric field induced was seen to be the 
linear superposition of seven contributions: The first expresses the conventional 
pressure-induced streaming field, which in the non-isothermal case was seen to be qualitatively identical to 
the one at uniform temperature. In the course of the derivation it was shown that, for this term and given the 
typically temperature-independent hydraulic radii of the ions, the temperature dependencies of the viscosity, the 
Fickian diffusion coefficient and the ion electromobility compensate each other and do not have an influence on 
this contribution to the streaming field. Consequently, only the values of the permittivity and of the salt 
concentration need to be adjusted according to the local value of temperature. Such modifications can be 
suitably accounted for by averaging the corresponding correction factor over the channel length, 
which, for a temperature difference of $\Delta T = 25\:\textrm{K}$, may alter the induced streaming field up to 
$10\:\%$ when compared to the (isothermal) Smoluchowski limit. However, given the opposing effects due to 
thermophoresis and permittivity variations, only a small net modification of the corresponding streaming field 
compared to isothermal conditions may be observable.

The second contribution to the overall induced field is the thermoelectric field commonly found in bulk electrolytes 
exhibiting different thermophoretic mobilities of each ion species and kept at a non-uniform temperature 
(Soret equilibrium). Herein, it was seen that this contribution is only marginally affected by the confinement of the 
electrolyte between charged walls.

Besides these two well-known and expected contributions to the induced field, in this work five other contributions 
have been identified. They were all shown to be confinement effects detectable only in narrow channels, which -similar 
to the conventional Soret thermovoltage- do not vanish if the externally applied pressure gradient is removed while 
the temperature gradient is maintained. The first among those is driven by the coupling between the temperature-dependent 
ion electromobility and the wall charge, giving rise to selective ion migration in the EDL, whose strength depends 
on the polarity of the ion. As highlighted in a recent journal publication \citep{Dietzel:PRL2016}, this is a non-advective 
mechanism. It is dominant for very small Debye parameters $\overline{\kappa}_0 = \kappa_0 h \rightarrow 0$ ($\kappa_0^{-1}$ denotes 
the nominal EDL thickness and $h$ indicates half of the channel width), while it vanishes rapidly for larger values 
of $\overline{\kappa}_0$. The remaining four additional contributions to the induced electric field were found to 
be caused by thermoosmotic fluid propulsion, i.e. they rely on advective effects occurring independent of an 
externally applied pressure gradient. In the present work, the corresponding contributions to the induced electric field 
were termed \textit{thermoosmotic streaming fields} (TOSFs).

While \citet{Derjaguin:NewYork1987} pointed out that the temperature dependence of the Korteweg-Helmholtz force (by means 
of the temperature dependence of the dielectric permittivity) gives rise to thermoosmosis and a corresponding induced electric 
field, the ion cloud in the EDL was implicitly assumed to remain in mechanical equilibrium, as it can be proven to 
be the case at isothermal conditions. By contrast, according to the present work, the ion cloud in the EDL is no longer in a 
state of mechanical equilibrium if an axial temperature gradient acts along the channel. Consequently, along with 
the mentioned axial gradients of the EDL potential, a pressure field of thermal origin develops inside the EDL, leading to a 
thermo-electroosmotically induced flow field and a corresponding TOSF. The mechanical imbalance of the ion cloud may be caused, 
apart from the thermophoretic ion motion and the temperature dependence of the ion electromobility, by the temperature dependence of 
the dielectric permittivity as well. To the best of our knowledge, such an origin of thermoosmosis has never been analyzed 
in detail before, despite being of similar order of magnitude as the thermoosmotically induced field expressed by Derjaguin 
and co-workers. For thermoosmosis triggered by thermophoretic ion motion, it was shown that only the arithmetic mean of the intrinsic Soret 
coefficients is relevant. In the light of this finding, the further investigation of the induced field was limited to the 
case where apart from a temperature-dependent ion mobility and permittivity, both ion species have the same thermophoretic 
mobility (Soret A). This implies that the conventional thermovoltage due to the Soret equilibrium vanishes. 

Solutions of the TOSF obtained within the DH approximation as a function of $\overline{\kappa}_0$ were compared to full 
numerical solutions, where the $\zeta$ potential was varied in the range of $25$-$125 \cdot 10^{-3}\:\textrm{V}$. 
At low $\zeta$ potentials, solutions obtained in the DH limit fully agree with the  numerical results. It was seen that, 
for any value of the $\zeta$ potential and confirmed by the numerical results, the TOSF is vanishingly small for both 
very small or very large values of $\overline{\kappa}_0$. The same observation was made when either a thermophoretic ion 
motion, a temperature-dependent ion electromobility or a temperature-dependent permittivity cause the TOSF. The largest 
values of the TOSF were attained in the vicinity of $\overline{\kappa}_0 \approx 1$, i.e. when the channel half-width is 
of the same order of magnitude as the EDL thickness, reaching up to $35\:\%$ of $\zeta/T$, with $T$ being the absolute temperature. 
Within the DH limit, the dependence of the TOSF on $\overline{\kappa}_0$ induced either by the thermophoretic ion motion 
or by the temperature dependence of the ion electromobility was found to be identical. The numerical solution 
indicated that at high $\zeta$ potentials, the potential related to the TOSF induced by the intrinsic Soret effect alone 
saturates at about $12\:\%$ of $\zeta/T$. This limitation was seen to be due to the electroosmotically induced backflow of 
ions driven by the TOSF itself, which was more pronounced in the (more accurate) numerical solutions than in those obtained 
with the DH approximation. Nevertheless, the saturation of the TOSF could not be observed in the numerical solutions if either 
the effect of the temperature-dependent electromobility or of the temperature-dependent permittivity were considered alone. 

In case of thermoosmosis due to the temperature dependence of the dielectric permittivity alone, the TOSF was found to be 
a more complicated function of $\overline{\kappa}_0$ than it is for the intrinsic Soret effect or the temperature dependence 
of the ion electromobility. Particularly, it was seen to change sign with increasing Debye parameter. This behavior was 
linked to two observations: firstly, the contribution to the TOSF by means of the temperature dependence of the Korteweg-Helmholtz 
force was found to have the opposite sign of the contribution to the TOSF driven by the mechanical imbalance of the ion cloud in the EDL. 
Secondly, the latter is more restricted to a finite interval of $\overline{\kappa}_0$ in the vicinity of $\overline{\kappa}_0 \approx 1$ 
than the former. Hence, for $\overline{\kappa}_0 \lesssim 2$ thermoosmosis due to the EDL dominates, while for larger 
values thermoosmosis due to the (additional) Helmholtz-Korteweg term prevails. The change in sign of the TOSF as a 
function of $\overline{\kappa}_0$ directly corresponds to the thermo-electroosmotically driven axial flow caused by a 
temperature-dependent permittivity, which -depending on the value of $\overline{\kappa}_0$- may change direction within 
the channel cross section. Such a flow reversal could neither be observed for thermoosmotically propelled flow due to 
thermophoretic ion motion nor due to a temperature-dependent ion electromobility.

From a fundamental point of view, the presented findings are useful to distinguish between the different sources of 
thermoosmotic propulsion of non-isothermal electrokinetic transport through micro- and nanochannels. In addition, the 
results may help to further understand and fine tune the ion selectivity and sensing properties of artificial nanopores 
and biological ion channels or to measure their zeta potential. With respect to technological applications, the results 
are relevant for the energy conversion by means of non-isothermal electrokinetic streaming, where a temperature gradient 
is superimposed to the externally applied pressure gradient. Also, the findings might provide further insight to the 
transport of ions through the nanoporous catalyst layer of a fuel cell, over which a temperature gradient exists. Along 
this line, the presented results supplement research efforts concerning small-scale fluidic waste-exergy (availability) 
recovery units and other low-cost energy sustainability devices, in which the induced voltage is generated thermally. 
Finally, while the results have been derived for domains of a small aspect ratio, the underlying physical mechanisms might 
as well be applicable for the stabilization and transport of charged particles embedded in a non-isothermal liquid electrolyte.

\appendix

\section{Thermo-diffusion potential}\label{sec:app_thermodiffpot}

In the following, a brief overview of the conventional theoretical treatment of thermal diffusion potentials 
observed in multi-component bulk electrolytes subject to a temperature difference is given. This summary is 
included in this study for the following reason: The starting point is classical nonequilibrium 
thermodynamics based on the phenomenological Onsager theory \citep{deGrootMazur:Dover1984}, which is applicable 
to a wide range of different problems. As remarked by \citet{Hartung:UBayreuth2007}, classical textbooks treat 
the Soret effect rather generally as one among many others, without going into great detail. Herein, a 
comprehensive outline of the matter is provided. As discussed in the book of \citet{Fitts:McGrawHill1962}, it is 
assumed that the open system contains one ("second-law") heat flux $\boldsymbol{q}$ and $P=K+1$ 
material fluxes $\boldsymbol{j}_k$. These vectorial fluxes are driven by $P+1$ conjugate forces $\boldsymbol{X}_k$, 
where $\boldsymbol{X}_0 = \bnabla \textrm{ln} (T)$ and $\boldsymbol{X}_i = \bnabla^{(T)} \mu'_i$ ($i=1,..,P$). 
The latter denotes the spatial gradient of the chemical potential at isothermal conditions. The material fluxes 
are not independent of each other but obey $\boldsymbol{j}_P = -\sum^K_{k=1} {\boldsymbol{j}_k}$. In 
electrochemical systems, it is useful to relate the material fluxes of the solutes to the motion 
of the solvent. For a system in mechanical equilibrium (i.e. the Gibbs-Duhem equation is valid), 
$\boldsymbol{q}$ and the $K$ solute fluxes can be described by [page 66 in \citet{Fitts:McGrawHill1962}]
\begin{equation}
\label{Eq:heat_flux} -\boldsymbol{q} = L'_{00} \bnabla \textrm{ln} (T) + \sum^K_{i=1} {L'_{0i} \bnabla^{(T)} \mu'_i},
\end{equation}
\begin{equation}
\label{Eq:mass_flux} -\boldsymbol{j}_k = L'_{k0} \bnabla \textrm{ln} (T) + \sum^K_{i=1} {L'_{ki} \bnabla^{(T)} \mu'_i},
\end{equation}
where $L'_{ki}$ are phenomenological coefficients. The qualitatively same equations are obtained if the system 
is not necessarily in mechanical equilibrium, but the chemical potential of the solvent varies only with temperature 
throughout the domain so that $\bnabla^{(T)} \mu'_P = 0$. Herein we have
\begin{equation}
\label{Eq:chempotgrad_isotherm} \bnabla^{(T)} \mu'_i = \bnabla \left(\mu_i\right)_{|T} + \frac{\nu_i F}{M_i} \bnabla \phi,
\end{equation}
with $\bnabla \left(\mu_i\right)_{|T} = \bnabla \mu_i - \left(\partial_T \mu_i \right)_{|p} \bnabla T$, and $\mu_i$ is the chemical potential per unit mass 
of component $i$. The Faraday constant as the specific ion charge is denoted by $F$, $M_i$ is the molar mass and $\phi$ is the electric potential. 
With $\rho_k$ as the mass density of species $k$, one has $\mu_i = \mu_i(T,p,\rho_1,...\rho_K)$ so that 
$\bnabla(\mu_i)_{|T} = \partial_p (\mu_i)_{|T,\rho_i} \bnabla p + \sum^K_{l=1} \partial_{\rho_l} (\mu_i)_{|T,p} \bnabla \rho_l$, with 
$\partial_p (\mu_i)_{|T,\rho_i} = \overline{V}_i$ as the partial volume $V_i$ per mass $m_i$. Hence,
\begin{equation}
\label{Eq:mass_flux2} -\boldsymbol{j}_k = L'_{k0} \bnabla \textrm{ln} (T) + 
\sum^K_{i=1} L'_{ki} \left[\overline{V}_i \bnabla p + \sum^K_{l=1} \partial_{\rho_l} (\mu_i)_{|T,p} \bnabla \rho_l + \frac{\nu_i F}{M_i} \bnabla \phi \right].
\end{equation}
The diffusion coefficients are defined by $D_{kl} = \sum^K_{i=1} L'_{ki} \partial_{\rho_l}(\mu_i)_{|T,p}$, while $L'_{k0} = \rho_k T D_{T,k}$ 
with $D_{T,k}$ being the thermodiffusion coefficient of component $k$ [page 79 and 102 in \citet{Fitts:McGrawHill1962}]. Neglecting pressure-induced diffusion and defining ionic mobilities $\omega_k$ by $e \nu_k \rho_k \omega_k = \sum^K_{i=1} L'_{ki} \nu_i F/M_i$ leads to
\begin{equation}
\label{Eq:mass_flux3} -\boldsymbol{j}_k = \rho_k D_{T,k} \bnabla T + \sum^K_{l=1} D_{n,kl} \bnabla \rho_l + e \nu_k \rho_k \omega_k \bnabla \phi.
\end{equation}
Given the typically large values of the electrostatic pressure within the EDL, recent work plausibly suggests that pressure-induced diffusion must 
not be neglected \citep{Dreyer:PCCP2013}. Generally this would lead to less steep and wider EDLs so that the effect can be expected to assist the 
thermo-electroosmotic fluid propulsion mechanism discussed in the main part of this work. However, since this is a rather recent discussion and 
the magnitude of the transport coefficient for pressure-induced diffusion appears to be unknown, this important issue will be left open for future 
investigations. In the case considered in this work $K=2$. From (\ref{Eq:mass_flux2}), expression (\ref{Eq:NPE_dim}) 
can be found by neglecting cross diffusion due to concentration gradients and inserting the result into $d_t n_k = -N_A/M_k \bnabla \cdot \boldsymbol{j}_k$, 
where $N_A$ is the Avogadro-constant. Note that $\rho_k = n_k M_k/N_A$ and $F=e N_A$. The intrinsic Soret coefficients are defined by 
$\rho_k D_{T,k} = \sum^K_{l=1} \rho_l S_l D_{kl}$ so that
\begin{equation}
\label{Eq:mass_flux4} -\boldsymbol{j}_k = \sum^K_{l=1} D_{kl} \left(\bnabla \rho_l + \rho_l S_l \bnabla T\right) + e \nu_k \rho_k \omega_k \bnabla \phi.
\end{equation}

Alternatively, each material flux expressed with (\ref{Eq:mass_flux}) is weighted with parameters $\tilde{Q}_k$ to be determined, summed over all 
$K$ and the result subtracted from (\ref{Eq:heat_flux}). This leads to
\begin{equation}
\label{Eq:heat_flux2} \boldsymbol{q} = \sum^K_{k=1} {\tilde{Q}_k \boldsymbol{j}_k}  - \left(L'_{00} - \sum^K_{k=1} {\tilde{Q}_k L'_{k0}} \right) 
\bnabla \textrm{ln} (T) - \sum^K_{i=1} \left(L'_{0i} - \sum^K_{k=1}{\tilde{Q}_k L'_{ki}}\right) \bnabla^{(T)} \mu'_i.
\end{equation}
The weighting factors $\tilde{Q}_k$, the so-called heat of transports, are now selected so that the last term in (\ref{Eq:heat_flux2}) vanishes, 
i.e.
\begin{equation}
\label{Eq:heat_transport} L'_{0i} = \sum^K_{k=1}{\tilde{Q}_k L'_{ki}}.
\end{equation}
Defining the overall heat conductivity $\lambda$ via 
\begin{equation}
\label{Eq:heat_conductivity} \lambda T = L'_{00} - \sum^K_{k=1} {\tilde{Q}_k L'_{k0}}
\end{equation}
leads to
\begin{equation}
\label{Eq:heat_flux3} \boldsymbol{q} = \sum^K_{k=1} {\tilde{Q}_k \boldsymbol{j}_k}  - \lambda \bnabla T.
\end{equation}
With (\ref{Eq:heat_transport}) and the Onsager-relation $L'_{ik} = L'_{ki}$ so that $L'_{0i} = L'_{i0}$, one has 
$L'_{k0} = \sum^K_{i=1} {L'_{ki} \tilde{Q}_i}$. Consequently, with (\ref{Eq:chempotgrad_isotherm}), the material fluxes can be expressed by
\begin{equation}
\label{Eq:mass_flux5} -\boldsymbol{j}_k = \sum^K_{i=1} {L'_{ki} \left[\tilde{Q}_i \bnabla \textrm{ln} (T) + \bnabla \left(\mu_i\right)_{|T} + \frac{\nu_i F}{M_i} \bnabla \phi\right]}.
\end{equation}
In solution chemistry, it is a common practice to express the dependence of the chemical potential on the composition in terms of the activities 
$a_i = \gamma_i \tilde{m}_i$, with $\gamma_i$ as the activity coefficients and $\tilde{m}_i = n_i/\rho_\textrm{solv}$ as the molarities, where 
$\rho_\textrm{solv}$ expresses the (constant) solvent density. One has $\bnabla(\mu_i)_{|T} = V_i \bnabla p + RT/M_i \bnabla\textrm{ln}(a_i)$, where $R$ 
denotes the ideal gas constant. This expression omits the dependence of the chemical potential of species $i$ on the concentration of species 
$k$, i.e. cross-diffusional effects are implicitly removed. This is reasonably accurate for very low concentrations of the solutes. One 
obtains \citep{Hill:Nature1957}
\begin{equation}
\label{Eq:mass_flux6} -\boldsymbol{j}_k = \sum^K_{i=1} {L'_{ki} \left[\tilde{Q}_i \bnabla \textrm{ln} (T) + V_i \bnabla p + 
\frac{RT}{M_i} \bnabla \textrm{ln}\left(\gamma_i \tilde{m}_i\right) + \frac{\nu_i F}{M_i} \bnabla \phi\right]}.
\end{equation}
Neglecting pressure-induced diffusion (i.e. ignoring the corresponding discussion when introducing (\ref{Eq:mass_flux3})), assuming constant 
activity coefficients (dilute limit) and with $R = N_A k_\textrm{B}$ as well as with $\tilde{Q}_i \equiv N_A Q_i/M_i$, one finds 
\begin{equation}
\label{Eq:mass_flux7} -\boldsymbol{j}_k = \sum^K_{i=1} {L'_{ki} \frac{k_\textrm{B} T N_A}{n_i M_i}\left[n_i \frac{Q_i}{k_\textrm{B} T^2} \bnabla T + 
\bnabla n_i + n_i \frac{\nu_i e}{k_\textrm{B} T} \bnabla \phi\right]}.
\end{equation}
As mentioned, this equation can only be derived by neglecting cross-diffusion between solute species, i.e. $L'_{ki} = 0$ for $k \neq i$. 
This, according to the definition of $D_{ki}$ introduced before, yields $D_{k} \equiv D_{kk} = L'_{kk} \partial_{\rho_k}(\mu_k)_{|T,p}$ with 
$(\mu_k)_{|T,p} = N_A k_\textrm{B} T/M_k \textrm{ln}\left(\gamma_k n_k/\rho_\textrm{solv}\right)$. Using $\partial_{\rho_k}(.) = N_A/M_k \partial_{n_k}(.)$ 
one finds $\partial_{\rho_k}(\mu_k)_{|T,p} = N_A^2 k_\textrm{B} T/(n_k M^2_k)$ and
\begin{equation}
\label{Eq:mass_flux8} -\boldsymbol{j}_k = {D_k\frac{M_k}{N_A}\left[n_k \frac{Q_k}{k_\textrm{B} T^2} \bnabla T + \bnabla n_k + n_k \frac{\nu_k e}{k_\textrm{B} T} \bnabla \phi\right]}.
\end{equation}
In the absence of cross diffusion between different ion species and equivalent to the treatment chosen in the main part of this work, 
(\ref{Eq:mass_flux8}) provides the non-isothermal diffusive ion flux in terms of the (ionic) heats of transport \citep{Wuerger:RepProgPhys2010}. 
From this equation one can also deduce that $S_k = Q_k/(k_\textrm{B} T^2)$.

For a (mass-) closed system at steady-state ($t \rightarrow \infty$), each material flux vanishes, i.e. $\boldsymbol{j}_k=0$ for all $k$, and thus 
$\nu_k F \boldsymbol{j}_k/(D_k M_k) = 0$ as well. With (\ref{Eq:mass_flux8}), summing over all $K$ leads to
\begin{equation}
\label{Eq:current} \sum^K_{k=1} e \nu_k n_k \left[\frac{Q_k}{k_\textrm{B} T} \left(\frac{\bnabla T}{T}\right)_{|t\rightarrow \infty} + \frac{e \nu_k}{k_\textrm{B} T} (\bnabla \phi)_{|t \rightarrow \infty} \right] + 
\bnabla \sum^K_{k=1}  (e \nu_k n_k) = 0
\end{equation}
The last term vanishes at electroneutral conditions present in the bulk electrolyte outside the EDL. The corresponding steady-state thermoelectric potential 
reads
\begin{equation}
\label{Eq:thermoelectric_pot_app} (\bnabla \phi)_{|t \rightarrow \infty} = -\frac{\sum^K_{k=1} e \nu_k n_k Q_k}{\sum^K_{k=1} e^2 \nu^2_k n_k} \left(\frac{\bnabla T}{T}\right)_{|t \rightarrow \infty}
\end{equation}
From this equation one can deduce that the bulk thermoelectric diffusion potential, equivalent to the Seebeck effect observed in metals and semi-conductors, should 
vanish for symmetric, dilute electrolytes, if the (ionic) heats of transport (or equivalently, the intrinsic Soret coefficients) of each ion species are identical.

For completeness, $\boldsymbol{j}_k=0$ for all $k$ implies $\sum^K_{k=1} \boldsymbol{j}_k=0$. Defining the average salinity with $2 n = \sum^K_{k=1} n_k$, 
with (\ref{Eq:mass_flux8}) one finds
\begin{equation}
\label{Eq:steady_state_mass_flux} \left(\frac{\bnabla T}{T}\right)_{|t \rightarrow \infty} \sum^K_{k=1} \frac{n_k}{n} \frac{Q_k}{2 k_\textrm{B} T} 
+ \frac{1}{2}\left(\frac{\sum^K_{k=1} \bnabla n_k}{n_0}\right)_{|t \rightarrow \infty} + (\bnabla \phi)_{|t \rightarrow \infty}\sum^K_{k=1} \frac{n_k}{n} \frac{e \nu_k}{2 k_\textrm{B} T} = 0
\end{equation}
Then, charge neutrality leads to the salinity gradient at steady-state (Soret equilibrium) \citep{Wuerger:RepProgPhys2010}:
\begin{equation}
\label{Eq:Soret-equilibrium} \left(\frac{\bnabla n}{n}\right)_{|t \rightarrow \infty} = 
-\alpha \left(\frac{\bnabla T}{T}\right)_{|t \rightarrow \infty},
\end{equation}
with $\alpha = \sum^K_{k=1} n_k Q_k/(2 k_\textrm{B} T n)$.

\section{Variation of EDL potential with temperature}\label{sec:app_variaEDLtemp}

\begin{figure}
  \centerline{\includegraphics[width=13cm]{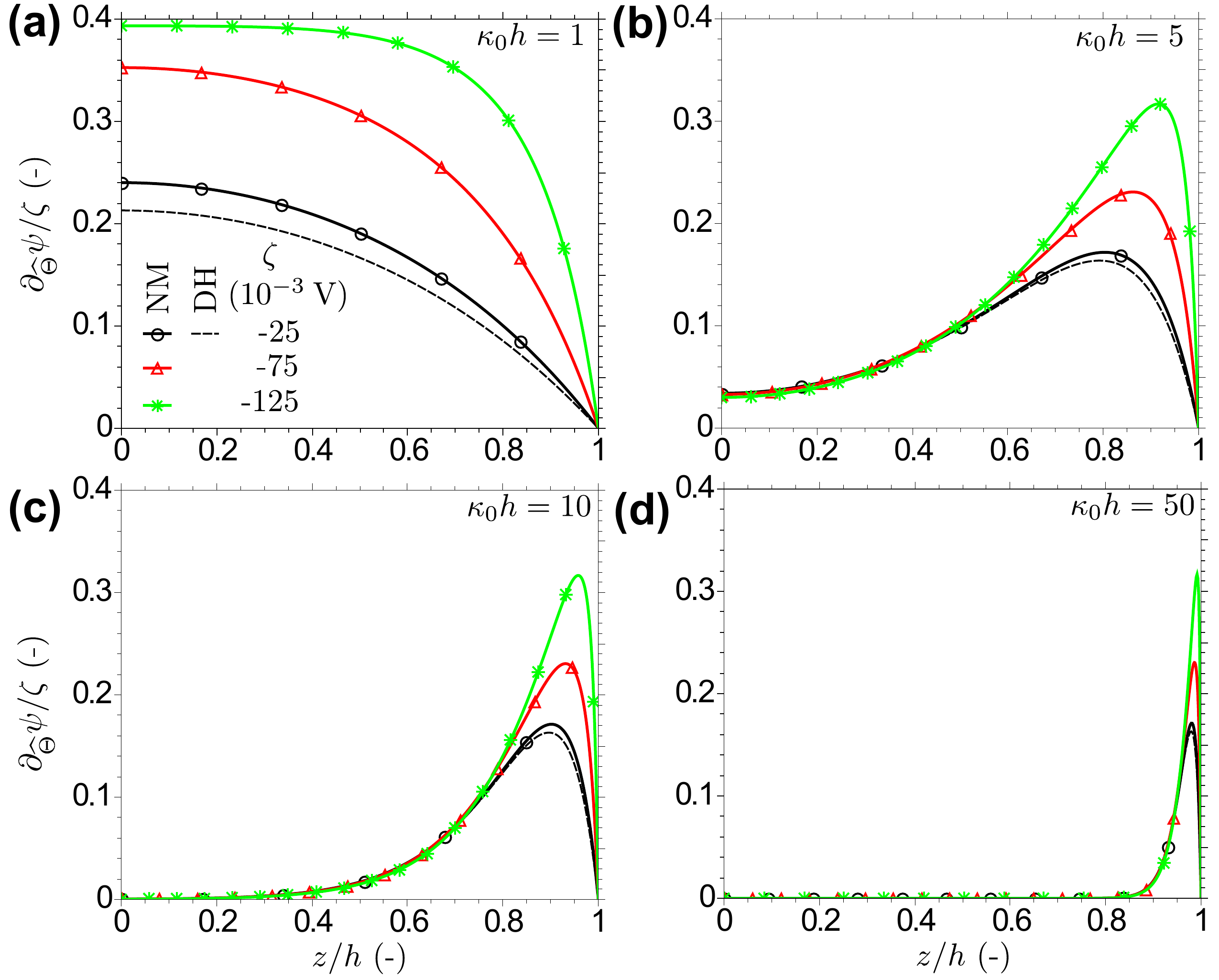}}
  \caption{Local change of EDL potential with temperature expressed by 
	$\partial_{\widehat{\Theta}} \Psi(Z)/\overline{\zeta} \equiv \partial_{\widehat{\Theta}} \psi(Z)/\zeta$ for 
	$\zeta = [-25,-75,-125] \cdot 10^{-3}\:\textrm{V}$. In all plots, thermophoretic ion motion is considered with equal 
	intrinsic Soret coefficients for each ion species (Soret A, $S_0 = 5 \cdot 10^{-3}\:\textrm{K}^{-1}$). A temperature-dependent 
	ion mobility and a temperature dependent dielectric permittivity is included ($M_0 = -5.1 \cdot 10^{-3}\:\textrm{K}^{-1}$) as well. 
	The temperature difference equals $\Delta T = 25\:\textrm{K}$, while the local temperature used for evaluation is $T = T_0 + \Delta T$. 
	Numerical solutions ('NM', bold lines with symbols) were computed by means of (\ref{Eq:ddThetaPsi_2}) and compared for the lowest 
	$\zeta$ potential with those obtained within the Debye-H\"uckel approximation ('DH, thin dashed lines) and given by 
	(\ref{Eq:ddThetaPsi_3}). In the plots (a)-(d), the nominal Debye parameter is varied  according to 
	$\overline{\kappa}_0 = \kappa_0 h = [1,5,10,50]$. The legend shown in (a) is valid for (b)-(d) as well.}
\label{Fig:ddThetaHat_Psi}
\end{figure}

In figure \ref{Fig:ddThetaHat_Psi}, the partial derivative 
$\partial_{\widehat{\Theta}} \Psi(Z)/\overline{\zeta} \equiv \partial_{\widehat{\Theta}} \psi(Z)/\zeta$ is plotted for 
the $\zeta$ potentials $\zeta = -25$, $-75$ and $-125 \cdot 10^{-3}\:\textrm{V}$, while the temperature difference is 
identical in each plot and equal to $\Delta T = 25\:\textrm{K}$. Thermophoretic ion motion is considered with equal intrinsic 
Soret coefficients for each ion species (Soret A, $S_0 = 5 \cdot 10^{-3}\:\textrm{K}^{-1}$); a temperature-dependent ion 
electromobility and a temperature-dependent dielectric permittivity is included ($M_0 = -5.1 \cdot 10^{-3}\:\textrm{K}^{-1}$) 
as well. The local temperature used for evaluation equals $T = T_0 + \Delta T$. The nominal Debye parameter is set either to 
$\overline{\kappa}_0 = \kappa_0 h = 1$ [shown in (a)], $5$ [part (b)], $10$ [part (c)] or equal to $50$ [part (d)]. The legend 
depicted in (a) is valid for all plots (a)-(d). Based on the procedure described in \S \ref{subsec:axial_flow_SoretEDL}, 
the numerical solutions ('NM', bold lines with symbols) were computed with (\ref{Eq:ddThetaPsi_2}), where the 
numerical evaluation of $\partial_{\overline{\kappa}} \widetilde{\Psi}$ and 
$\partial_{\widetilde{\overline{\zeta}}} \widetilde{\Psi}$, respectively, employed a 5-point stencil finite-difference (FD) 
scheme. This scheme is accurate to third order in $\Delta \widehat{\Theta}$ \citep{Fletcher:Springer1991}. After a grid 
independence study, $\Delta \overline{\kappa} = \Delta \widetilde{\overline{\zeta}} = 10^{-3}$ was chosen, but values up to 
${\cal O}(10^{-1})$ and higher give practically indistinguishable results. To obtain the needed functions 
$\widetilde{\Psi}$ for each $\overline{\kappa}_w$, respectively for each $\widetilde{\overline{\zeta}}_w$, (\ref{Eq:Poisson_1D}) 
was solved by collocation with the BVP4C-function implemented in Matlab (see previous discussion of figure \ref{Fig:PsifZ}). 
In figure \ref{Fig:ddThetaHat_Psi}, the numerical solutions for the lowest $\zeta$ potential are compared to those obtained 
within the Debye-H\"uckel approximation ('DH, thin dashed lines) and given by (\ref{Eq:ddThetaPsi_3}), indicating good 
agreement for the case of non-overlapping EDLs.

For channels without EDL overlap ($\overline{\kappa}_0 \gtrsim 5$), $\partial_{\widehat{\Theta}} \Psi/\overline{\zeta}$ vanishes 
further away from the EDL, particularly along the channel center plane (see plots (c) and (d)). For any value of the $\zeta$ potential, 
the maximum of $\partial_{\widehat{\Theta}} \Psi/\overline{\zeta}$ occurs along the channel center plane if $\overline{\kappa}_0 = 1$, 
while for the larger $\overline{\kappa}_0$ values considered, it occurs at $Z \gtrsim 0.8$. Along the channel wall, 
$\partial_{\widehat{\Theta}} \Psi/\overline{\zeta}$ is zero for all cases due to the given $\zeta$ potential. 
Hence, the thermal modification of the EDL potential described by (\ref{Eq:ddThetaPsi_2}), respectively by (\ref{Eq:ddThetaPsi_3}), 
indeed leads to an axial electric field, which is restricted to the EDL only.

\section{Verification of propulsion by the EDL potential}\label{sec:app_streamfct}

In the main text it is shown that -under the sole presence of a thermal gradient- a thermoosmotic velocity emerges 
in the channel, which under confinement contributes to the thermoelectric potential, even if the thermodiffusive 
mobilities of the ion species are identical (Soret A). Essentially, this is caused by an axial gradient of the EDL potential which propels 
the fluid by means of the electro-hydrostatic pressure and the electrostatic body force (Maxwell stress). For an isothermal electrokinetic 
system, it is well-known that the EDL itself does not set the fluid into motion. It was shown by \citet{Levich:Prentice1962}, page 484, that, 
for such systems, the electric body force due to the EDL and the osmotic pressure contribution exactly cancel each other and that 
only the externally applied electric field is relevant for the fluid propulsion \citep{Pascall:JFM2011}. Therefore, the phenomenon 
described in this work may be suspected to be an artifact of the lubrication approximation (LA) used in the analysis. In the following, 
without relying on the LA, it is shown that a non-isothermal EDL can indeed propel the fluid (while an isothermal EDL cannot).

According to \citet{Fitts:McGrawHill1962}, page 43, the incompressible Newtonian Navier-Stokes equation can 
be written as
\begin{equation}
\label{Eq:Stokes_Fitts} \rho d_t \boldsymbol{v} - \eta \nabla^2 \boldsymbol{v} = -\bnabla p + \rho_0 \boldsymbol{Y},
\end{equation}
where $\boldsymbol{Y}$ denotes the sum of all body forces. The RHS of (\ref{Eq:Stokes_Fitts}) 
can be expressed by the general form of the Gibbs-Duhem equation for a multicomponent system (\citet{Fitts:McGrawHill1962}, page 44)
\begin{equation}
\label{Eq:Gibbs-Duhem} \bnabla p - \rho \boldsymbol{Y} = \sum^P_{k=1} \rho_k \bnabla^{(T)} \mu'_k,
\end{equation}
The mass of component $k$ per total Volume $V$ is denoted by $\rho_k$. The sum contains all components $P = K + 1$, 
where $K$ is the number of solutes. The spatial gradient at constant temperature of the overall chemical potential 
is defined according to
\begin{equation}
\label{Eq:grad_chemppot_isoth} \bnabla^{(T)} \mu'_k = \bnabla \mu'_k - \partial_T \left(\mu_k\right)_{|p} \bnabla T,
\end{equation}
where $\mu_k = \mu'_k - \nu_k F/M_k \phi$ ($M_k$ is the molar mass, $F = e N_A$ is the Faraday-constant with $N_A$ as the Avogadro number). 
It is assumed that the gradient of the chemical potential of the solvent at constant temperature is negligibly small so that 
the effective upper limit of the sum in (\ref{Eq:Gibbs-Duhem}) is $K$. No external potential is applied herein and $\phi \equiv \psi$. 
The Nernst-Planck equations for the ion concentrations $n_k$ ($k=1,..,K$) read $d_t n_k = -\bnabla \bcdot \boldsymbol{j}_{k}$, where in agreement with 
(\ref{Eq:NPE_dim}) the diffusion flux density is given by
\begin{equation} 
\label{Eq:diff_flux_dens_app} \boldsymbol{j}_k = -D_k\bnabla n_k - n_k D_{T,k}\bnabla{T} - e\nu_k n_k \omega_k\bnabla \psi.
\end{equation}
In the latter equation, $D_k$ and $D_{T,k}$ are the Fickian and the thermal diffusion coefficients, respectively, 
of the ion species $k$, while $\omega_k = D_k/(k_\textrm{B} T)$ are the electrophoretic mobilities. On the one hand, a vanishing flux 
density, indicating chemical equilibrium, is identical to
\begin{equation} 
\label{Eq:zero_flux_dens_app} e\nu_k \bnabla \psi + k_\textrm{B} T \bnabla \left[\textrm{ln}(n_k) + S_k T\right] = 0,
\end{equation}
where $S_k = D_{T,k}/D_k$. On the other hand, for a chemical potential $\mu^*_k = \mu^*_k(\psi,n_k,T)$ 
per number of ions, chemical equilibrium implies that
\begin{equation} 
\label{Eq:differential_chempot_app} \bnabla \mu^*_k = (\partial_\psi \mu^*_k)_{|n_k,T} \bnabla \psi 
+ (\partial_{n_k} \mu^*_k)_{|\psi,T} \bnabla n_k + (\partial_T \mu^*_k)_{|\psi,n_k} \bnabla T = 0,
\end{equation}
Expression (\ref{Eq:zero_flux_dens_app}) can be derived from (\ref{Eq:differential_chempot_app}) if $\mu^*_k$ is 
defined according to
\begin{equation} 
\label{Eq:chempot_app} \mu^*_k = e\nu_k \psi + k_\textrm{B} T \left[\textrm{ln}(n_k) + S_k T\right].
\end{equation}
With the mass-specific chemical potential $\mu'_k = N_A \mu^*_k/M_k$ and (\ref{Eq:grad_chemppot_isoth}) one finds
\begin{equation}
\label{Eq:grad_chemppot_isoth2} \bnabla^{(T)} \mu'_k = \frac{\nu_k F}{M_k} \bnabla \psi + \frac{k_\textrm{B} T N_A}{M_k} \frac{\bnabla n_k}{n_k}.
\end{equation}
Employing $n_k = \rho_k N_A/M_k$ and $\sum^P_{k=1} e \nu_k n_k = \sum^K_{k=1} e \nu_k n_k = \rho_f$ (for the neutral solvent, $\nu_P = 0$) 
yields
\begin{equation}
\label{Eq:Gibbs-Duhem2} \sum^K_{k=1} \rho_k \bnabla^{(T)} \mu'_k = \rho_f \bnabla \psi + k_\textrm{B} T \bnabla n,
\end{equation}
where $n = \sum^K_{k=1} n_k$. The last term on the RHS is the gradient in osmotic pressure, i.e. $\bnabla p_\textrm{osm} = k_\textrm{B} T \bnabla n$. 
At constant temperature, the RHS of (\ref{Eq:Gibbs-Duhem2}) is identical to zero and the system 
is in mechanical equilibrium \citep{Levich:Prentice1962, Squires:JFM2004}.

In the present non-isothermal case, for Soret A ($S_- = S_+ = \overline{S} \approx S_0$, $\Delta S = 0$), 
according to expressions (\ref{Eq:SalinitySoret}) and (\ref{Eq:charge_dens}), the charge density can be replaced by
\begin{equation} 
\label{Eq:charge_dens_SoretA_app} \frac{\rho_f}{e \nu n_0} = -2 N \textrm{sinh}(\widetilde{\Psi}).
\end{equation}
while by summation of (\ref{Eq:ion_distribution}), one finds
\begin{equation} 
\label{Eq:number_concentration_app} \frac{n}{n_0} = 2 N \textrm{cosh}(\widetilde{\Psi}).
\end{equation}
Hence, with the definitions $\Psi = \psi e \nu/(k_\textrm{B} T_0)$ and $\widetilde{\Psi} = \Psi/(1+\widehat{\Theta})$, one has 
$\bnabla \widetilde{\Psi} = \bnabla \Psi/(1+\widehat{\Theta}) - \widetilde{\Psi} \bnabla \widehat{\Theta}/(1+\widehat{\Theta})$. 
With this, (\ref{Eq:Gibbs-Duhem2}) can be evaluated to read
\begin{equation}
\label{Eq:Gibbs-Duhem3} \sum^K_{k=1} \rho_k \bnabla^{(T)} \mu'_k = 2 k_\textrm{B} T_0 n_0 
\left[(1+\widehat{\Theta})\textrm{cosh}(\widetilde{\Psi}) \bnabla N 
- N \widetilde{\Psi} \textrm{sinh}(\widetilde{\Psi}) \bnabla \widehat{\Theta}\right].
\end{equation}
In the present study, $N$ is a function of the local temperature $\widehat{\Theta} = T/T_0$ so that both terms 
on the RHS of the latter equation differ from zero under non-isothermal conditions. This is also the case 
even if $N \equiv 1$ (no thermophoretic ion motion). The last term on the RHS of (\ref{Eq:Gibbs-Duhem3}) is due 
to a temperature-dependent ion mobility, i.e. this effect alone leads to a thermoosmotic fluid propulsion in the EDL. 
Similarly, if $\bnabla \widehat{\Theta} \equiv 0$, $\bnabla N$ is not necessarily zero, e.g. if an axial gradient 
of the salt concentration is imposed. This resembles the classical case of solvent transport in osmotic capillaries. 
Finally, the system is in mechanical equilibrium only if $\bnabla N = \bnabla \widehat{\Theta} = 0$.

The last term on the RHS of (\ref{Eq:Gibbs-Duhem3}) vanishes outside the EDL where $\widetilde{\Psi} \rightarrow 0$. 
In the same limit, $\textrm{cosh}(\widetilde{\Psi}) \rightarrow 1$ so that the RHS of (\ref{Eq:Gibbs-Duhem3}) is 
equal to $\bnabla p_\textrm{drift} = 2 k_\textrm{B} T n_0 \bnabla N$. This is the gradient in osmotic pressure of the 
ions due to the axial gradient in salt concentration (without interaction with the EDL). Herein it is assumed 
that this gradient is caused by the Soret thermodiffusion alone. This gradient is uniform across the channel 
and causes a constant drift of the ion cloud but no charge separation. Hence, the effective force the ion 
cloud exerts on the solvent can be expressed by the sum of $\bnabla p_\textrm{drift}$ and the contribution due to the 
EDL. In linearized form, the latter reads
\begin{align}
\label{Eq:eff_bodyforce_EDLcontr} -\boldsymbol{F}_\textrm{EDL} &= \rho_f \bnabla \psi + k_\textrm{B} T \bnabla n - \bnabla p_\textrm{drift} \nonumber \\
&\approx 2 k_\textrm{B} T_0 n_0 \left\{[\textrm{cosh}(\Psi^{(e)}) - 1]\bnabla N 
- \Psi^{(e)} \textrm{sinh}(\Psi^{(e)}) \bnabla \widehat{\Theta}\right\},
\end{align}
with $\Psi^{(e)}$ being the EDL potential at isothermal conditions. 

Thus, the mechanical propulsion observed in the non-isothermal case can be traced back to the circumstance that 
the divergence of the EDL Maxwell stress is not irrotational, while it appears to be under isothermal conditions. If a 
temperature-dependent ion mobility is included, the gradient of the osmotic pressure of the ion cloud itself is 
also not curl-free while it is if only a thermophoretic ion motion is considered. In this context, in the present derivations 
discussed in the main text based on the LA and unlike in other studies of electrokinetics, an expression for the 
osmotic pressure of the ion cloud is never explicitly added to the governing equations as a body force. In the LA, 
a corresponding term automatically emerges from the momentum equation in $z$ direction, (\ref{Eq:momentum_z}). 
Its combination with the electromotive force in the EDL gives, under non-isothermal conditions, a non-vanishing 
force contribution. Note that within this derivation based on (\ref{Eq:Gibbs-Duhem}), the propulsion by a 
temperature-dependent dielectric permittivity is not captured. This is simply due to the Maxwell stress 
$-\rho_f \bnabla \psi$ considered herein, which is not the full expression of the Korteweg-Helmholtz electric 
body force used in the main text.

\bibliographystyle{jfm}

\bibliography{references}

\end{document}